\documentclass[a4paper,useAMS,fleqn,usenatbib]{mnras}

\usepackage{graphicx}
\usepackage{natbib}
\usepackage{amsmath}
\usepackage{amssymb}
\usepackage{wasysym}
\usepackage{txfonts}
\usepackage[T1]{fontenc}
\usepackage{ae,aecompl}
\usepackage{cancel}
\usepackage{latexsym,longtable,epsf,ulem}
\usepackage{enumerate}
\usepackage{array, multirow, float, xcolor}

\newcommand{\EQ}{\begin{equation}}
\newcommand{\EN}{\end{equation}}
\newcommand{\kmin}{k_{\rm min}}
\newcommand{\gcm}{\,{\rm gm\,cm^{-3}}}
\newcommand{\kms}{\,{\rm km\,s^{-1}}}

\newcommand{\cmn}{\,{\rm cm^{-3}}}
\newcommand{\gyr}{\,{\rm Gyr}}
\newcommand{\sigmafd}{\sigma_{\rm FD}}
\newcommand{\xx}{\mbox{\boldmath $x$}{}}
\newcommand{\kk}{\mbox{\boldmath $k$} {}}
\newcommand{\BB}{\mbox{\boldmath $B$} {}}
\newcommand{\FF}{\mbox{\boldmath $F$} {}}
\newcommand{\mkG}{\,{\rm \umu G}}
\newcommand{\mJy}{\rm {\umu Jy}}

\newcommand{\kpc}{\, {\rm kpc}}
\newcommand{\radm}{\,{\rm rad\,m^{-2}}}

\newcommand{\ghz}{\,{\rm GHz}}
\def\Rm{\rm Rm}

\def\Rey{{\rm Re}}
\def\Pm{\rm Pm}

\title[Polarization of ICM]  
{Properties of polarized synchrotron emission from fluctuation-dynamo action -- I. Application to galaxy clusters} 
\author[S.~Sur, A.~Basu and K.~Subramanian]{Sharanya Sur,$^{1}$\thanks{E-mail: sharanya.sur@iiap.res.in (SS)}
Aritra Basu,$^{2,3,4}$\thanks{abasu@tls-tautenburg.de (AB)}
and Kandaswamy Subramanian$^{5}$\thanks{kandu@iucaa.in (KS)} \\
$^1$Indian Institute of Astrophysics, 2nd Block, Koramangala, Bangalore 560034, India\\
$^2$Fakult{\"u}t f{\"u}r Physik, Universit{\"a}t Bielefeld, Postfach 100131, D-33501 Bielefeld, Germany\\
$^3$Max-Planck-Institut f{\"u}r Radioastronomie, Auf dem H\"ugel 69, D-53121 Bonn, Germany\\
$^4$Th\"uringer Landessternwarte, Sternwarte 5, D-07778 Tautenburg, Germany\\
$^5$Inter-University Centre for Astronomy and Astrophysics, Post Bag 4, Ganeshkhind, Pune 411007, India\\
}

\begin{document}

\date{Accepted 2020 December 1. Received 2020 December 1; in original form 2020 September 11}

\pagerange{\pageref{firstpage}--\pageref{lastpage}} \pubyear{2002}

\maketitle

\label{firstpage}

\begin{abstract}
Using magnetohydrodynamic simulations of fluctuation dynamos, we 
perform broad-bandwidth synthetic observations to investigate the 
properties of polarized synchrotron emission and the role that Faraday 
rotation plays in inferring the polarized structures in the intracluster 
medium (ICM) of galaxy clusters. In the saturated state of the dynamo, 
we find a Faraday depth (FD) dispersion $\sigma_{\rm FD} \approx 100$ 
rad m$^{-2}$, in agreement with observed values in the ICM. Remarkably, 
the FD power spectrum is qualitatively similar to $M(k)/k$, where $M(k)$ 
is the magnetic spectrum and $k$ the wavenumber. However, this 
similarity is broken at high $k$ when FD is obtained by applying RM 
synthesis to polarized emission from the ICM due to poor resolution 
and complexities of spectrum in FD space. Unlike the Gaussian probability 
distribution function (PDF) obtained for FD, the PDF of the synchrotron 
intensity is lognormal. A relatively large $\sigma_{\rm FD}$ in the ICM 
gives rise to strong frequency-dependent variations of the pixel-wise mean 
and peak polarized intensities at low frequencies ($\lesssim 1.5\,{\rm GHz}$). 
The mean fractional polarization $\langle p \rangle$ obtained at the resolution 
of the simulations increases from $<0.1$ at 0.5~GHz to its intrinsic value of
$\sim0.3$ at 6~GHz. Beam smoothing significantly affects the polarization
properties below $\lesssim 1.5\,{\rm GHz}$, reducing 
$\langle p \rangle$ to $\lesssim 0.01$ at 0.5~GHz. At frequencies $\gtrsim 5\,{\rm GHz}$, 
polarization remains largely unaffected, even when recovered using RM 
synthesis. Thus, our results underline the need for high-frequency 
($\gtrsim 5\,{\rm GHz}$) observations with future radio telescopes to 
effectively probe the properties of polarized emission in the ICM. 

\end{abstract}


\section{Introduction}
\label{sec:intro}

Observations of Faraday rotation measure (RM) of polarized radio sources
located either inside or behind galaxy clusters suggest that the intracluster
medium (ICM) is magnetized. The observed fields are of $\mkG$ strength
correlated on several kiloparsec (kpc) scales \citep{CKB01,CT02,GF04, 
Kale+16, Roy+16, Kierdorf+17,weere19}. In the absence of any large-scale 
rotation, {\it fluctuation} dynamos \citep{K68, ZRS90} are ideally suited 
for amplifying dynamically insignificant seed magnetic fields to observable 
strengths, on time scales much shorter than the age of the cluster. 
There is also observational evidence for subsonic turbulence with velocities 
of order of a few hundred km\,s$^{-1}$, conditions that are required for such 
dynamo action, from measurements of pressure fluctuations in X-ray emission 
\citep{Schu04,Churazov+12,Zhu2019} and widths of X-ray lines 
\citep{SFS11,SF13,Hitomi18a}.  

While RM provides information about the line-of-sight (LOS) component of 
the field, synchrotron emission and its polarization are the other two 
complimentary observables that furnish information about the magnetic field 
in the plane of the sky. The observed emission is partially linearly polarized, 
and the Stokes parameters $I, Q$ and $U$ at GHz frequencies can be readily 
measured by a radio telescope. However, due to a combination of low 
surface brightness, Faraday depolarization, and steep radio continuum 
spectra of the ICM, detecting polarized emission from radio halos has so far 
been difficult \citep{Vacca+10,Govoni+13}, being observed only in bright 
filaments of galaxy clusters A2255 \citep{Govoni+05}, MACS J0717+1345 
\citep{Bonafede+09} and in relics \citep{Kierdorf+17}. Detailed analysis further 
shows that Stokes parameters in turn are related to the magnetic field 
components in a non-linear fashion \citep{WSE09} and are thus sensitive to 
the structure of these fields in the ICM. Early attempts to compare the 
Faraday RM and depolarization of cluster radio sources derived from 
numerical models with the observed data relied on assumption that cluster 
magnetic fields are Gaussian. This allowed the power spectrum of the field 
to be expressed in a simple power law form \citep{Murgia+04,Laing+08, B+10,Vacca+10}.  
However, these assumptions are in contrast to two important characteristics 
of a fluctuation-dynamo generated field, namely, their spatially intermittent 
nature, and that the field components are non-Gaussian \citep{HBD04,
Schek04,BS05,Vazza+18,Seta+20}. Therefore, to make a meaningful 
comparison between theoretical predictions and observations, it is only 
logical to explore and extract information about the Faraday RM, synchrotron 
emissivity, and polarization signals directly from numerical simulations of 
fluctuation dynamos. 

In view of the above arguments, we focus on probing the properties of 
polarized emission from magnetic fields that are self-consistently generated 
by the fluctuation dynamo in the context of galaxy clusters. Consequently, 
we make use of the simulation data from a run reported in \citet{S19} 
where the steady-state turbulent velocity is subsonic. The key questions 
that we intend to address here concern the Faraday depth (FD), how one 
can relate the power spectrum of FD to that of the magnetic field, the 
statistical nature of the total and polarized synchrotron emission, and how 
these are affected by frequency dependent Faraday depolarization. 
Any realistic comparison between synthetic and real astronomical 
observations must take into account the effects of the finite resolution of a 
radio telescope. With this in mind, we also aim to understand the effects 
of beam smoothing on the polarized intensity, the fractional polarization 
as well as on the Stokes $Q$ and $U$ parameters. As will become clear 
in the subsequent sections, our analysis allows us to draw certain 
unambiguous inferences and make predictions for upcoming new 
generation of radio telescopes. 

The paper is organized as follows. In Section~\ref{sec:simsetup}, we 
discuss in brief the initial conditions and the set-up of the simulation, 
whose data we analyze here. Using the simulation data as an input, 
we next outline the basic methodology used to perform synthetic 
observations in Section~\ref{sec:synth_obs}. Thereafter, Section~\ref{sec:results} 
deals with the results obtained from our study, spread over different 
subsections. Starting with a brief discussion on the characteristics of 
turbulence and dynamo-generated field as evidenced from their power 
spectra, we focus on a number of topics dealing with the Faraday 
depth, the structure and probability distribution function (PDF) of total 
synchrotron intensity, and the frequency dependence of the polarization 
parameters. In Section~\ref{sec:smooth_pol} we discuss the effects 
of beam smoothing on the properties of the polarized emission 
followed by a discussion on FD and fractional polarization computed 
using the technique of RM synthesis in Section~\ref{sec:rmsynth}.
Finally, we conclude with a summary of the important findings of our 
work in Section~\ref{sec:conc} and discuss the relevance of our 
results in the context of future radio observations and possible 
future research directions in Section~\ref{sec:discuss}.

\section{Simulation Set-up} \label{sec:simsetup} 

We use the data from a non-helically driven fluctuation-dynamo simulation 
that was reported earlier in \citet{S19} using the publicly available compressible
MHD code {\scriptsize FLASH}\footnote{\url{http://flash.uchicago.edu/site/flashcode/}} 
\citep{Fry+00}. The magnetohydrodynamics (MHD) equations are solved in 
a three-dimensional box of unit length having $512^{3}$ grid points with 
periodic boundary conditions. The equations include explicit viscosity and 
resistivity.  An isothermal equation of state is adopted with the initial density 
and sound speed set to unity. In {\scriptsize FLASH}, turbulence is driven as 
a stochastic Ornstein-Uhlenbeck (OU) process with a finite time correlation 
\citep{EP88, Benzi+08} through a forcing term ($\FF$) in the Navier-Stokes 
equation. To maximize the efficiency of the dynamo, we use only solenoidal 
modes {(i.e. $\nabla\cdot\FF = 0$)} for the turbulent driving. In the run 
presented here, we have chosen the forcing wavenumber of driven turbulence 
to be between $1 \leq |\kk|\, L/2\,\upi \leq 3$ such that the average forcing 
wavenumber $k_{\rm f}\,L/2\,\upi \sim 2$, where $L$ is the length of the box. 
The strength of the solenoidal driving is adjusted so that the resulting root 
mean square (rms) Mach number of the flow 
$\mathcal{M} = u_{\rm rms}/c_{\rm s} \approx 0.18$ (subsonic), where 
$u_{\rm rms}$ is the turbulent rms velocity and $c_{\rm s}$ is the isothermal 
sound speed. In our simulations, turbulence is fully developed by about a 
two-eddy turnover time. The magnetic Reynolds number $\Rm$ based on the 
forcing scale $l_{\rm f} = 2\pi/k_{\rm f}$ is $1080$ and the magnetic Prandtl 
number $\Pm =1$. Magnetic fields are initialized as weak seed fields of the 
form $\BB = B_{0}[0,0,\sin(10\,\upi\,x)]$ with the amplitude $B_{0}$ adjusted 
to a value such that the plasma $\beta\sim 10^{6}$. 
Further, to maintain the condition $\nabla \cdot \BB = 0$ to machine precision 
level, we use the unsplit staggered mesh algorithm in {\scriptsize FLASH} with 
a constrained transport scheme \citep{LD09, Lee13} and Harten-Lax-van 
Leer-Discontinuities (HLLD) Riemann solver \citep{MK05}. The simulation is 
run till one obtains many realizations of the saturated state of the fluctuation 
dynamo for our analysis. At saturation, the rms value of the magnetic field is 
$b_{\rm rms} = 0.08\,c_{\rm s}$. Table~\ref{sumsim} highlights the important 
dimensionless parameters of the run.

\begin{table}
\centering 
\setlength{\tabcolsep}{8.0 pt}
\caption{Key parameters of the subsonic simulation used in this study. 
$\mathcal{M}$ and $b_{\rm rms}$ are the average value of the 
rms Mach number and the magnetic field obtained in the steady state.}
\begin{tabular}{cccccc} \hline
$N^{3}$ & $k_{\rm f}\,L/2\upi$ & $\mathcal{M}$ & $b_{\rm rms}$ & $\Pm$ & 
$\Rey = u\,l_{\rm f}/\nu$ \\ \hline
$512^{3}$ & 2.0 & $\approx 0.18$ & $\approx 0.08$ & 1 & 1080 \\  \hline        
\end{tabular}
\label{sumsim}
\vspace{-0.78em}
\end{table}

Although the simulation used here is in terms of dimensionless variables, to
make connections to observations it is imperative to express the relevant
length and time-scales together with the values of the physical variables
obtained from the simulation in terms of characteristic values typical of the 
ICM. In this spirit we first renormalize the length of the simulation domain to 
$L = 512\kpc$ in each dimension, which implies a resolution of 
$\Delta x = \Delta y = \Delta z = 1\kpc$. Assuming that the underlying 
turbulence has resulted from previous episodes of mergers, the scale of 
turbulent motions for the above domain size is $l_{\rm f} = 2\upi/k_{\rm f} = 256\kpc$. 
Next, we assume $\langle n_{\rm e}\rangle = 10^{-3}\cmn$ and 
$c_{\rm s} = 10^{3}\kms$ as the typical mean number density of free 
thermal electrons and sound speed of the ICM, respectively \citep{Sar88}. 
For $\mathcal{M} \approx 0.18$, this implies a turbulent rms velocity 
$u_{\rm rms} \approx 180\kms$ with an eddy turnover time, 
$t_{\rm ed} = l_{\rm f}/u_{\rm rms} \approx 1.38\gyr$ on the forcing scale. 
Considering the fact that the ICM is fully ionized, the mean mass density 
$\langle \rho\rangle = \langle n_{\rm e}\rangle\,\mu_{\rm e}\,m_{p} \approx 1.97\times 10^{-27}\gcm$, 
where $\mu_{\rm e} = 1.18$ is the mean molecular weight per free electron.
Thus, our simulation domain can be thought of as a local patch of the ICM. 
We further define the unit of the magnetic field strength as 
$\sqrt{4\,\upi\,\rho\, c_{\rm s}^{2}}$. Using the above mentioned values of 
$\rho$ and $c_{\rm s}$ we obtain $b_{\rm unit} \approx 15.7\,{\umu\rm G}$, 
suggesting that the steady value of $b_{\rm rms} \approx 0.08\,c_{\rm s}$ 
in the subsonic case corresponds to $\approx 1.3\,\mkG$. If the magnetic 
and turbulent energy densities are in equipartition, the equipartition 
magnetic field strength 
$B_{\rm eq} = \sqrt{4\,\upi\,\rho\, u_{\rm rms}^{2}} \approx 2.8\,\mkG$.  
Thus, $b_{\rm rms} \sim B_{\rm eq}/2$ in our simulation. Furthermore, 
using the above values of the mass density and sound speed, the initial 
$\beta \sim 10^{6}$ implies an initial magnetic field strength 
$B_{0}\approx 22.2\,{\rm nG}$.

\section{Synthetic observations} 
\label{sec:synth_obs}

In order to address the aims of this paper we use the data obtained from 
the run as input to compute a variety of observables that characterizes 
the nature of the polarized emission observed in the ICM. For this purpose, 
we perform synthetic observations using the {\scriptsize COSMIC} package 
developed by \citet{basu19b}. Depending on the type of simulations and the 
ancillary data, {\scriptsize COSMIC} allows a user to freely choose from 
different methods for computing 2D maps of Faraday depth (FD) and total 
synchrotron intensity ($I_{\rm sync}$), and Stokes~$Q$ and $U$ parameters 
at user-specified frequencies. Here, we have used {\scriptsize COSMIC} to 
generate synthetic observations between 0.5 and 6~GHz divided into 1024 
spectral channels. In the later sections, we will discuss our results at three 
representative frequencies of 0.5, 1 and 6~GHz. These frequencies are 
chosen to gain insight into what can be expected from observations using the 
Square Kilometre Array's (SKA)\footnote{https://www.skatelescope.org/} 
LOW and MID frequency components.

The MHD simulation outputs dimensionless values of the mass density $\rho(\xx)$, 
where $\xx$ is the three-dimensional position vector. For the purpose of our 
analysis, this needs to be expressed in terms of the electron number density in 
$\cmn$ to compute the Faraday depth. In {\scriptsize COSMIC}, we achieve 
this by computing the local $n_{\rm e}(\xx) = \rho(\xx)/\mu_{\rm e}\,m_{p}$.
Similarly, the dimensionless values of the components of the magnetic field
obtained from the run are expressed in gauss by scaling them with the unit of
magnetic field strength, $b_{\rm unit}$. We did not include any cosmic rays in 
our simulations. However, computing the total synchrotron emission and the 
Stokes $Q$ and $U$ parameters of the linearly polarized emission depends 
on the number density of cosmic ray electrons, $n_{\rm CRE}$.  
Here, we assume that $n_{\rm CRE}$ is constant at each mesh point and 
that the cosmic ray electrons (CREs) follow a power-law energy spectrum, 
$n_{\rm CRE}(E)\,{\rm d}E = n_0\,E^{\gamma}\,{\rm d}E$, where, $n_{\rm CRE}(E)$ 
is the number density of CREs in the energy range $E$ and $E+{\rm d}E$, 
$\gamma = -3$ is the constant energy index in all mesh points, and the 
normalization $n_0$ is chosen such that the simulated volume gives rise to 
a total synchrotron flux density of 1~Jy at 1~GHz \citep[see][for details]{basu19b}. 
The chosen flux density is similar to that observed for the Coma cluster 
\citep{weere19}. Thus, the total synchrotron intensity of the medium has 
a frequency spectrum given by $S(\nu) = 1\,\textrm{Jy}\,(\nu/\nu_0)^\alpha$, 
where $\nu_0 = 1\ghz$. For our calculations, $\gamma=-3$ corresponds 
to spectral index of synchrotron emission $\alpha =-1$, typical of the 
observed radio spectra in galaxy clusters \citep{Fer+12}. Details of the 
numerical calculations are provided in Appendix~\ref{sec:Appcosmic}. 
Note that all values of flux densities, and corresponding sensitivities, 
computed in this work can be scaled depending on the flux density at 
$\nu_0$ and $\alpha$ that are representative of another galaxy cluster.

We note in passing that, due to synchrotron and inverse-Compton (IC) 
cooling, CREs in the diffuse regions undergo energy-dependent losses, 
which results in steepening of the integrated radio continuum spectrum 
of galaxy clusters towards higher frequencies.  This implies $n_0$ should 
vary with frequency, especially in regions where diffusive shock acceleration 
is weak. However, since we are mainly interested in the effect of Faraday 
rotation on the structural properties of the polarized synchrotron emission, 
we do not consider steepening of the radio continuum spectrum, which 
will require detailed treatment of diffusion-loss equation for CREs in the 
MHD simulations. Furthermore, we have not added noise from telescope 
and/or source confusion, and systematics which can arise when observing 
over large bandwidths or with incomplete $u\textrm{--}v$ coverages. 
Table~\ref{tab:pars} provides a summary of the important physical parameters 
of the simulated volume and of the synthetic observations.
\begin{table} \centering
\caption{Summary of physical parameters of the simulated medium and of the 
synthetic observations.}
\begin{tabular}{cc} \hline 
Parameter name & Value \\ \hline 
Mean electron density & $\langle n_{\rm e}\rangle = 10^{-3}\cmn$ \\
Isothermal sound speed & $c_{\rm s}= 10^{3}\kms$ \\ 
Turbulent rms velocity & $u_{\rm rms} \approx 180\kms$ \\
Unit of field strength & $b_{\rm unit} \approx 15.7\,\mkG$ \\ 
rms field strength & $b_{\rm rms} \approx 1.3\mkG$ \\
Equipartition field strength & $B_{\rm eq} \approx 2.8\mkG$ \\ 
Box size & $512\times512\times 512\kpc^{3}$ \\ 
Turbulence driving scale & $256\kpc$ \\
Mesh resolution & $1\times1\times 1\kpc^{3}$ \\
Spectral index & $\alpha = -1$ \\ 
Spectral curvature & None \\
Frequency range & $\nu_{\rm min}=0.5\ghz$, $\nu_{\rm max}=6\ghz$ \\
Number of channels & $n_{\rm chan}=1024$ \\
Total flux density & 1~Jy at $1\ghz$ \\ \hline 
\end{tabular}
\label{tab:pars}
\end{table}

\section{Results} 
\label{sec:results}

\begin{figure}
{\mbox{\includegraphics[width=\columnwidth]{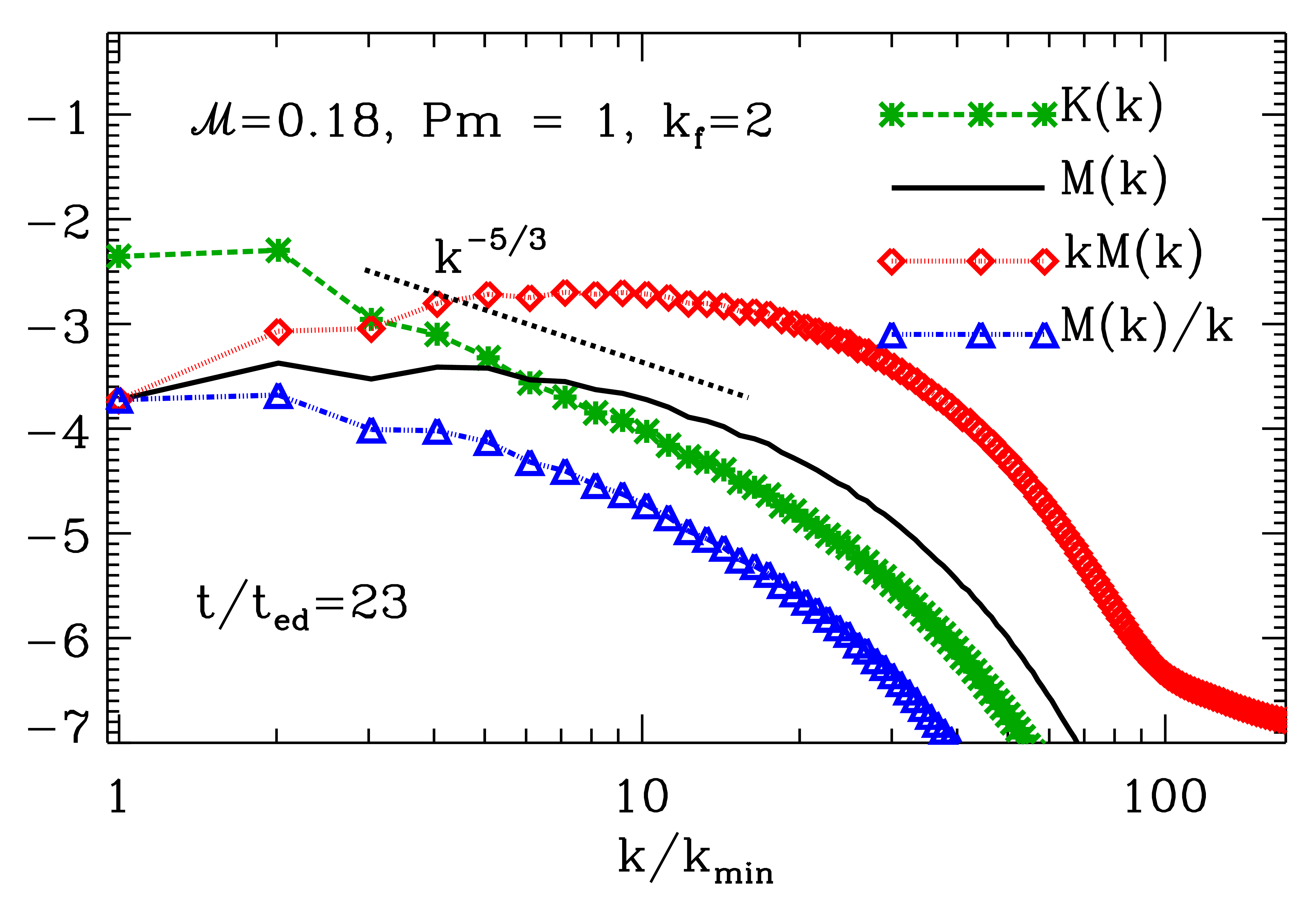}}}
\caption{Power spectra of kinetic energy, $K(k)$ (green, dashed); magnetic
energy, $M(k)$ (black, solid); $k\,M(k)$ (red, dotted); and $M(k)/k$ (blue,
dashed-dotted) from a snapshot in the saturated state at $t/t_{\rm ed} = 23$. 
The black dotted line shows the $k^{-5/3}$ slope for comparison. Here, 
the wavenumber is normalized in units of $\kmin = 2\upi/L$.}
\label{fig:spec_sub}
\end{figure}

In this section we present the results obtained by applying 
{\scriptsize COSMIC} to the data obtained from our run where 
the steady-state rms $\mathcal{M} \approx 0.18$. In addition 
to the time-series data of various physical variables, we also 
output three-dimensional snapshots at user-defined intervals 
of time in both the kinematic and saturated phases of the dynamo. 
These snapshots contain information about the three components 
of the velocity and magnetic field and the gas density on a $512^{3}$ 
Cartesian grid. In this work, we focus on the saturated phase of the 
dynamo at a time $t/t_{\rm ed} = 23$, as representative of the 
physical conditions in this phase. For comparison, we have also 
used a snapshot during the kinematic stage $t/t_{\rm ed} = 2$, 
and several additional snapshots in the saturated stage separated 
by least one $t_{\rm ed}$ to determine the statistical robustness of 
our results. For the purpose of our analysis, we choose the $x$- 
and $y$-axes to be in the plane of the sky and $z-$axis along the 
LOS. Thus, $B_x$ and $B_y$ gives rise to the polarized synchrotron 
emission and the magnetic field strength in the plane of the sky is 
$B_\perp = (B_x^2 + B_y^2)^{1/2}$. The magnetic field component 
parallel to the LOS is given by $B_\| = B_z$ and is thus responsible 
for Faraday rotation and frequency-dependent Faraday depolarization. 

Before we delve into the details of the Faraday depth and the associated
properties of the polarized emission, we briefly assess the characteristics 
of the turbulence and the dynamo-generated magnetic fields obtained 
from the simulation. We show in Fig.~\ref{fig:spec_sub} the power spectra 
of (a) kinetic energy, $K(k)$ (green dashed with asterisks); (b) magnetic 
energy, $M(k)$ (black solid); (c) spectra of $k\, M(k)$ (red, dotted with 
diamonds), i.e. the largest energy carrying scale of the field; and (d) that 
of $M(k)/k$ (blue, dash-dotted with triangles) from a snapshot in the 
saturated phase of the dynamo. Here, $k$ is the wave number. It is evident 
from the plot that $M(k)$ exceeds $K(k)$ on all but the largest scales. 
The peak of $M(k)$ lies at $\sim 1/4 - 1/6$-th of the box size corresponding 
to physical scales of $\sim 128 - 85\kpc$, while the peak of $k\, M(k)$ 
occurs on scales even smaller than that of $M(k)$ (at $\sim 51\kpc$).  
On the other hand, the peak of $M(k)/k$ occurs on a scale similar to the 
forcing scale of $K(k)$. In the next subsection, we focus on the Faraday 
depth, and discuss the qualitative similarity of its spectrum to the spectrum 
of $M(k)/k$. 

\begin{figure}
\centering
{\mbox{\includegraphics[width=\columnwidth]{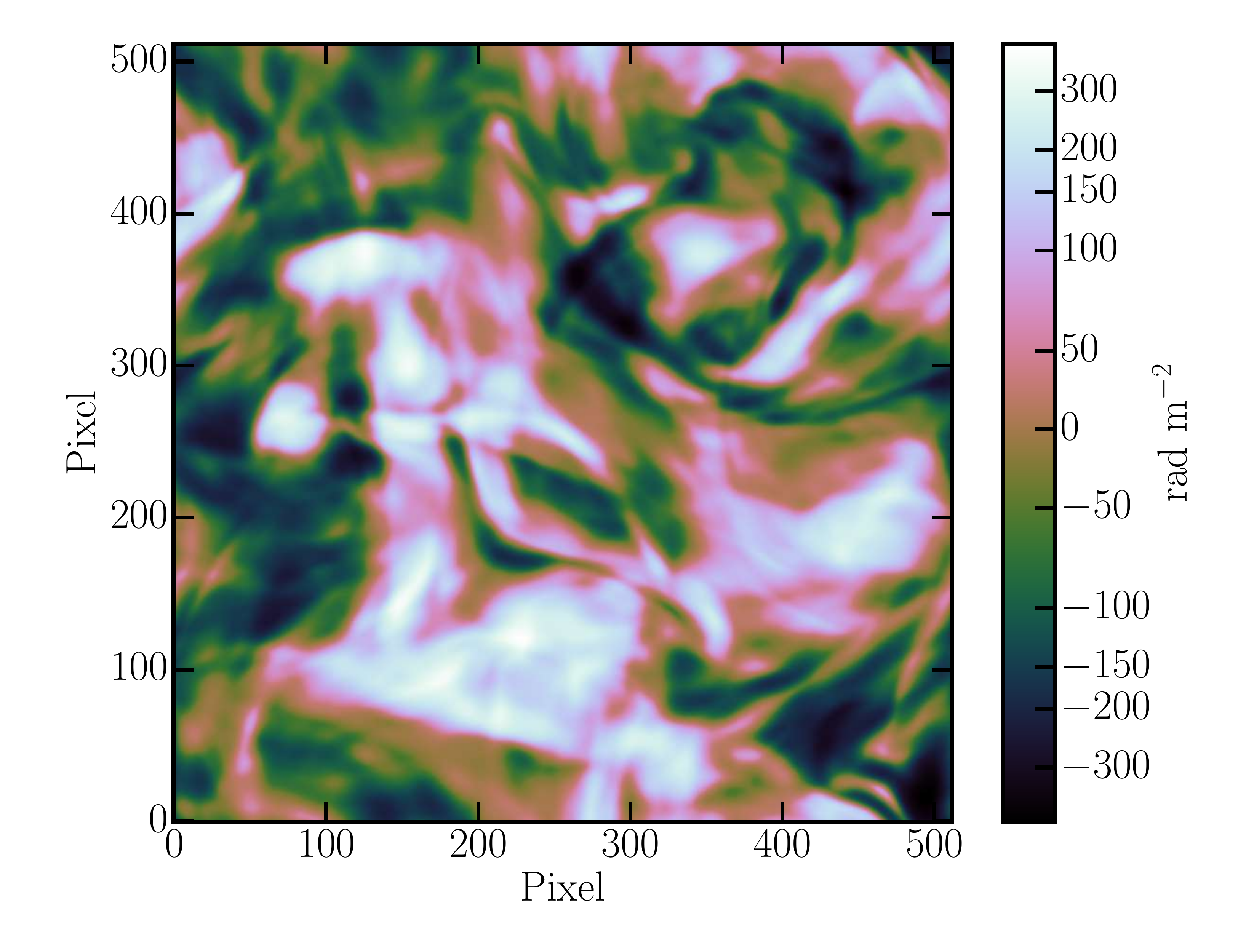}}}\\
{\mbox{\includegraphics[width=8.1cm]{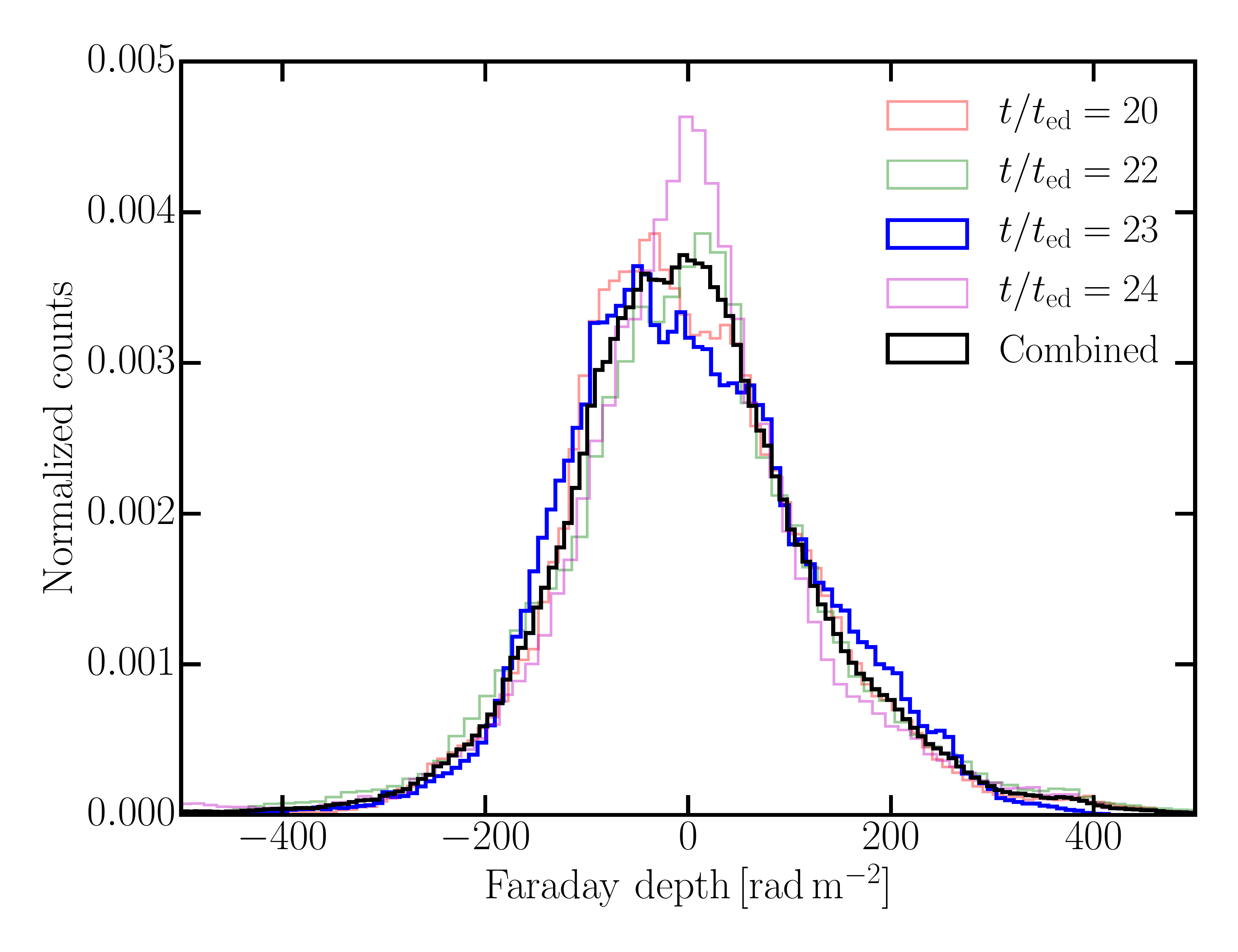}}}
\caption{Top: Faraday depth map at $t/t_{\rm ed} = 23$. 
Bottom: Distributions of Faraday depth for different snapshots in time 
in the saturated stage. Each snapshot is separated by at least one
eddy turnover time and shows the general stability of the statistics for 
$t/t_{\rm ed} = 23$, the highlighted blue histogram corresponding to the
top panel, used for analysis in this work. The black histogram shows the 
combined distribution for all the runs.}
\label{fig:fd_subsonic}
\end{figure}

\subsection{Faraday depth} 
\label{sec:fd}

Much of what we know about magnetic fields in the ICM is derived from
observations of the Faraday depth (FD) \citep{CKB01, VE03, B+10, B+15, BCK16}:
\EQ
\label{FD}
{\rm FD} = K \int n_{\rm e}\,B_\|\,dl \,, 
\EN
where the integration is along the LOS from the source to the observer, with 
$K = 0.812$ rad\,m$^{-2}$\,cm$^{3}$\,$\mkG^{-1}$\,pc$^{-1}$.
Following the procedure outlined in Section \ref{sec:synth_obs}, the mean 
number density of thermal electrons in the volume is 
$\langle n_{\rm e} \rangle = 10^{-3}\cmn$. Because of the incompressible 
nature of our run, we find over densities 
$\delta\,n_{\rm e}/n_{\rm e} \sim \mathcal{M}^2 \sim 3$ per cent 
for $\mathcal{M} \approx 0.18$, implying that $n_{\rm e}$ is roughly 
uniformly distributed in the simulated volume. Therefore, the Faraday depth 
and depolarization studied here predominantly originate from fluctuations 
in the magnetic field. The top panel in Fig.~\ref{fig:fd_subsonic} shows the 
FD map in the plane of the sky obtained from a snapshot in the saturated 
state of the dynamo at $t/t_{\rm ed}=23$. The solid blue histogram in the 
bottom panel of Fig.~\ref{fig:fd_subsonic} shows the corresponding PDF 
of FD. The detailed procedure to compute the FD maps from the simulation 
data is described in Appendix A and in \citet{basu19b}. We find that the 
value of FD lies in the range $-438$ to $+415\radm$ with a mean of almost 
zero and a dispersion, $\sigmafd \approx 118\radm$. This relatively large 
dispersion in FD is a consequence of fluctuations in the magnetic field 
component along the LOS.

In order to account for the fact that the above snapshot is not special in 
any way, we also computed $\sigmafd$ for three additional snapshots at
different times in the saturated state. The PDFs of the FD from these 
snapshots are also shown in the bottom panel of Fig.~\ref{fig:fd_subsonic}. 
Note that each one of these snapshots corresponds to a random realization 
of the non-linear state of the dynamo. We find that 
$\sigmafd \approx 110 \textrm{--} 130\radm$. 
These values are similar to the observational estimates of $\sigmafd$ in 
the ICM determined using FD measured towards background polarized 
sources \citep{CKB01} and Faraday depolarization measured in polarized 
relic embedded in a cluster medium \citep{Kierdorf+17}.
Overall, the shape of the individual distributions are very similar to each other,
implying that the non-linear saturated states of the dynamo at different times 
are statistically equivalent. Even though the individual components of the field 
are expected to have a non-Gaussian distribution, the FD is a sum of $B_\|=B_{z}$ 
over several independent magnetic correlation cells. The PDF of FD is also 
calculated over several independent areas in the $x-y$ plane, and over 
several independent snapshots. Consequently, the PDF of the sum is likely 
to tend to a Gaussian distribution. Indeed, the Gaussian nature of the PDF 
is clearly confirmed by the thick solid black histogram which represents the 
combined distribution of the four snapshots. 

Earlier work by \citet{CR09} and \citet{BS13} has shown that the LOS 
integral of the magnetic field, has a variance which is related to the magnetic 
integral scale $L_{{\rm int}, M}$. In the subsonic turbulence considered here, 
where $n_{\rm e}$ is roughly constant, this variance is also proportional to 
$\sigmafd$, which for a statistically homogeneous and isotropic random 
magnetic field is given by
\EQ
\label{sigmaFD}
\sigmafd = K \,\langle n_{\rm e}\rangle \,\frac{b_{\rm rms}}{2} 
\sqrt{L \, L_{{\rm int}, M}}\,, \qquad L_{{\rm int}, M} = 
\frac{2\,\upi \int (M(k)/k)\, dk}{\int M(k)\, dk}. 
\EN
For the saturated state at $t/t_{\rm ed}=23$, we get $b_{\rm rms} \approx 1.3\mkG$
and $L_{{\rm int}, M} = 112.4\kpc$. Adopting  $L = 512\kpc$ as the path length, 
$\langle n_{\rm e} \rangle = 10^{-3}$ cm$^{-3}$, we get from equation (\ref{sigmaFD})
$\sigmafd= 126.6\radm$, which is in good agreement with $\sigmafd$ determined 
directly from the PDF of the FD using the simulation data.

\begin{figure}
\centering
\begin{tabular}{c}
{\mbox{\includegraphics[width=\columnwidth]{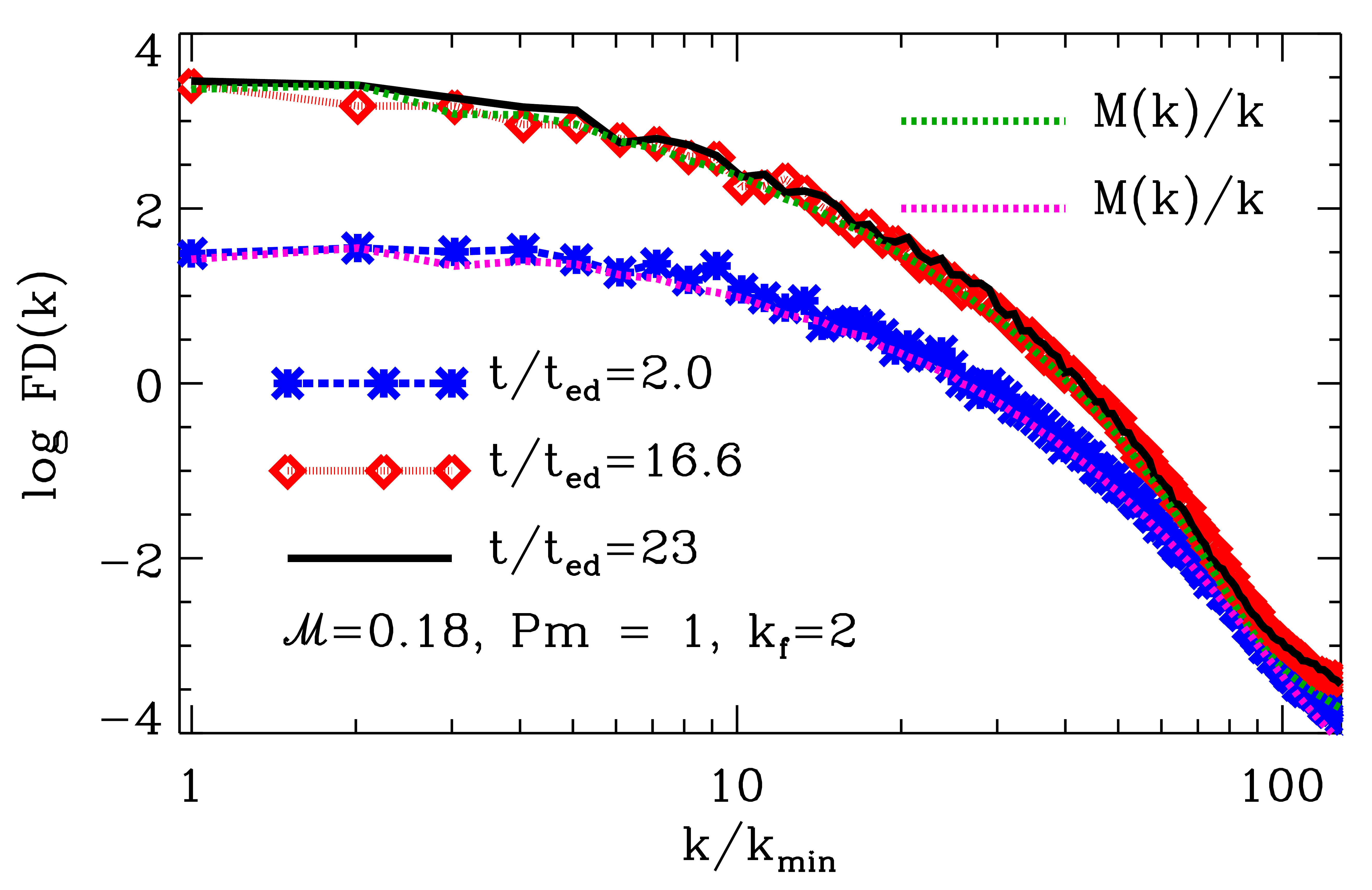}}}\\
\end{tabular}
\caption{Power spectra of the Faraday depth. The red dotted, blue dashed, 
and the solid black curve correspond to the spectra in the kinematic and 
saturated phases, respectively. For comparison, the magenta and green 
dashed curves show the scaled spectra of $M(k)/k$ at $t/t_{\rm ed} = 2$ 
and $23$, respectively. Here, the wavenumber is normalized in units of 
$\kmin = 2\upi/L$. }
\label{fig:fdPS_sub}
\end{figure}

More importantly, equation (\ref{sigmaFD}) suggests that the FD power spectra 
could be qualitatively similar to the power spectra of $M(k)/k$ shown in 
Fig.~\ref{fig:spec_sub}. 
To check this, we use the 2D maps of the Faraday depth to compute the power 
spectra of FD at three different times -- one in the kinematic phase ($t/t_{\rm ed} = 2$) 
and at two intervals ($t/t_{\rm ed} = 16.6, 23$) in the saturated phase of the dynamo. 
The results are shown in Fig.~\ref{fig:fdPS_sub}. We find that the spectra remains 
flat on large scales in the range $k\sim 1-3$ with minor fluctuations on much smaller 
scales $k > 8$. Due to the subsonic nature of the turbulent driving, the structures 
seen in the 2D map and thereby in the power spectra at any given time arises 
purely from the spatial fluctuations of $B_\|$. In order to explore how the FD power 
spectra compares with that of $M(k)/k$, we over plotted the spectra of $M(k)/k$ at 
$t/t_{\rm ed} = 2, 23$ by scaling it by a factor such that it overlaps with the FD power 
spectra at the forcing scale.\footnote{{We note that the FD power spectrum and 
$M(k)/k$ have different dimensions and so their amplitudes will be different. 
The scaling is performed to check how closely their shapes match.}} 
In the figure these are shown by the dashed magenta and green curves, respectively. 

Remarkably, the power spectrum of FD is strikingly similar to that of $M(k)/k$ 
in both the kinematic and saturated phases of the dynamo over the entire range of 
wave numbers shown here. Thus, the observationally determined power spectrum 
of FD can be used to directly infer the power spectrum of the random magnetic field 
in the ICM provided fluctuations in $n_{\rm e}$ are small, and values of FD are 
robustly estimated (see Section \ref{sec:rmsynth} below). The plot further shows 
that, compared to the kinematic phase, the FD power spectrum and that of $M(k)/k$ 
have more power on all but the very smallest scales in the saturated phase. This 
is related both to the smaller field strength and smaller integral scale during the 
kinematic stage of the fluctuation dynamo. Thus the power spectrum of Faraday 
depth in galaxy clusters contains crucial information even on the evolutionary stage 
of turbulent dynamo operating in them. Moreover, the top panel of Fig.~\ref{fig:fd_subsonic} 
shows that the maximum scale of the structures seen in the 2D map of FD are 
comparable to the forcing scale of turbulence, and so this scale can also then be 
inferred from FD maps.

\subsection{Total synchrotron intensity} 
\label{sec:sync}

\begin{figure}
{\mbox{\includegraphics[width=0.95\columnwidth, trim=0mm 0mm 20mm 0mm, clip]{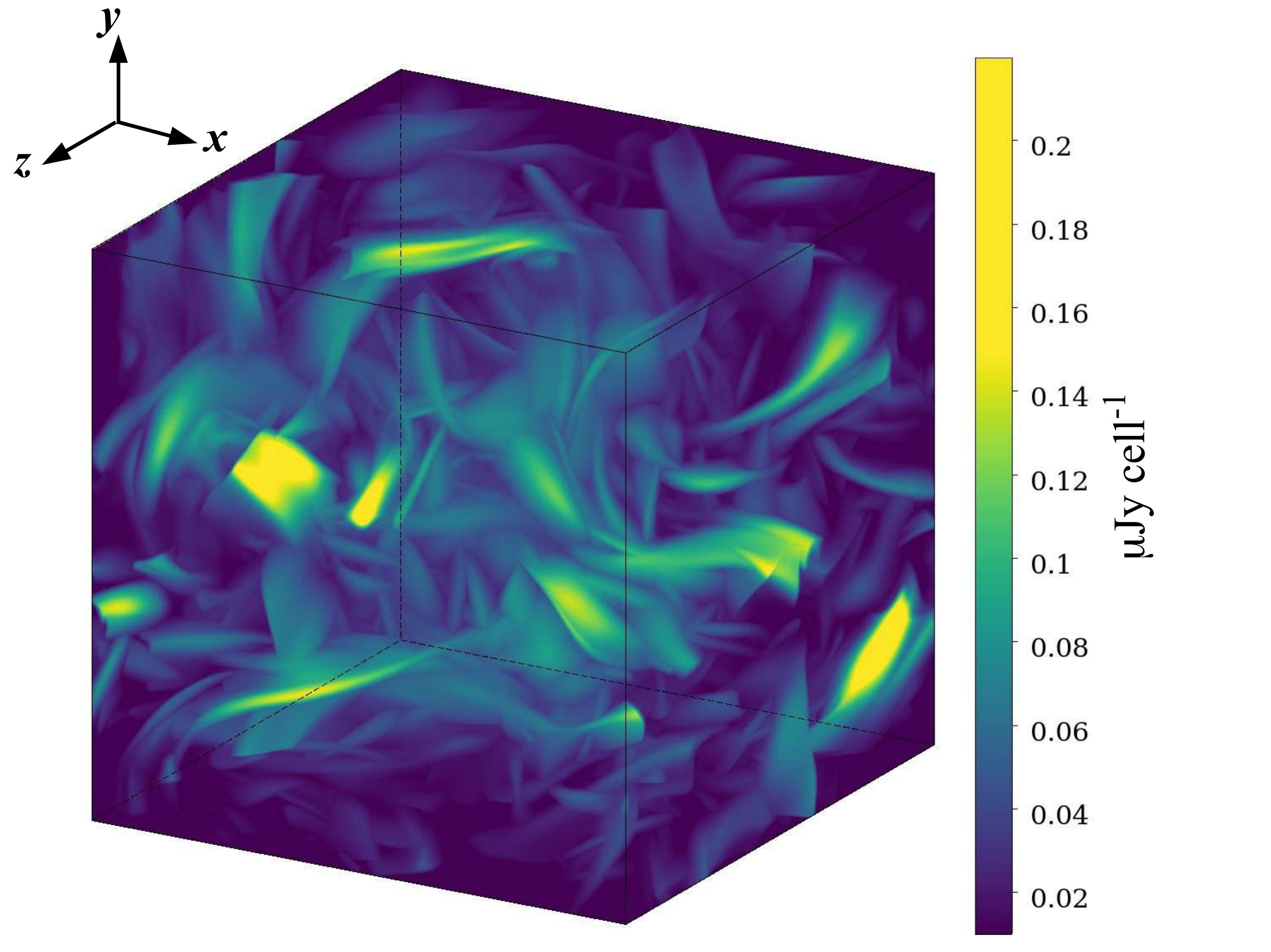}}}\\
{\mbox{\includegraphics[width=0.95\columnwidth]{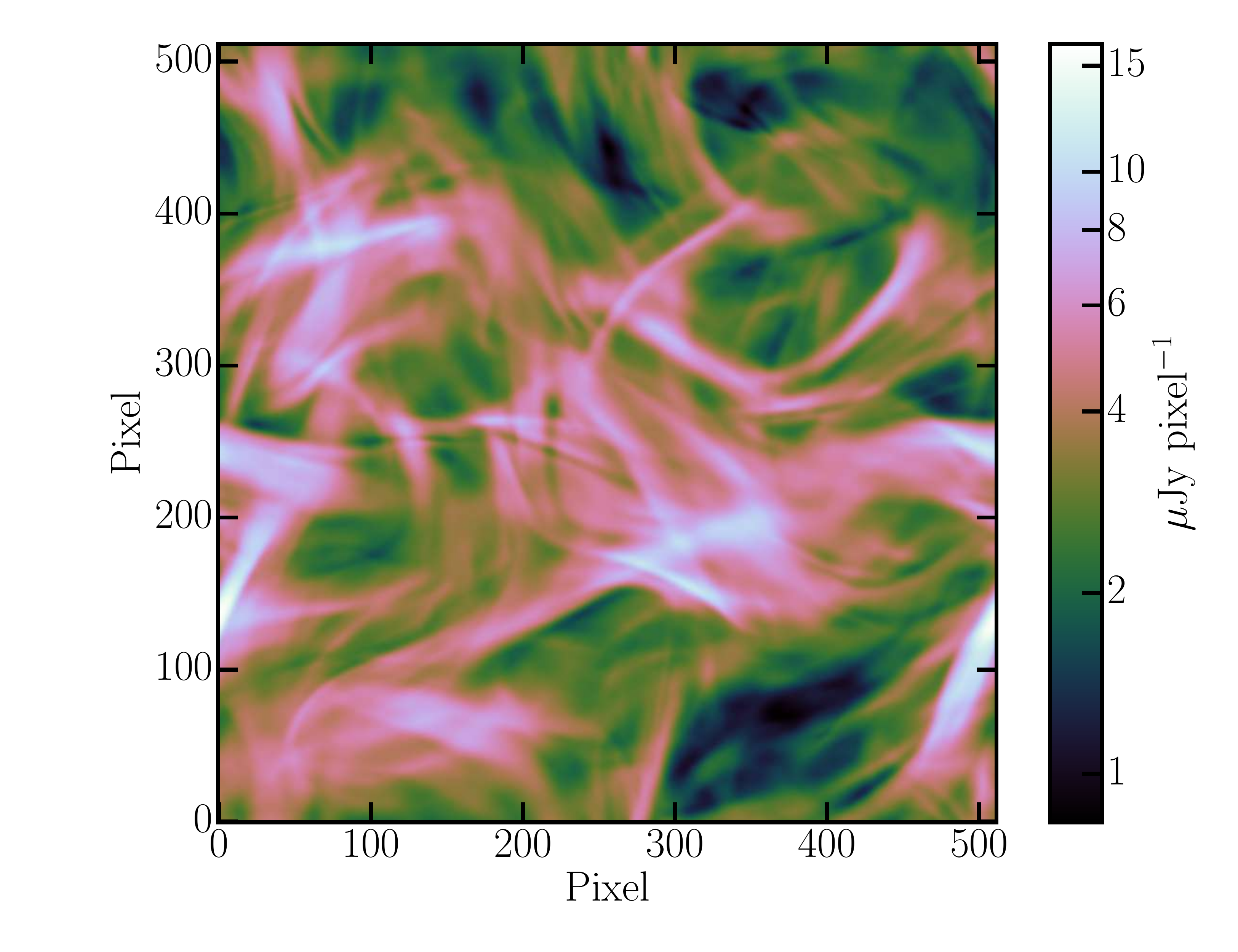}}}\\
{\mbox{\includegraphics[width=0.95\columnwidth]{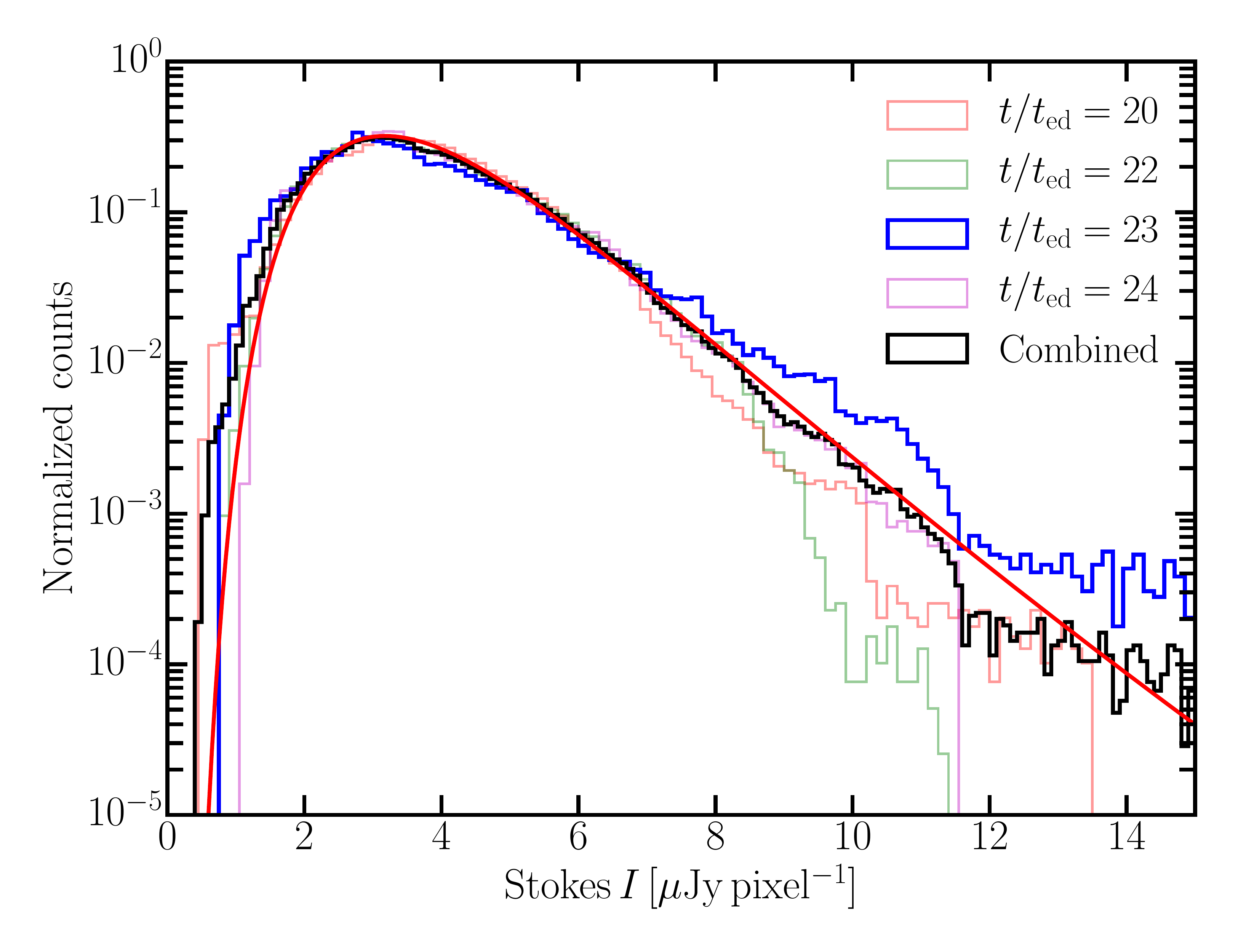}}}\\
\caption{Top: 3D volume rendering of synchrotron emissivity at $1\ghz$ at 
$t/t_{\rm ed} = 23$ in the saturated phase of the dynamo. 
Middle: 2D map of synchrotron intensity in the plane of the sky at $1\ghz$
obtained by integrating the 3D emissivity shown in the top panel along the LOS
($z$-axis). Bottom: Distributions of the total intensity for four different 
snapshots in the saturated stage. The solid black histogram is the combined 
distribution of the different snapshots and the solid red curve is the best-fitting 
lognormal distribution.}
\label{fig:sync_sub}
\end{figure}
We calculate the total synchrotron intensity ($I_{\rm sync}$) at a frequency
$\nu$  from the LOS integration of the synchrotron emissivity
($\varepsilon_{\rm sync, \nu}$) along the $z$-axis following equation 
(\ref{eq:sync_intensity}). The top panel of Fig.~\ref{fig:sync_sub} shows the 
3D volume rendering of the synchrotron emissivity in units of $\mJy\,{\rm cell}^{-1}$, 
while the middle panel shows the 2D map of the synchrotron intensity 
($I_{\rm sync}$) integrated along the LOS at 1~GHz in the saturated state of 
the dynamo at $t/t_{\rm ed} = 23$. In accordance with the normalization 
defined in Section \ref{sec:synth_obs}, the 2D map has a total flux density 
of 1~Jy. Both the 3D volume rendering and the 2D map shows bright 
structures extending to about $1/2$ the scale of the box that corresponds to 
the size of the turbulent cells. Since we have assumed a constant distribution 
of $n_{\rm CRE}$, $\varepsilon_{\rm sync} \propto B_{\perp}^{1-\alpha}$, 
and the total intensity $I_{\rm sync} \propto \int\,B_{\perp}^{2}\,dl$ (for $\alpha = -1$), 
the structures seen here essentially arises due to the magnetic fields, which 
are being randomly stretched and twisted due to turbulent driving. 

In the bottom panel of Fig.~\ref{fig:sync_sub}, we show the PDF of $I_{\rm sync}$ 
at four different times during the saturated phase. It is clear from the plots that,
unlike the distribution of FD, the histograms of each of these are well represented 
by lognormal distribution, and the best fit to the combined black distribution is 
shown by the solid red curve in the figure. Since the total synchrotron intensity 
in spatially resolved objects depends non-linearly on the magnetic field, the 
distribution is not expected to be a Gaussian. It is interesting to note that, when 
the small-scale features are resolved in the synthetic total intensity maps obtained 
at the native resolution of the simulations, the lognormal distribution has long tails. 
However, the extent of the tail towards higher flux values goes down drastically,
making the distributions more symmetric when synthetic observations are 
smoothed to mimic observations performed using a telescope. This is perhaps 
the reason why small-scale structures seen in the $I_{\rm sync}$ map in 
Fig.~\ref{fig:sync_sub} (middle panel) have not yet been observed in astronomical 
observations of ICM. Smoothing of the observable quantities introduced by a 
telescope has strong implications on polarization and we discuss that in detail in 
Section \ref{sec:smooth_pol}.

\begin{figure}
\centering
\begin{tabular}{c}
{\mbox{\includegraphics[width=\columnwidth]{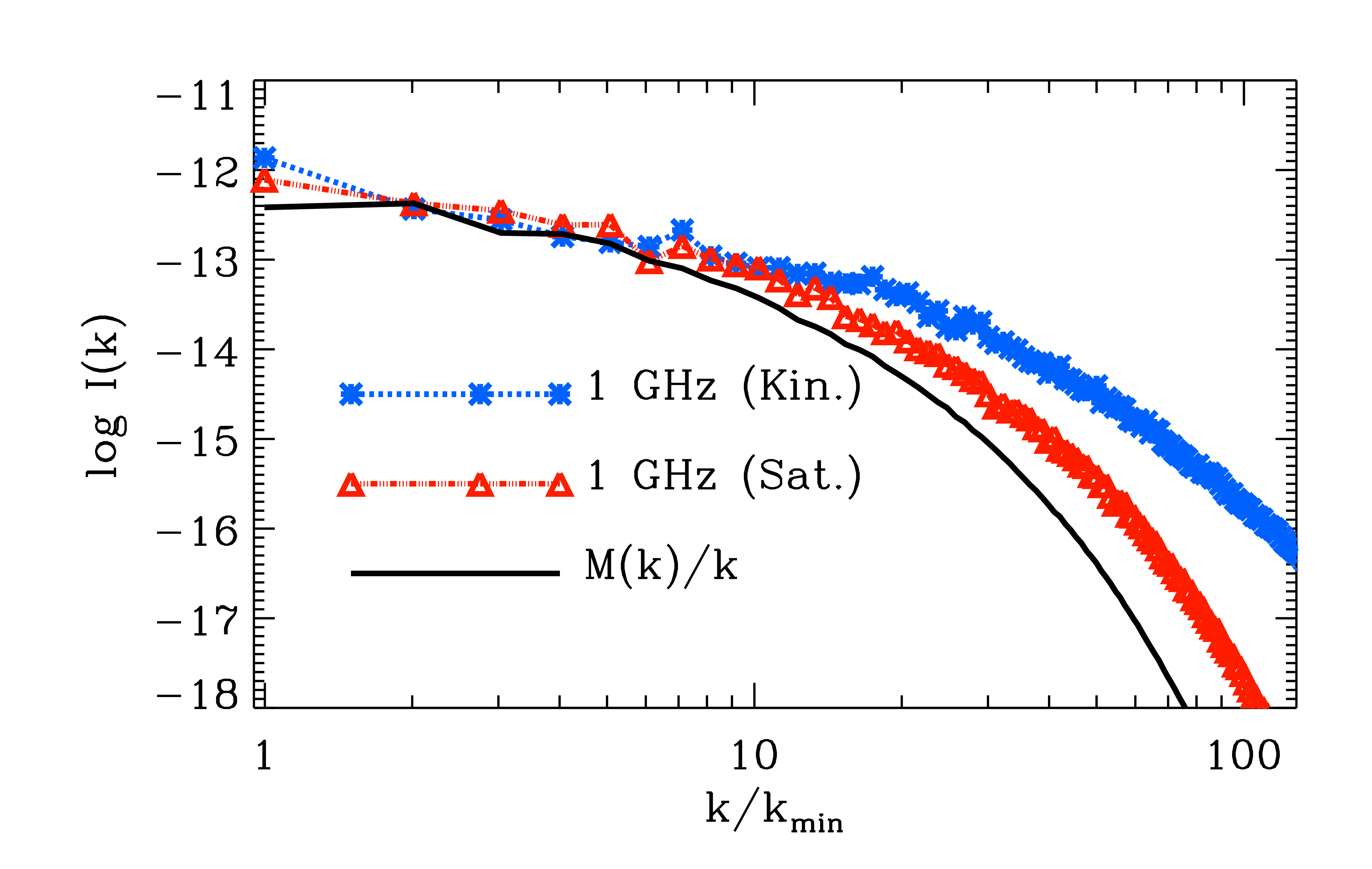}}}
\end{tabular}
\caption{Power spectra of the total synchrotron emission ($I$) at $1\ghz$ in 
the kinematic (blue, dotted line with asterisks) and in the saturated phase (red, 
dashed line with open triangles). The solid black line shows the scaled spectra 
of $M(k)/k$. The wavenumber is normalized in units of $\kmin = 2\upi/L$.}
\label{fig:totIPS_sub}
\vspace{-0.5em}
\end{figure}

\begin{figure*}
\centering
\begin{tabular}{ccc}
\large{0.5~GHz} & \large{1~GHz} & \large{6~GHz} \\
 & & \\
 & \large{Polarized intensity} &  \\
{\mbox{\includegraphics[height=4cm, trim=1mm 2mm 2mm 0mm, clip]{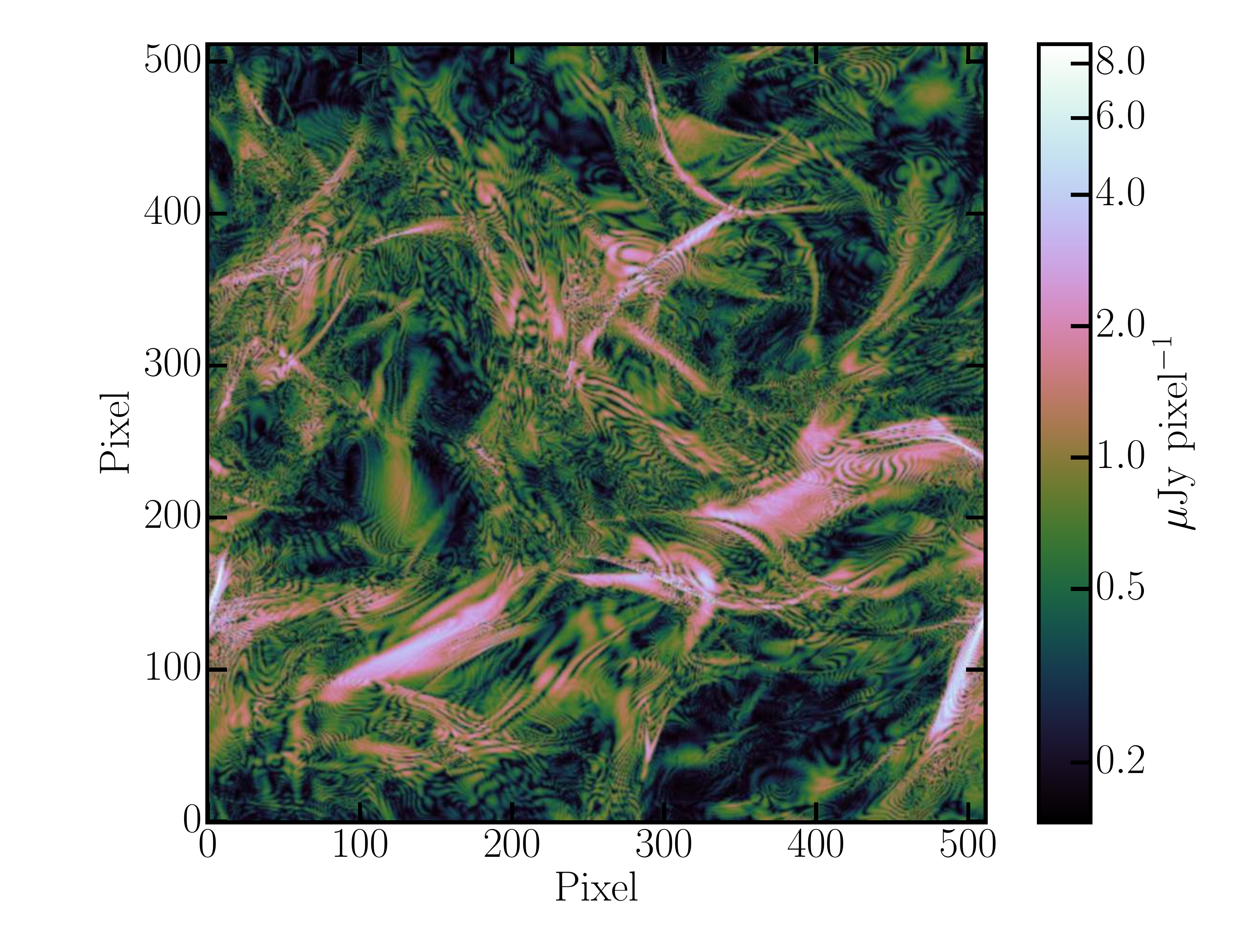}}}&
{\mbox{\includegraphics[height=4cm, trim=1mm 2mm 2mm 0mm, clip]{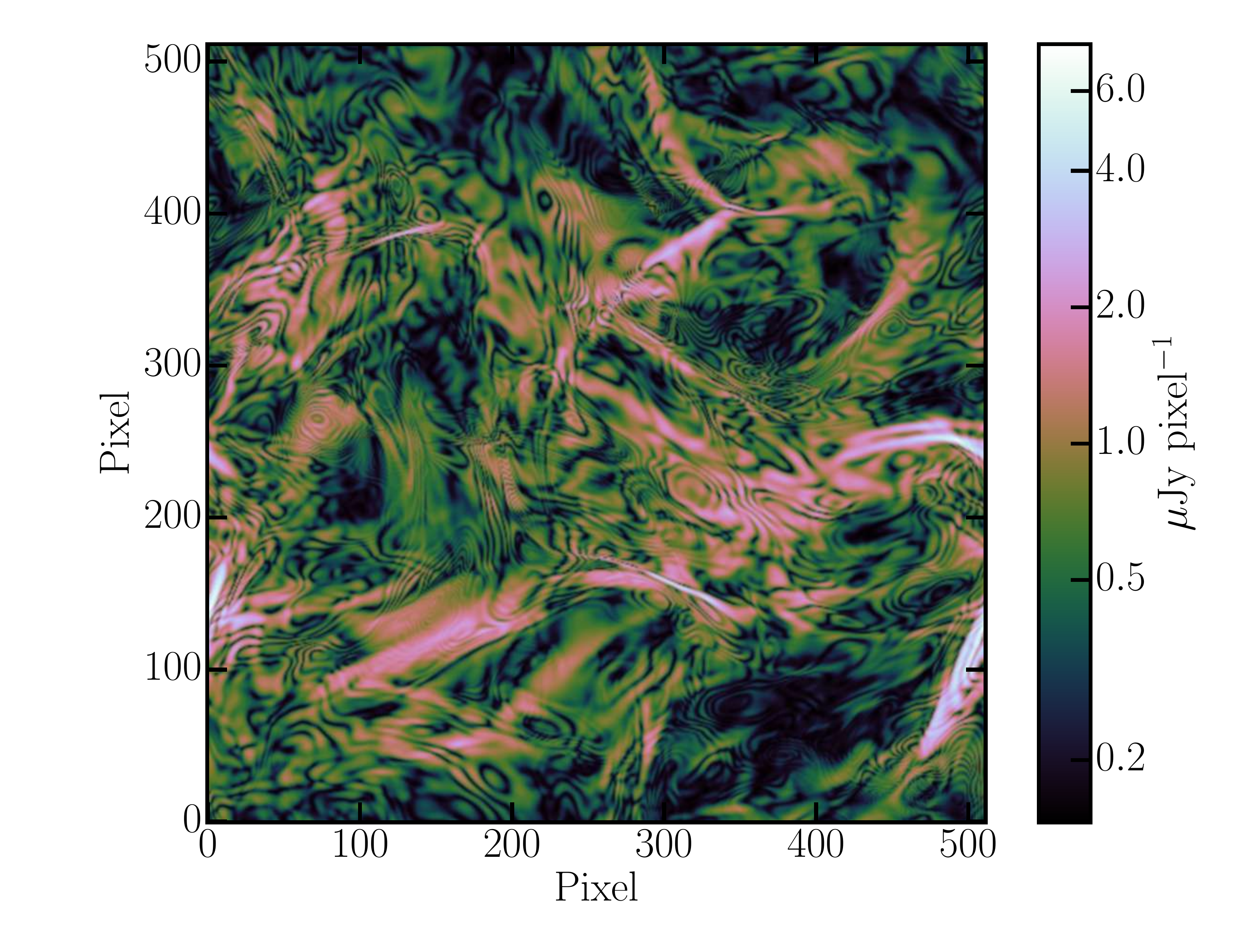}}}&
{\mbox{\includegraphics[height=4cm, trim=1mm 2mm 2mm 0mm, clip]{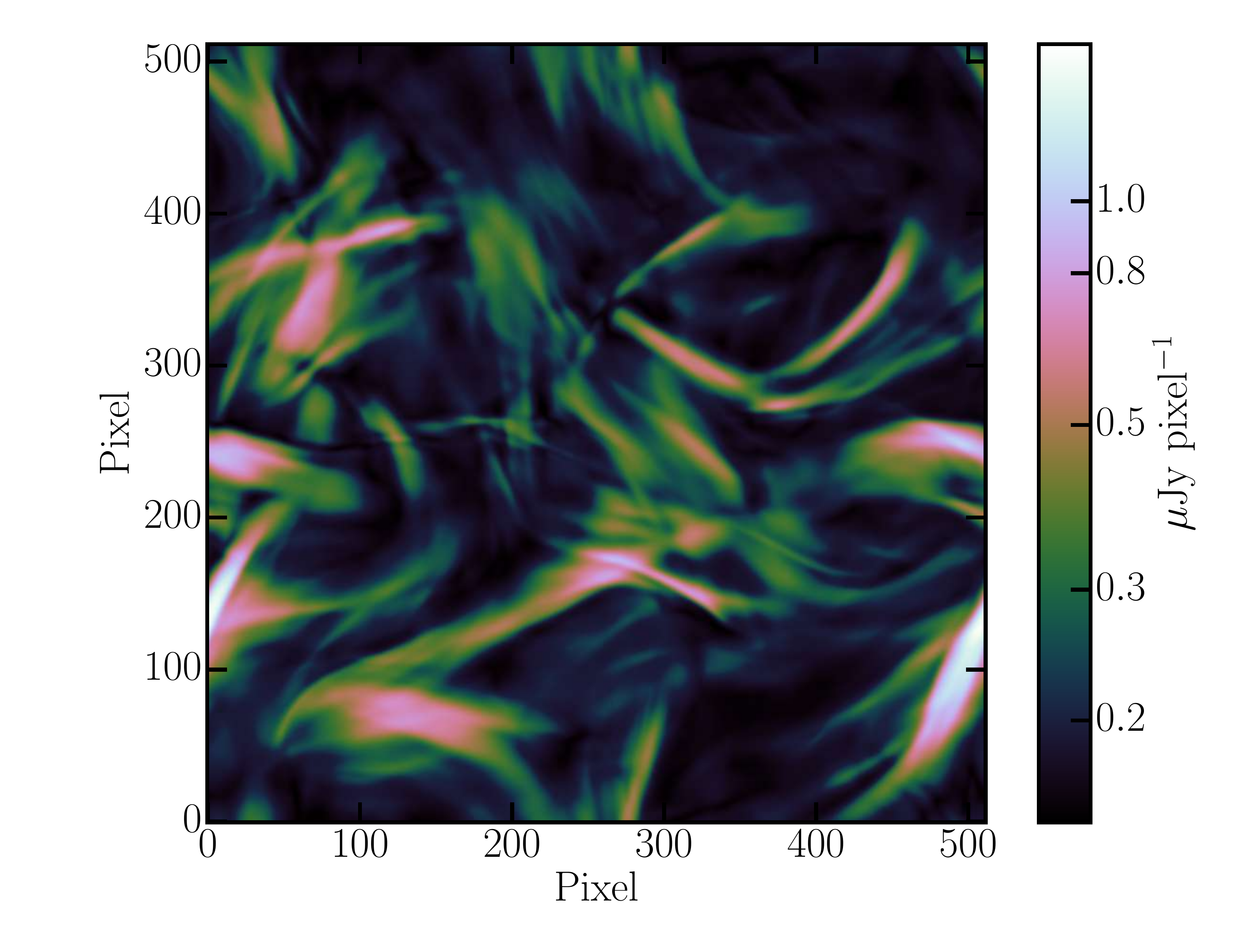}}}\\
 & & \\
 & \large{Fractional polarization} &  \\
{\mbox{\includegraphics[height=4cm, trim=1mm 2mm 2mm 0mm, clip]{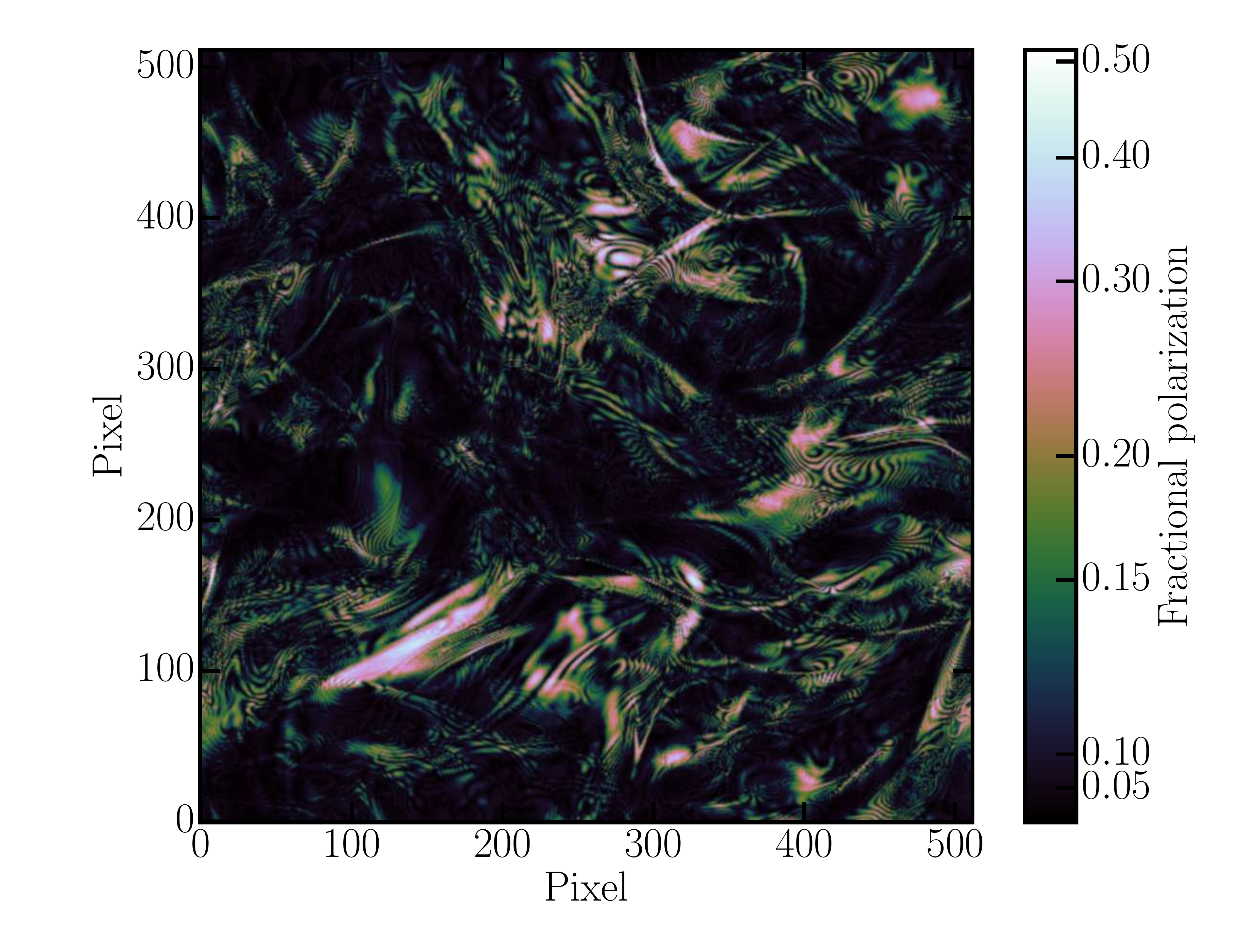}}}&
{\mbox{\includegraphics[height=4cm, trim=1mm 2mm 2mm 0mm, clip]{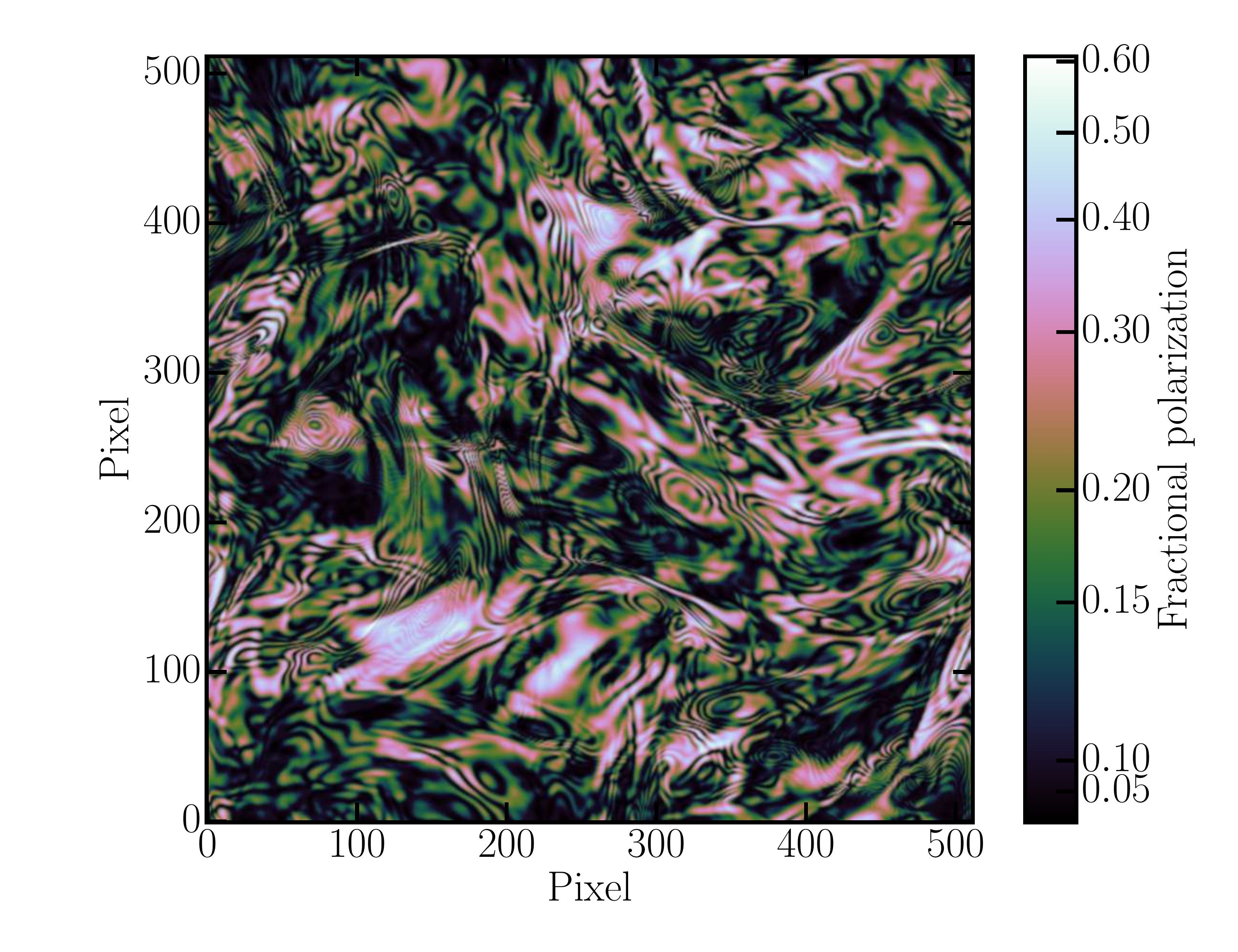}}}&
{\mbox{\includegraphics[height=4cm, trim=1mm 2mm 2mm 0mm, clip]{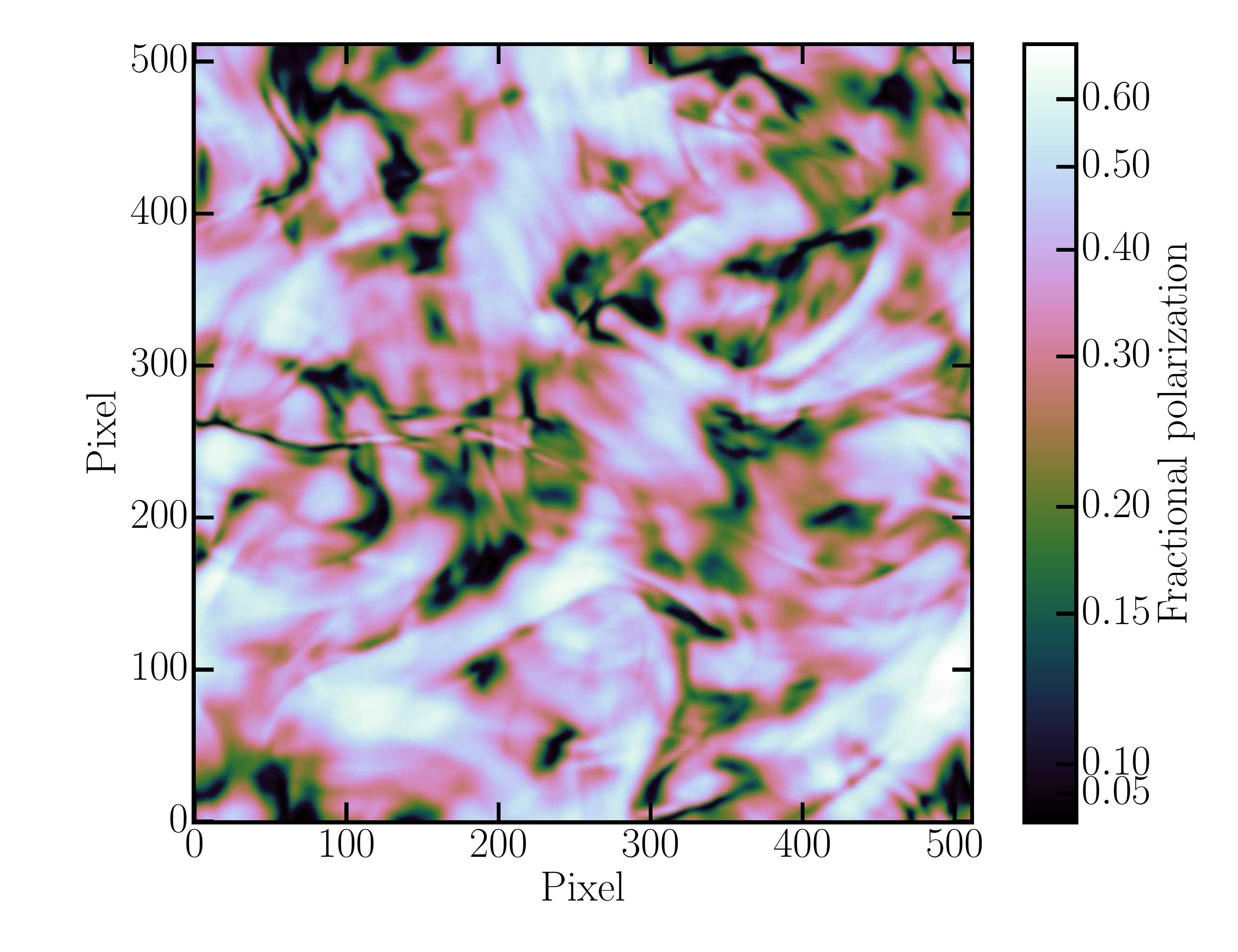}}}\\
 & \large{Polarized intensity (smoothed 10 pixels)} &  \\
{\mbox{\includegraphics[height=4cm, trim=1mm 2mm 2mm 0mm, clip]{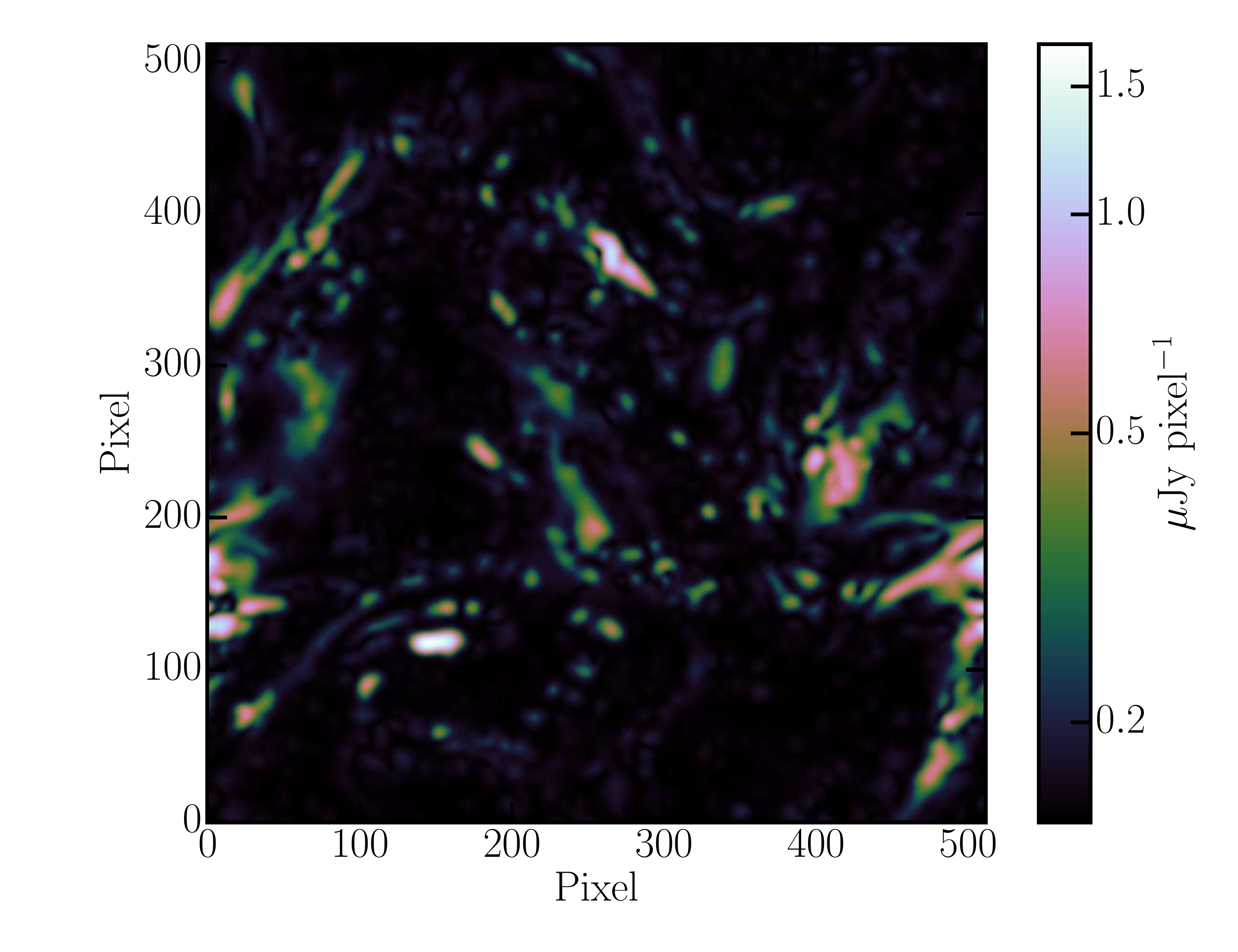}}}&
{\mbox{\includegraphics[height=4cm, trim=1mm 2mm 2mm 0mm, clip]{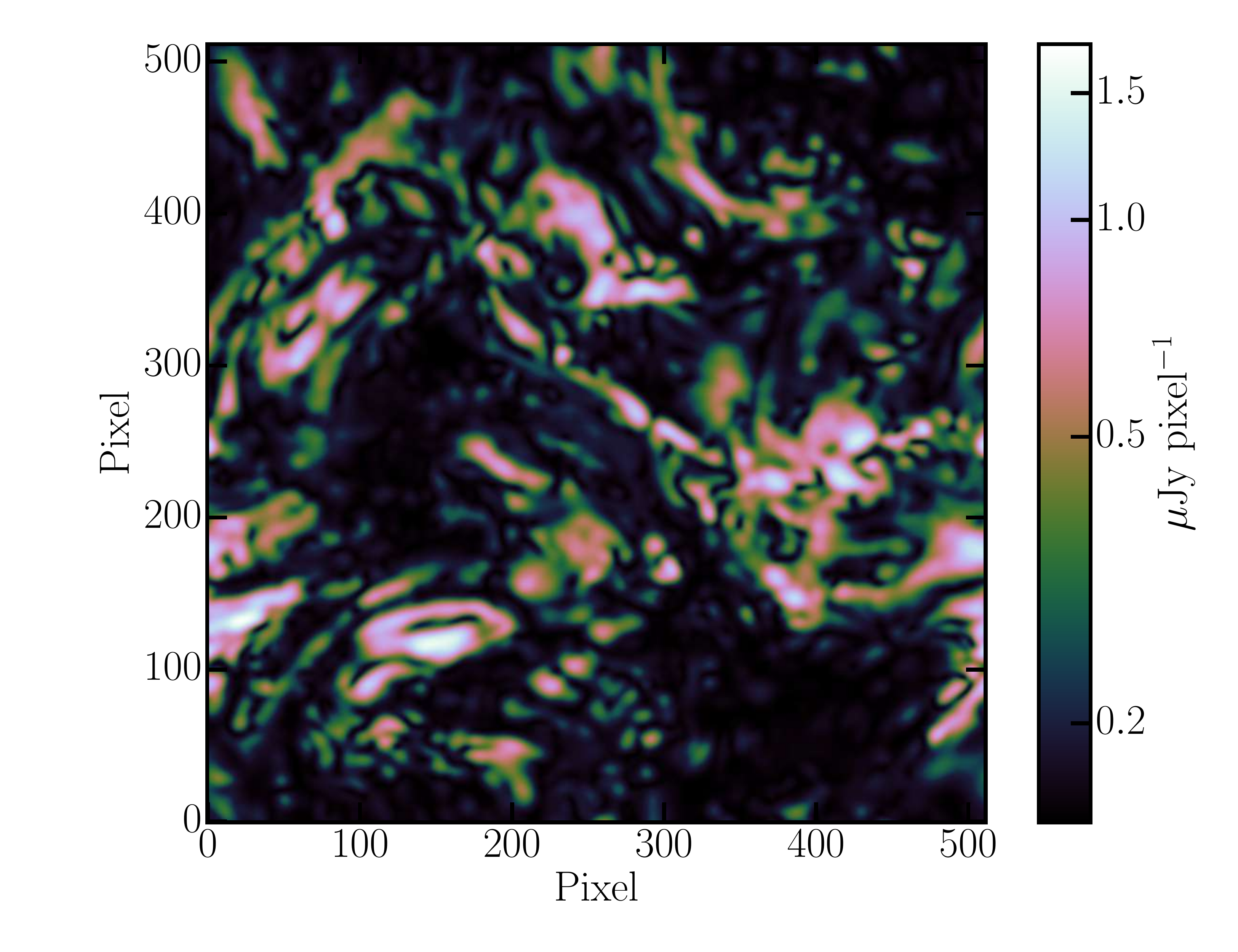}}}&
{\mbox{\includegraphics[height=4cm, trim=1mm 2mm 2mm 0mm, clip]{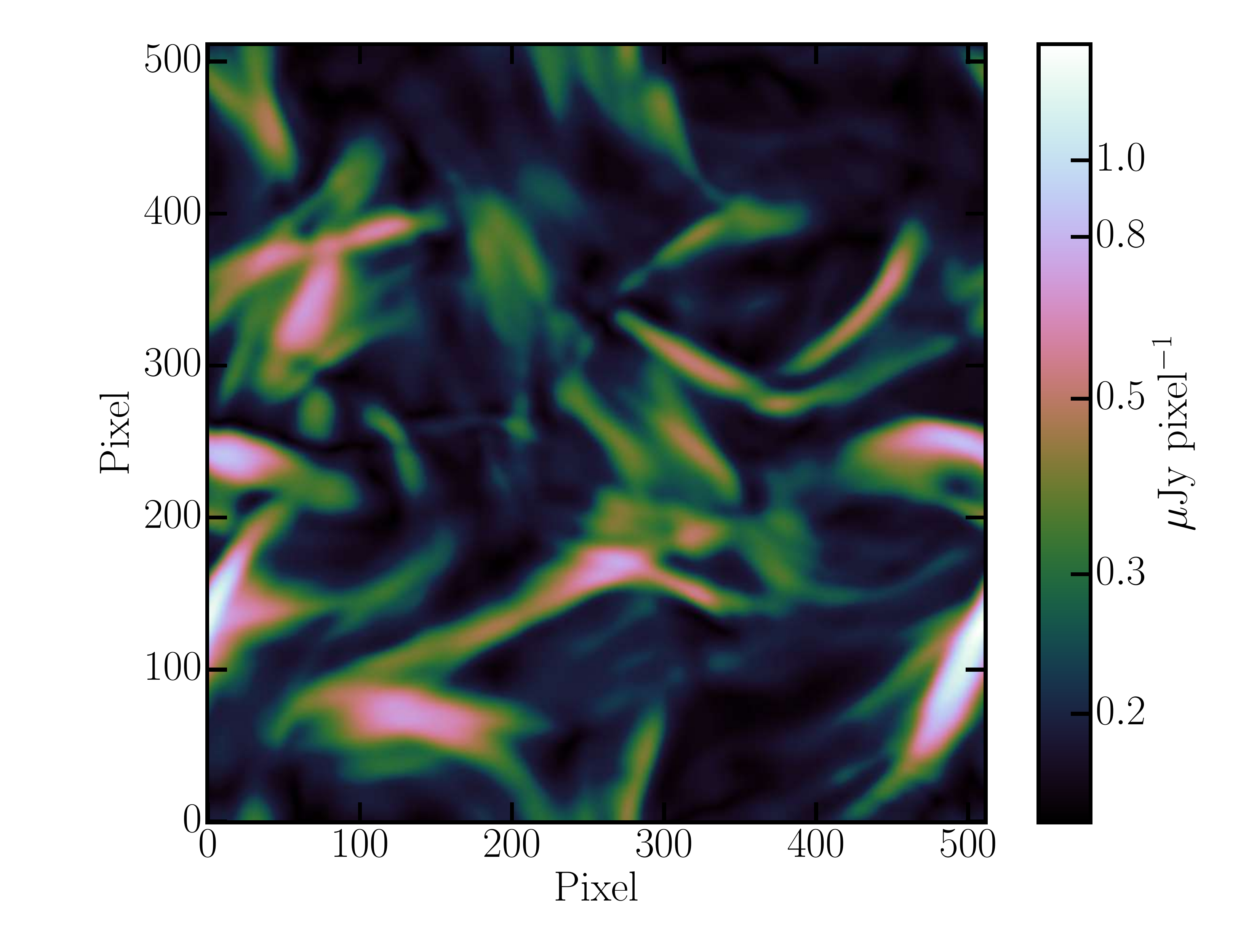}}}\\
 & & \\
 & \large{Fractional polarization (smoothed 10 pixels)} &  \\
{\mbox{\includegraphics[height=4cm, trim=1mm 2mm 2mm 0mm, clip]{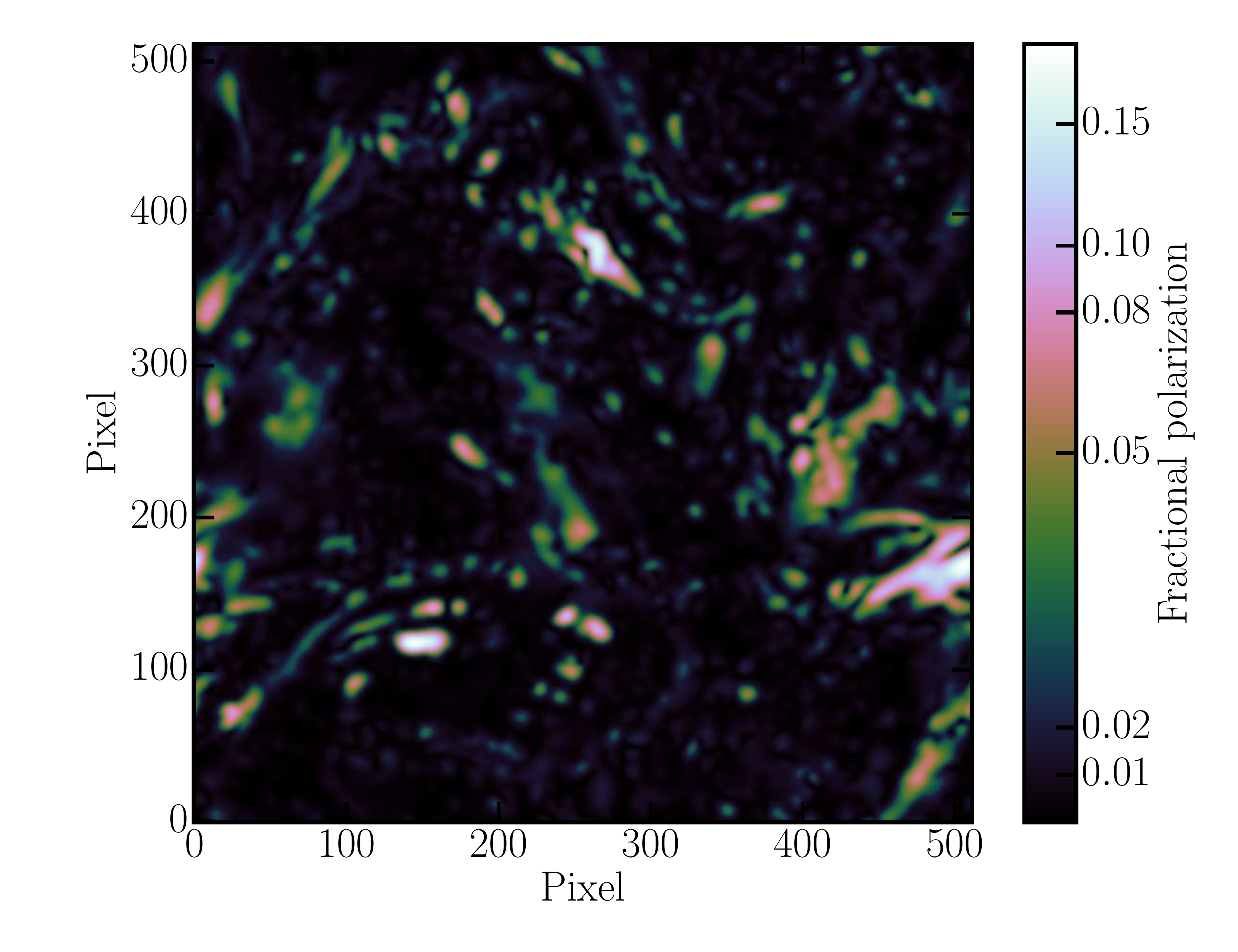}}}&
{\mbox{\includegraphics[height=4cm, trim=1mm 2mm 2mm 0mm, clip]{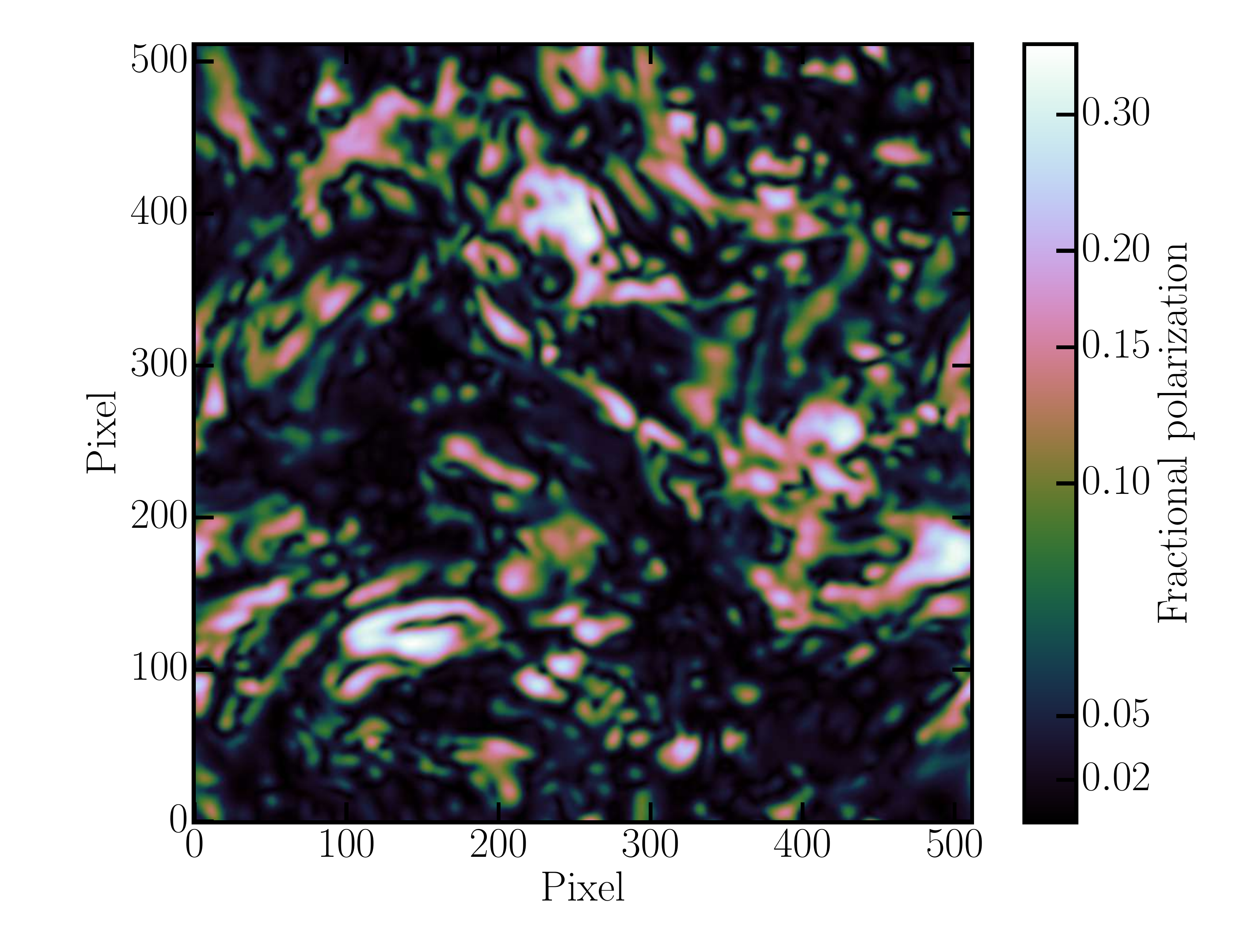}}}&
{\mbox{\includegraphics[height=4cm, trim=1mm 2mm 2mm 0mm, clip]{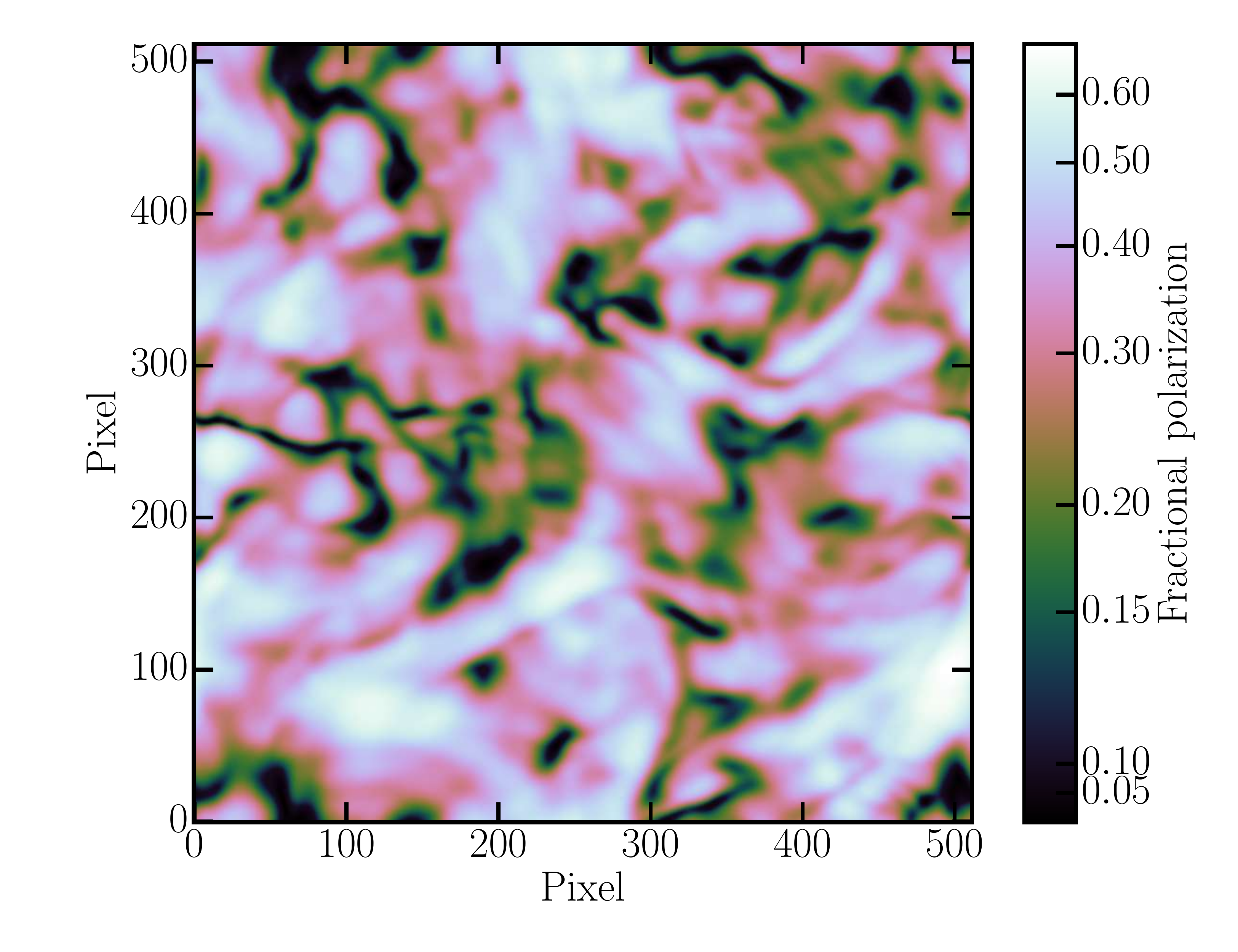}}}\\
\end{tabular}
\caption{First row: Polarized intensity maps in units of Jy~pixel$^{-2}$. Second row: Fractional polarization 
maps. The third and fourth rows show the quantities after smoothing by a Gaussian kernel with FWHM 10\,pixels.
The left-hand, middle, and right-hand columns are at 0.5, 1 and 6~GHz, respectively, at $t/t_{\rm ed} = 23$.}
\label{fig:pol_subsonic}
\end{figure*}

In Fig.~\ref{fig:totIPS_sub}, we show the power spectra of the total intensity at 
$1\ghz$. The spectrum at $t/t_{\rm ed} = 2$ in the kinematic phase is shown in 
blue dashed line with asterisks while that at $t/t_{\rm ed} = 23$ in the saturated 
phase is shown by the red dotted curve with open triangles. Both the curves 
show that the amplitude at wave numbers $k \leq 10$ is independent of the 
evolutionary stage of the dynamo and is peaked on the scale of the simulation 
domain (i.e. at $512\kpc$).\footnote{Note that, although magnetic field strengths 
are weaker in the kinematic phase as compared to that in the saturated phase, 
the amplitudes at $k=1$ are similar. This is because we always generate the 
synthetic observations normalized to the same total flux value of 1~Jy at $1~\ghz$ 
for each snapshot.} However, the spectrum in the saturated state falls off with 
a steeper slope on scales $k > 10$, compared to the kinematic phase suggesting 
the presence of more structures on these scales, similar to what was seen for 
FD power spectra. The spectra at two other representative frequencies 
corresponding to $0.5$ and $6\ghz$ (not shown in the figure) also show 
similar behavior. Since the total intensity $I \propto \nu^{-\alpha}$, the amplitude 
of the spectra decreases as the frequency increases from $0.5-6\ghz$. 
We also find that the overall shape of the scaled spectrum of $M(k)/k$ in the 
saturated state resembles that of the red curve to a fair degree in the range 
$2 \leq k \leq 6$, corresponding to physical scales $256\kpc \geq l \geq 85\kpc$.

\subsection{Polarization parameters} \label{sec:polarization}

In this subsection, we focus on extracting the polarization parameters from the
simulation data and exploring how these parameters depend on the frequency.  
To this end, we computed the frequency dependence of the linearly polarized
intensity (${\rm PI}_{\nu}$) from the Stokes $Q$ and $U$ using
equations~\eqref{eq:compQU}--\eqref{eq:compPI} and the fractional polarization
$p_\nu = {\rm PI}_{\rm \nu}/I_{\rm sync,\nu}$ from a single snapshot of the 
magnetic field in the saturated state. 

In Fig.~\ref{fig:pol_subsonic}, the first two rows show the 2D maps of the
total polarized intensity (first row) and the fractional polarization (second
row) at three different frequencies: $0.5\ghz$ (first column), $1\ghz$ (second
column), and $6\ghz$ (third column). The corresponding Stokes $Q$ and 
$U$ parameters are presented in the top two panels of Fig.~\ref{fig:qu_subsonic} 
in the Appendix. Basic pixel-wise statistics of the polarization parameters at
these three representative frequencies are presented in Table~\ref{tab:stats}, 
in surface brightness units of Jy~pixel$^{-1}$.  Note that this is equivalent
to performing observations with a sufficiently small telescope resolution that
is same as that of the pixel size of these MHD simulations.\footnote{For
reference, for an image made at an angular resolution of $10\times 10$~arcsec$^2$
(FWHM) sampled using pixels of size $1\times1$~arcsec$^2$, the values with
$\mJy$~pixel$^{-1}$ units are to be multiplied by a conversion factor of 113.31 
to obtain values in units of $\mJy$~beam$^{-1}$.}

The pixel-wise mean and maximum polarized intensities shown in the first 
row of Fig.~\ref{fig:pol_subsonic} progressively decrease from $0.5$ 
to $6\ghz$ (see Table~\ref{tab:stats}). This trend arises due to a mix of 
frequency-dependent Faraday depolarization and the spectral dependence 
of total synchrotron emission. The PDF of polarized intensity at 6 GHz is 
found to be similar to lognormal distribution as seen for total intensity. 
However, at lower frequencies, due to structures introduced by Faraday 
depolarization, the PDFs show extended power-law tails. Spectral effects 
of Faraday depolarization are seen in the frequency variation of the 
fractional polarization, the maps of which are shown in the second row of 
Fig.~\ref{fig:pol_subsonic}. In contrast to ${\rm PI}_{\nu}$, these maps 
show the opposite trend with mean $p$ increasing significantly from 
$\langle p \rangle = 0.09$ at $0.5\ghz$ to $\langle p \rangle = 0.345$ at $6\ghz$. 
Note that, although the strengths of magnetic field components are random 
and there are no mean fields in the simulated turbulent media (due to the 
non-helical nature of turbulence and lack of scale separation), the mean 
fractional polarization at $6\ghz$ is relatively high where the frequency-dependent 
Faraday depolarization is low. This is because, fields that are stretched and 
twisted by dynamo action can be ordered locally on the scale of turbulent 
driving. Moreover, due to relatively low number of magnetic field integral 
scales (of the order of five) in the simulation volume used here, the 
cancellation of the polarized emission along the path-length is low and, 
along with locally ordered fields, partially contributes towards the high 
fractional polarization. At low frequencies, depolarization due to Faraday 
rotation along the LOS play a significant role in reducing the observed 
fractional polarization. 

For a longer path-length $L$ these predictions are likely to decrease as 
$L^{-1/2}$. Therefore, for a longer path-length of, say, a few Mpc, the 
decrease in polarization would be a factor of the order of two. However, 
magnetic field strength, $n_{\rm CRE}$, and $n_{\rm e}$ in galaxy clusters 
are all likely stratified, decreasing away from cluster core over scales of 
order a few hundred kpc, their core radii. So, the relative contribution to 
total and polarized intensities from longer path-lengths will also decrease 
along the LOS as one moves away from the cluster core, and the cancellation 
of polarization due to random fields and Faraday depolarization will be 
dominated by the domain with largest magnetic field strengths, cosmic 
ray and free electron densities. In this case, the estimated polarization 
using a box scale of 512\,kpc, is expected to be reasonable up to factors 
of the order of two.

\begin{table*}
\centering 
\setlength{\tabcolsep}{8.0 pt}
\caption{Summary of statistics of Stokes ${\rm PI}, Q, U$ and the fractional
polarization ($p$) at native-pixel and 10-pixel FWHM resolutions. 
}
\begin{tabular}{lcccccccccccc} 
\hline
\multicolumn{1}{l}{Quantity} & 
\multicolumn{1}{c}{Resolution} & 
\multicolumn{3}{c}{Mean} & 
&
\multicolumn{3}{c}{Median} & 
&
\multicolumn{3}{c}{Dispersion} \\
         &       & 
\multicolumn{3}{c}{} &
&
\multicolumn{3}{c}{} &
&
\multicolumn{3}{c}{} \\
         &       & {0.5 GHz} & {1 GHz} & 6 GHz & & 0.5 GHz & 1 GHz & 6 GHz & & 0.5 GHz & 1 GHz & 6 GHz\\
\hline
Stokes $PI$ & Native & 0.670 & 0.638 & 0.228 & & 0.521 & 0.519 & 0.192 & & 0.571 & 0.496 & 0.163\\
($\rm \umu Jy\,pixel^{-1}$)  & 10 pixels & 0.099 & 0.227 & 0.222 & & 0.053 & 0.156 & 0.190 & & 0.130 & 0.218 & 0.151\\
Stokes $Q$ & Native & 0.005 & 0.007 & -0.038 & & 0.002 & 0.000 & -0.030 & & 0.625 & 0.572 & 0.180\\
($\rm \umu Jy\,pixel^{-1}$)  & 10 pixels & 0.005 & 0.007 & -0.038 & & 0.001 & 0.001 & -0.031 & & 0.120 & 0.224 & 0.171\\
Stokes $U$ & Native & 0.000 & -0.003 & -0.035 & & 0.003 & -0.006 & -0.027 & & 0.620 & 0.570 & 0.209\\
($\rm \umu Jy\,pixel^{-1}$) & 10 pixels & 0.000 & -0.003 & -0.035 & & 0.001 & -0.004 & -0.026 & & 0.111 & 0.221 & 0.200\\
$p$ & Native & 0.090 & 0.169 & 0.345 & & 0.077 & 0.155 & 0.346 & & 0.061 & 0.094 & 0.129\\
            & 10 pixels & 0.013 & 0.060 & 0.337 & & 0.008 & 0.044 & 0.336 & & 0.015 & 0.052 & 0.126\\
\hline        
\end{tabular}
\label{tab:stats}
\end{table*}

\begin{figure}
\centering
\begin{tabular}{c}
{\mbox{\includegraphics[width=0.4\textwidth, trim=0mm 0mm 40mm 0mm, clip]{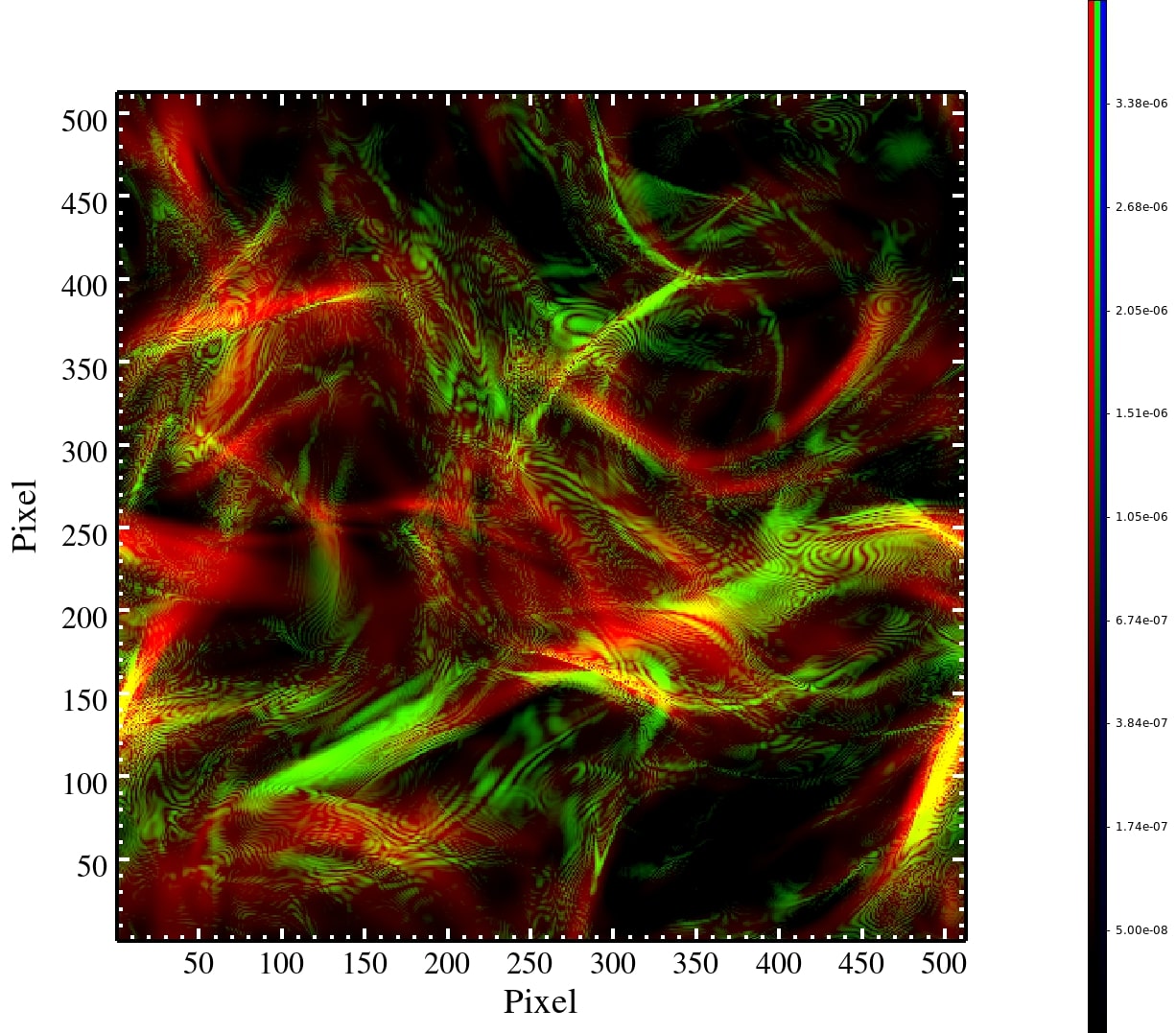}}}\\
\end{tabular}
\caption{Overlay of the total synchrotron intensity (in red) and the polarized
intensity at $0.5\ghz$ (in green). Both the maps are saturated towards lower
surface brightness to show the close correspondence between bright filamentary
structures observed in polarization with that of total intensity (see text in
Section \ref{sec:polarization}).}
\label{fig:overlay}
\end{figure}

\begin{figure*}
\centering
\begin{tabular}{c}
{\mbox{\includegraphics[width=\textwidth]{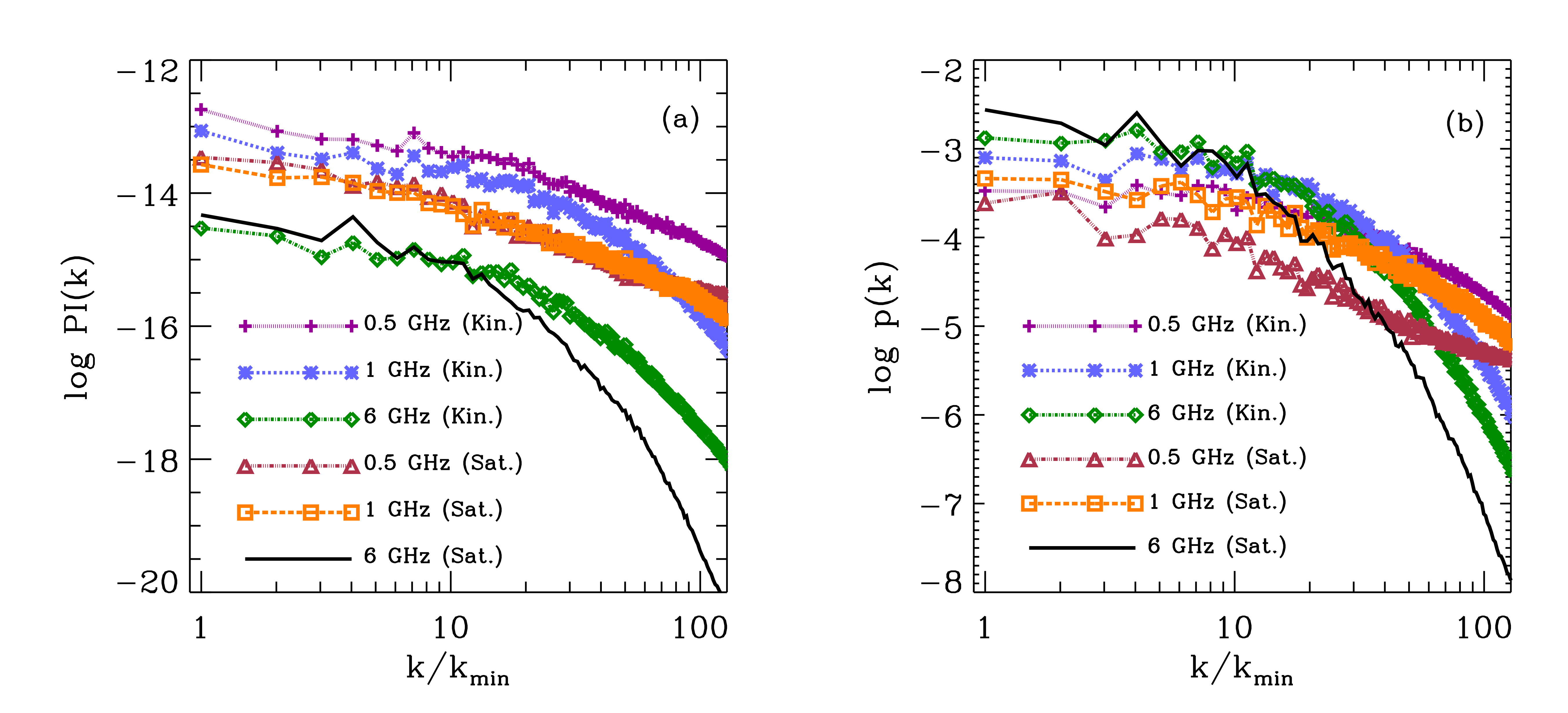}}}\\
\end{tabular}
\caption{Power spectra of the total polarized intensity (left) and the fractional polarization 
(right) at 0.5, 1, and 6~GHz in the kinematic ($t/t_{\rm ed} = 2$) and saturated phase 
($t/t_{\rm ed} = 23$). The wavenumber is normalized in units of $\kmin = 2\upi/L$.}
\label{fig:spec_pol_sub}
\end{figure*}

In our simulated volume, of particular importance is the fact that strong 
frequency-dependent Faraday depolarization at lower frequencies ($\lesssim 1.5\ghz$) 
gives rise to small-scale structures on scales much smaller than the driving 
scale of turbulence. These small-scale structures appear sharper at lower 
frequencies (near $\sim0.5\ghz$) resembling \textit{filament-like} features 
similar to the polarization filaments observed in the Galactic interstellar 
medium \citep{shuku03,FS06, Zar+15, Jelic+15, Jelic+18}. Although these 
features are not readily visible in the total synchrotron intensity emission 
(Fig.~\ref{fig:sync_sub} middle panel) or in the Faraday depth (Fig.~\ref{fig:fd_subsonic}) 
maps, with a cursory look, they somewhat correspond to sharp edges 
observed in the total synchrotron intensity (see Fig.~\ref{fig:overlay}).

In the left- and right-hand panels of Fig.~\ref{fig:spec_pol_sub}, we present
the power spectra of the ${\rm PI}_\nu$ and $p_\nu$, respectively. The 
effects of the small-scale structures introduced due to Faraday depolarization 
towards low frequencies are easily discernible as larger power at large $k$ 
in the power spectra of the polarized intensities shown in Fig.~\ref{fig:spec_pol_sub} 
and the Stokes $Q$ and $U$ parameters shown in Fig.~\ref{fig:spec_qu_sub}.  
At higher frequencies, the power spectra of the polarization parameters 
appear similar to that obtained for the total synchrotron intensity and the 
Faraday depth map. A direct comparison of the power spectra of 
${\rm PI}_{\nu}$ with that of the Stokes $Q$ and $U$ (see Appendix~\ref{sec:spec_qu}) 
reveals that at low frequencies near $0.5\ghz$ the power spectra of $Q$ 
and $U$ parameters remain flat over the entire range of $k$, while in 
contrast the spectrum of ${\rm PI}_{\nu}$ falls off at large $k$. This is 
because strong Faraday rotation and depolarization give rise to strong 
fluctuations in the values of Stokes $Q$ and $U$ parameters that are 
changing signs on scales of pixel resolution of these MHD simulations.

An interesting feature is noticed in the power spectra of Stokes $Q$ and $U$
parameters. In the saturated phase, due to stronger magnetic field strengths,
Faraday rotation effects at low frequencies (around $0.5\ghz$) significantly
wipe out any resemblance to the power spectra of ${\rm PI}$ and FD on all 
scales. This is, however, not the case during the kinematic stage and correlated
structures in Stokes $Q$ and $U$ are observed on scales of few tens of pixels.
Such a trend, if seen in astronomical observations, could allow us to identify
the evolutionary stage of a fluctuation dynamo operating in galaxy clusters.

\subsubsection{Correlation scales}

\begin{table}
\centering
\setlength{\tabcolsep}{5.0 pt}
\caption{Values of the integral scales (in kpc) of the velocity ($L_{{\rm int},V}$), 
magnetic fields ($L_{{\rm int},M}$), Faraday depth ($L_{\rm int, FD}$), the total 
synchrotron intensity ($L_{{\rm int},I}$) at $1\ghz$, and the polarized intensity 
($L_{{\rm int},{PI}}$) at $0.5, 1$ and $6\ghz$ at two different times in the steady 
state.}
\begin{tabular}{ *{9}{c}}
    \hline
$t/t_{\rm ed}$  & $L_{{\rm int},V}$ &  $L_{{\rm int},M}$ & $L_{\rm int, FD}$ &
$L_{{\rm int},I}$ & \multicolumn{3}{c}{$L_{{\rm int},PI} ~(\kpc)$}  \\
      & $(\kpc)$ & $(\kpc)$  & $(\kpc)$ & $(\kpc)$ & $0.5\ghz$ & $1\ghz$ & $6\ghz$  \\ \hline
16.6 & 320  &  106  &  212.5  &  224 &  122 &  155 &  199  \\
23 & 340 & 112.4 & 216 & 227.6 & 138 & 128 & 182 \\ \hline
\end{tabular}
\label{tab:intgscales}
\end{table}

It is often of interest to relate the correlation scales or the integral scales of 
observable quantities like the total and polarized intensity of the synchrotron 
emission to those corresponding to the random magnetic and velocity fields. 
We can measure these directly from our simulation, using the power spectra 
discussed above. We define these from the 1D spectra of various quantities, 
as for the magnetic integral scale $L_{{\rm int},M}$ defined in equation 
(\ref{sigmaFD}). Table~\ref{tab:intgscales} shows the values of these integral 
scales (in $\kpc$) computed from two different snapshots in the saturated 
phase of the dynamo corresponding to $t/t_{\rm ed} = 16.6$ and $23$. This 
allows us to check the sensitivity of these scales to random fluctuations in 
the field. In both cases, we find that the velocity integral scale, 
$L_{{\rm int},V} \sim 3\,L_{{\rm int},M}$. It is further evident that, although 
there is some random scatter from one realization to another, the integral 
scales associated with all the observables, FD, $I_{\rm sync}$ and the 
high-frequency ${\rm PI}$ are all comparable and larger than the magnetic 
integral scale by a factor of about two (Table~\ref{tab:intgscales}). However, 
for the polarized intensity these are frequency dependent due to the effect 
of Faraday depolarization. The integral scales at lower frequencies are 
generally smaller, due to small-scale structures introduced by Faraday 
depolarization.

\section{Smoothed polarization parameters} \label{sec:smooth_pol}

So far we have presented results from synthetic broad-band observations
obtained at the native-pixel resolution of the simulations. In order to make
any meaningful connection between the synthetic observations and astronomical
observations performed using a radio telescope with finite resolution, it is
necessary to smooth the synthetic maps. A finite telescope resolution, 
referred to as the \textit{beam}, has a significant impact on the nature of 
structures that are observed in the plane of the sky, especially in the case 
when the corresponding spatial resolution is significantly larger than the 
intrinsic scale of emission. For positive definite quantities, such as the 
total intensity, the telescope beam smoothes out fine-scale structures 
while generally preserving larger-scale features. This, however, is not the 
case for polarized intensity because smoothing of the Stokes $Q$ and 
$U$ parameters that frequently change sign spatially leads to two types 
of depolarization phenomena, namely, \textit{beam} and \textit{Faraday depolarization}.

\textit{Beam depolarization} is caused by turbulent magnetic fields on scales
smaller than the beam, and is independent of the frequency of observations. 
In contrast, \textit{Faraday depolarization} is strongly dependent on the
frequency of observations, and is caused by local fluctuations in FD along 
the LOS as well as in the plane of the sky when linearly polarized signal 
propagates through magneto-ionic media \citep[see e.g.][]{burn66, tribb91, sokol98}.
This means that rapid spatial variations of Stokes $Q$ and $U$ parameters 
when smoothed can result in recovering polarized structures that are 
significantly different compared to the intrinsic structures. Stronger 
Faraday rotation towards frequencies below $\sim 1.5\ghz$ also means 
that the observed polarized emission can appear significantly different at 
different frequencies. Hence, it is crucial to investigate the effects of beam 
smoothing in combination with Faraday depolarization to determine 
optimum frequency of observations so as to gain maximum insight into 
the magnetic field properties of galaxy clusters with future observations.

Here we investigate the effects of smoothing by convolving the synthetic 
images obtained at native resolution of simulated data with a 2D Gaussian 
kernel. Details of the convolution process performed in {\scriptsize COSMIC} 
is given in Appendix~\ref{sec:kernel}.  In the bottom two rows of 
Fig.~\ref{fig:pol_subsonic}, we show the smoothed maps of ${\rm PI}_\nu$ 
and $p_\nu$. Smoothing was performed using a symmetric kernel size with 
full width at half-maximum (FWHM) of $10\times 10$~pixel$^2$ corresponding 
to Gaussian spatial smoothing on $4.25\kpc$ scale. We note that, at the 
distance of the Coma cluster of about 100 Mpc, a physical scale of 1 kpc 
corresponds to an angular scale of 2 arcsec, and thus this smoothing would 
be over a beam with FWHM of 20~arcsec. The corresponding 2D maps of 
the Stokes $Q$ and $U$ are shown in bottom two rows of Fig.~\ref{fig:qu_subsonic}. 

In Fig.~\ref{fig:pol_subsonic}, comparing the smoothed maps with the maps 
at native resolution, noticeable differences in the appearance of polarization 
features are seen, especially at 0.5 and 1~GHz due to substantial Faraday 
rotation and depolarization-induced spatial fluctuations in Stokes $Q$ and 
$U$ parameters (see Fig.~\ref{fig:qu_subsonic}). It should be noted 
that we have saturated the colour scales of polarized intensity and fractional 
polarization in Fig.~\ref{fig:pol_subsonic} towards lower values, roughly 
corresponding to typical sensitivities achievable with current radio telescopes. 
Fractional polarization maps are saturated below 0.02. Most of the bright 
filamentary structures and diffuse polarized emission that are seen at native 
resolution are lost in the smoothed maps at lower frequencies. Bright 
polarized emission at frequencies below $\sim1.5\ghz$ are confined as 
\textit{clumpy} structures that cannot be readily identified with features in 
either the total synchrotron intensity nor in the Faraday depth maps. 
These clumps are locally polarized at a level of up to 
$\sim0.2$ at $0.5\ghz$, and, up to $\sim0.4$ level at 1~GHz. However, with 
increasing frequency, the regions with large values of $p$ become more 
volume filling. At a higher frequency of $6\ghz$, where the effects of 
Faraday rotation are low, both the native resolution and smoothed maps 
show similar polarized structures. This brings to light an important aspect; in 
the ICM where Faraday rotation and synchrotron emission are mixed, 
higher frequency ($\gtrsim 5\ghz$) observations are better suited to infer 
magnetic properties from observations of polarized synchrotron emission.

The effects of beam smoothing in the presence of Faraday depolarization 
on the statistical properties of polarized quantities at different frequencies 
can be seen quantitatively in Table~\ref{tab:stats}. The magnitude to which 
a telescope beam affects polarization is best studied in the context of
fractional polarization because the values of other polarization parameters
depend on the spectral variation due to both Faraday depolarization and the
overall synchrotron spectrum. This can be seen in the statistical values of 
${\rm PI}_\nu$ of the smoothed maps (in Table~\ref{tab:stats}), which, unlike 
the maps at native resolution, has comparatively higher values at $1\ghz$ 
with respect to those at $0.5$ and $6\ghz$.\footnote{We would like to point 
out that this increase of polarized flux at $1\ghz$ is a feature of the physical 
properties of the set-up used for the simulations.} Also note that the mean 
and median values of Stokes $Q$ and $U$ parameters do not change 
significantly after smoothing across frequencies and remain close to zero. 
This is because, in the absence of a mean magnetic field in the simulated 
volume, their values are intrinsically distributed around zero mean. However, 
a telescope beam smooths out large spatial fluctuations in the values of 
the Stokes parameters, which manifests as reduced dispersion, except at 
a higher frequency of $6\ghz$ where Faraday rotation affects polarized 
structures the least (see Table~\ref{tab:stats}). This again points to the fact
that observations at frequencies above $\sim 5\ghz$ better represent the 
intrinsic statistical properties of polarized emission from ICM.

Smoothing introduced by a telescope beam has a drastic effect on the 
observable fractional polarization from the diffuse ICM. Although the 
median fractional polarization from the diffuse ICM obtained at native 
resolution decreases substantially with decreasing frequency from 
$p_{\rm 6\,GHz} \approx 0.35$ to $p_{\rm 0.5\,GHz} \lesssim 0.1$ 
(see Table~\ref{tab:stats}), they are still at a level such that polarization 
from fainter diffuse regions could be measured. However, the smoothed 
maps at lower frequencies have significantly lower fractional polarization 
with median $p_{\rm 0.5\,GHz} \lesssim 0.01$ and $p_{\rm 1\,GHz} \lesssim 0.1$. 
In fact, smoothing increases the relative fluctuations of $p$ with respect 
to its mean at lower frequencies, i.e. at $0.5\ghz$, $\sigma_p/\langle p\rangle$ 
increases from $\sim0.7$ at the native resolution to $\sim2$ after smoothing. 
Similar to other polarized quantities, the statistical properties of fractional 
polarization at $6\ghz$ are not significantly affected by beam smoothing.

To assess the severity of the beam smoothing, we further performed 
smoothing with kernels having FWHM of $30 \times 30$ and $50 \times 50$\,pixel$^2$. 
These represent Gaussian smoothing on linear scales of $12.75$ and $21.23\kpc$, 
respectively. Keeping the angular resolution of $10\times 10$\,arcsec$^2$ 
(FWHM) the same, these smoothing scales correspond to galaxy clusters 
roughly at redshifts $0.06$ and $0.12$ using cosmological parameters from 
\textit{Planck} 2018 results \citep{Planck2018VI}. We find that smoothing on 
larger physical scales of $\sim20$\,kpc decreases $\langle p\rangle$ at lower 
frequencies by factors of 10--20 as compared to native resolution, with values 
significantly lower than 0.01 at $0.5\ghz$; at $1\ghz$, $\langle p\rangle$ 
is expected only at a level of $\sim0.02$. In such cases, the clumpy polarized 
structures fill much less sky area as compared to smoothing by 
$10\times 10$\,pixel$^2$ discussed above. Larger smoothing scales, however, 
affects the polarization properties at $6\ghz$ less severely as compared to 
those at 0.5 and $1\ghz$ with $p$ decreasing by up to $\sim20$--25\, per cent 
on smoothing by $50\times 50$\,pixel$^2$. Therefore, for distant galaxy clusters, 
our study suggests high-frequency observations can circumvent beam smoothing 
issues. In addition, to minimize spatial correlation within the beam for power 
spectra analyses, high angular resolution $\lesssim1\times 1$\,arcsec$^2$ 
is required, which will also help to infer intrinsic spatial structures of polarized 
emission.

\section{Application of RM synthesis}
\label{sec:rmsynth}

Magnetic fields in the ICM have been inferred via FD measured towards 
polarized sources located behind galaxy clusters. Alternatively, they could also 
be probed if the galaxy cluster itself has polarized synchrotron emission. 
Here we assume this to be the case to compute FD and fractional polarization 
from synthetic broad-bandwidth observations of this emission by applying the 
commonly used techniques of RM synthesis \citep{brent05} and 
{\scriptsize RM\,CLEAN} \citep{heald09}.

\begin{figure*}
\centering
\begin{tabular}{cc}
{\mbox{\includegraphics[width=0.45\textwidth]{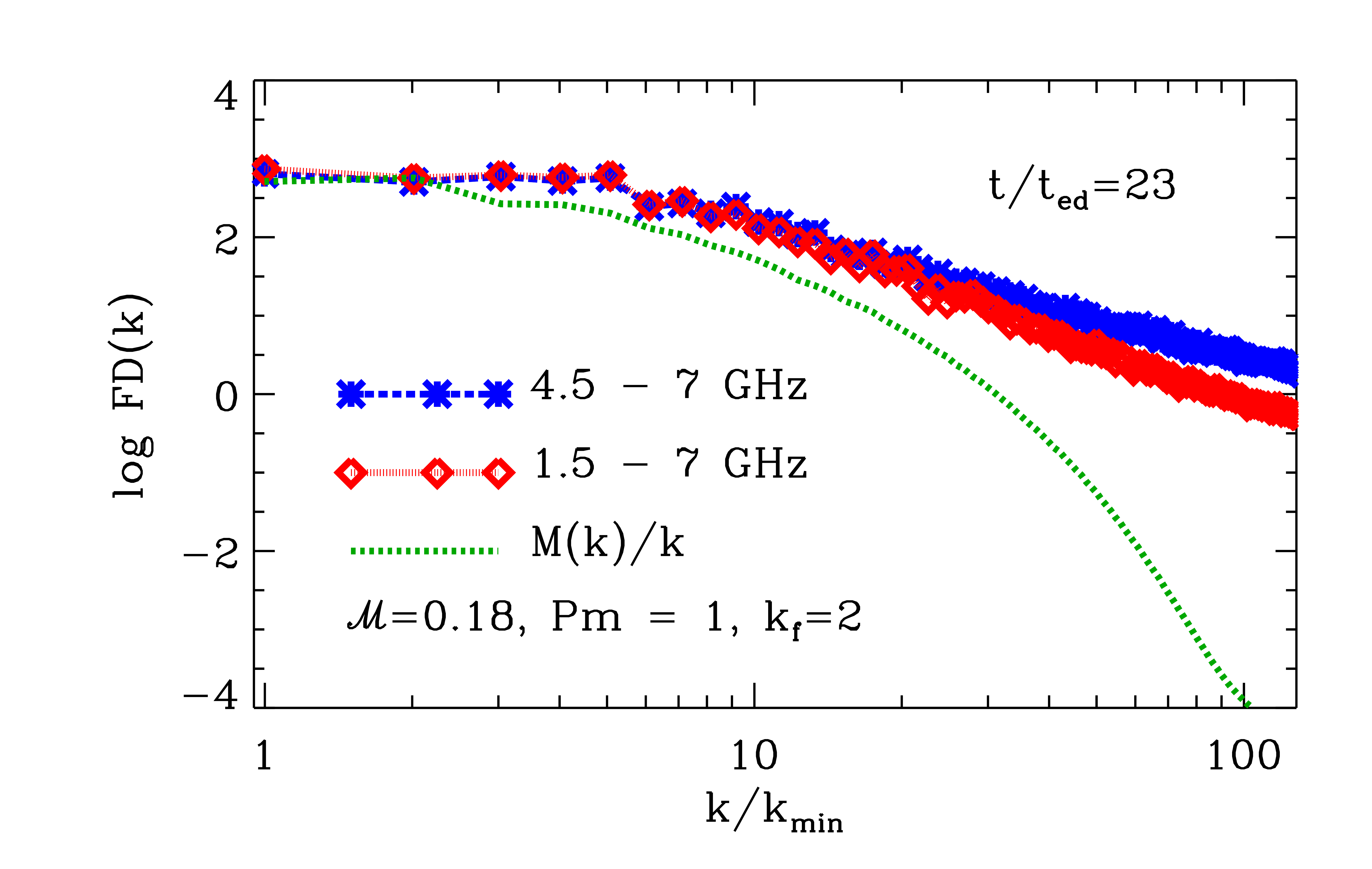}}}&
{\mbox{\includegraphics[width=0.45\textwidth]{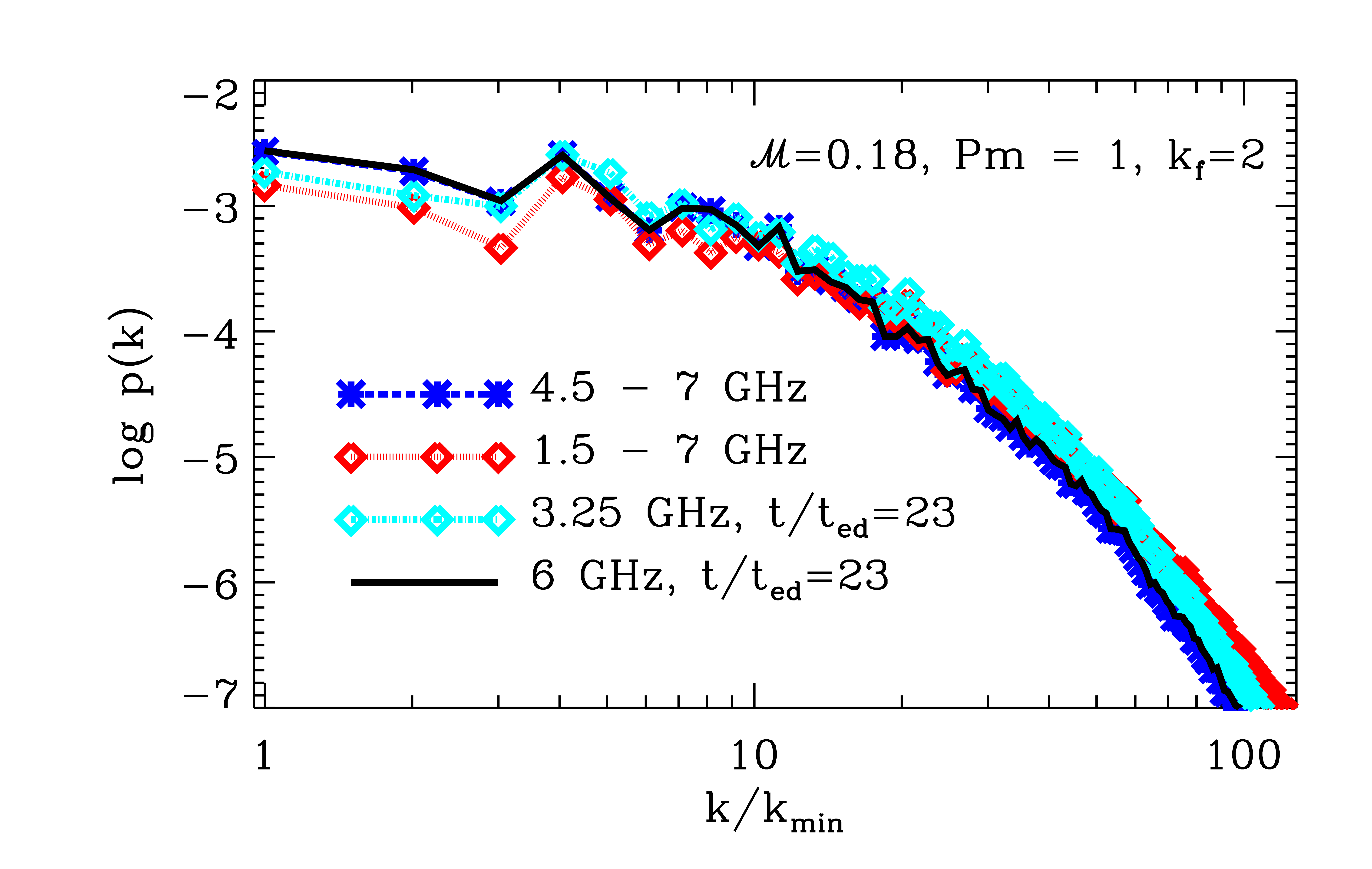}}}\\
\end{tabular}
\caption{Power spectra of the $\rm FD_{RM\,synth}$ (left) and the fractional 
polarization (right) maps obtained by applying RM synthesis to synthetic 
spectro-polarimetric observations in the frequency range $1.5  \textrm{--} 7\ghz$ 
(red points) and $4.5  \textrm{--} 7\ghz$ (blue points). The power spectrum 
of $M(k)/k$ is shown in green (same as Fig.~\ref{fig:fdPS_sub}), and the 
power spectra of the fractional polarization at $3.25$ and $6\ghz$ obtained 
directly from the synthetic maps are shown as the light-blue points and black 
curve, respectively (see text for details). The wavenumber is normalized in 
units of $\kmin = 2\upi/L$.}
\label{fig:spec_rmsynth}
\end{figure*}

As seen in Section \ref{sec:polarization}, Faraday depolarization affects the spatial 
structure of ${\rm PI}_\nu$ and $p_\nu$ at frequencies below $1.5\ghz$. Therefore, 
we explore RM synthesis and {\scriptsize RM\,CLEAN} applied to broad-bandwidth 
synthetic observations for two representative frequency coverages, between 1.5 
and 7\,GHz and between 4.5 and 7\,GHz. For these two frequency ranges, the 
sensitivities to maximum observable |FD| and extended FD structures are similar, 
$\sim6\times10^5$ and $\sim1700\radm$, respectively. However, the FWHM of the 
rotation measure spread function (RMSF), which determines the resolution in FD 
space, differs by one order of magnitude, having values of $91$ and $1330\radm$ 
for the $1.5$--$7\ghz$ and $4.5$--$7\ghz$ ranges, respectively.

We have used fractional Stokes parameters, $Q/I_{\rm sync}$ and $U/I_{\rm sync}$, 
for performing RM synthesis, which provides the so-called \textit{Faraday depth spectrum} 
--- fractional polarization as a function of FD, for each spatial pixel. Maps of Faraday 
depth ($\rm FD_{\rm RM\,synth}$) and fractional polarization ($p_{\rm RM\,synth}$) 
were computed from the {\scriptsize RM\,CLEAN}ed Faraday depth spectrum at each 
pixel. We refer the reader to Appendix~\ref{sec:rmsynthApp} for details and only 
present the results here. In the left-hand panel of Fig.~\ref{fig:spec_rmsynth} we show 
the power spectra of $\rm FD_{RM\,synth}$ derived from its 2D map and, in the 
right-hand panel, the corresponding power spectra of $p_{\rm RM\,synth}$. 
These have been obtained for the two frequency ranges mentioned above. 
Since RM synthesis provides fractional polarization parameters at frequencies 
corresponding to the mean wavelength ($\lambda$)-squared of the frequency 
coverage \citep{brent05}, we also show for comparison the power spectra of 
$p$ at $3.25$ and $5.6\ghz$ computed directly as in Section \ref{sec:polarization}.

Unlike the FD power spectra shown in Fig.~\ref{fig:fdPS_sub}, the power spectra 
of the recovered $\rm FD_{RM\,synth}$ maps for both the frequency coverages 
deviate significantly in shape from $M(k)/k$ (the green curve in the left-hand panel 
of Fig.~\ref{fig:spec_rmsynth}). This is due to a combination of poor resolution in FD 
space and limitations while interpreting Faraday depth spectra, which introduces 
spurious structures in maps of $\rm FD_{RM\,synth}$. These limitations arise due 
to complicated Faraday depth spectra in the presence of intermittent magnetic 
fields, with multiple peaks at FDs where the cumulative polarized emission binned 
in FD is high \citep[see][for details]{basu19b}. Therefore, $\rm FD_{RM\,synth}$ at 
the location of the highest peak in Faraday depth spectrum may not correspond to 
the FD along the entire LOS of the synchrotron-emitting volume. Irrespective of 
the frequency coverage and/or observational noise, this underscores the limitation 
of extracting $\rm FD_{RM\,synth}$ and calls for sophisticated techniques of 
determining the FD of an LOS from Faraday depth spectra obtained for media 
that are both synchrotron emitting and Faraday rotating. Until such techniques are 
available, FD in galaxy clusters can be probed by measuring the FD of background 
sources that have stronger polarized emission than the cluster itself, provided one 
can adequately disentangle the FD structure of the background sources themselves 
from that of the ICM, and adequately account for discrete sampling of LOS probed 
by these sources.

In contrast, there is an excellent match of the power spectra of $p_{\rm RM\,synth}$ 
with the corresponding power spectra of $p$ for both the frequency ranges. This is 
because, at the effective frequencies of $3.25$ and $5.6\ghz$ of $p_{\rm RM\,synth}$ 
obtained from RM synthesis, Faraday depolarization is rather low. Moreover, as the 
Faraday depth spectra for these two frequency coverages remains mostly unresolved, 
their peaks correspond well to the expected polarization fraction. This further 
emphasizes the requirement for studying polarization from the ICM of galaxy 
clusters at high frequencies in order to extract meaningful information on their 
magnetic field properties.

\section{Conclusions} \label{sec:conc} 

Statistical studies of Faraday RM in a number of galaxy clusters reveal that the 
ICM is magnetized with $\mkG$-level magnetic fields. As argued in Section \ref{sec:intro}, 
fluctuation dynamos are now acknowledged to be the main driver of amplification 
and maintenance of magnetic fields in these systems. In this paper we have 
explored in detail the properties of polarized emission, particularly the synchrotron 
emissivity and the polarization signals that arises from such turbulent dynamo 
action. Even though,  a great deal of work has been done on Faraday RM on 
both the observational and theoretical fronts
\citep{CKB01,CT02,VE03,SSH06, CR09,B+10,BS13,BCK16, Mari+18, SBS18, 
Vazza+18, On+19}, little attention has been paid to understanding the emissivity 
and polarization signals generated by the intermittent magnetic field structures 
produced by fluctuation dynamos. This is primarily because detection of polarized 
signals from radio halos has so far proved to be an arduous task with current radio 
interferometers \citep{Vacca+10}. More recent works \citep{Govoni+13, Govoni+15, 
Loi+19b} explore the possibility of their detection with upcoming radio interferometers 
like the SKA and its precursors. The issue is non-trivial to say the least as the 
observables themselves such as the synchrotron intensity ($I_{\rm sync}$) and 
the Stokes $Q$ and $U$ parameters are related to the components of the field 
in a non-linear manner.  This in turn makes them sensitive to the underlying 
structures of the magnetic field. 

The approach that we have adopted here is to use data from a numerical
simulation of fluctuation dynamos as an input to the {\scriptsize COSMIC} 
package of \citet{basu19b} to perform synthetic observations between $0.5$ 
and $7\ghz$ in the non-linear saturated state of the dynamo. Apart from the 
2D and 3D maps of FD and $I_{\rm sync}$ and their associated power 
spectra, we paid particular attention to the frequency dependence of the 
Stokes $Q$ and $U$ parameters and the effects of both Faraday depolarization 
and beam smoothing. We further computed the power spectra of FD and $p$ 
by applying the technique of RM synthesis to the simulation data. We have 
not added any complications introduced due to observation systematics or 
noise. In what follows, we summarize and discuss the main findings of our 
work. 
\begin{enumerate}[(i)]
\item By shooting LOS through the simulation volume we obtained 
$\sigmafd \sim 100\radm$ across four statistically independent realizations 
of the non-linear saturated state of the dynamo. These values are in close 
agreement with the observed values in the ICM. The PDF of FD (in the 
non-linear steady state) is well represented by a Gaussian distribution in 
each of the snapshots (see Section \ref{sec:fd}). While previous theoretical 
studies had hinted at a possible relation between the LOS integral of the 
magnetic field and the magnetic integral scale, here we confirm for the 
first time that the power spectrum of FD is strikingly similar to that of 
$M(k)/k$ over the entire range of scales. This implies that one can in 
principle reconstruct the power spectrum of the random magnetic fields 
by observationally determining the FD power spectrum, at least in bulk of 
the ICM where turbulence is believed to be subsonic. Moreover, in the ICM 
where turbulence is driven by a combination of mass accretion from 
filaments and major and minor mergers, the scale of structures in FD can 
be correlated at the most on the scale of turbulent driving. 

\item Analysis of the total synchrotron emission reveals that both 3D volume
rendering and the 2D map also show bright structures extending to about 
half the scale of the box. Because of the assumed constant distribution of 
cosmic ray electrons, these structures essentially represent the effects of 
random stretching and twisting of the field lines due to random turbulent 
driving. However, unlike the Gaussian PDF obtained for the FD, we find 
that the PDFs of total synchrotron intensity show a distinct lognormal
distribution. 

\item In accordance with the stated objectives of this work we probed in 
detail the frequency dependence and effects of finite size of the telescope 
beam on the polarization parameters (${\rm PI}_{\nu}$, $p_{\nu}$, Stokes 
$Q$ and $U$) at three representative frequencies, $0.5, 1$ and $6\ghz$. 
Our key observations are as follows : 

\begin{enumerate}

\item Frequency-dependent Faraday depolarization significantly affects
polarized structures observed at low radio frequencies ($\lesssim 1.5\ghz$).
This leads to the emergence of small-scale structures on scales much 
smaller than the driving scale of turbulence (see e.g. the top row in 
Fig.~\ref{fig:pol_subsonic}). The presence of these small-scale structures 
is further confirmed by larger power at large $k$ in the power spectra of 
${\rm PI}_{\nu}$ shown in Fig.~\ref{fig:spec_pol_sub}.  Concomitantly, 
fractional polarization is relatively high where frequency-dependent 
Faraday depolarization is low and vice versa. 

\item Map of ${\rm PI}$ at low frequencies of $0.5\ghz$ shows strong 
spatial variations on small scales. This results in flattening of the power 
spectra of Stokes $Q$ and $U$ on all scales as seen in 
Fig.~\ref{fig:spec_qu_sub}. This poses a serious challenge while 
inferring magnetic field properties from low frequency observations.

\item Careful analysis of the power spectra of Stokes $Q$ and $U$ 
reveal that, in contrast to the kinematic stage, Faraday rotation 
effects at $0.5\ghz$ erase any resemblance to the power spectra of 
${\rm PI}_{\nu}$ and FD on all scales. This provides an opportunity 
to diagnose the evolutionary stage of the dynamo in astronomical 
observations. 

\item Comparison of the smoothed maps of ${\rm PI}_{\nu}$ and 
$p_{\nu}$ with those at the native-pixel resolution of the simulations 
reveals that higher-frequency observations ($\gtrsim 5\ghz$) should 
be ideally preferred as they better represents the statistical properties 
of the polarized emission of the ICM and are therefore more suited to 
infer magnetic field structures in the ICM.

\end{enumerate}

\item The power spectra of FD obtained from RM synthesis
applied to polarized emission in frequency bands $4.5  \textrm{--} 7\ghz$ 
and $1.5  \textrm{--} 7\ghz$ deviate significantly in shape from the 
power spectrum $M(k)/k$. On the other hand, the power spectra 
of $p$ obtained using the same technique match accurately those 
computed directly from the simulated data for both frequency ranges.

\end{enumerate}

\section{Discussion and Outlook} \label{sec:discuss}

Our key findings outline the importance of and need for high-frequency 
observations in order to gain an understanding of the properties of 
polarized emission together with the structural properties of random 
magnetic fields in the ICM. In this regard, Band\,5 of SKA ($4.6 \textrm{--} 15.3\ghz$) 
will be ideally suited to detect polarized emission from the ICM. 
Our work clearly demonstrates that at frequencies below $1\ghz$ 
a combination of low sensitivity and observation noise, will result 
in polarized emission to be detected only from bright filamentary 
structures that could originate either due to shock compression or 
from Faraday depolarization. Thus, such detections may not provide 
adequate information on the global properties of turbulent magnetic 
fields in the ICM.

We find that the telescope beam drastically affects the properties of 
polarized emission in the presence of Faraday rotation at low frequencies. 
At frequencies below $\sim1\ghz$, depolarization within the beam is 
dominated by fluctuations introduced by Faraday rotation and the 
polarized structures are confined as clumps. For radio-frequency 
observations of ICM performed at frequencies near $0.5\ghz$ with 
an angular resolution of $10\times10$~arcsec$^2$ sampled with 
1~arcsec pixels, our synthetic maps suggests that these clumps have 
median fractional polarization at the level of a few per cent, reaching 
a maximum of $\sim 20$ per cent with a polarized intensity of 
$\sim20\,\mJy$\,beam$^{-1}$. This means that, in order to detect 
polarization at more than $5\sigma$ \citep[to reduce Ricean bias;][]{wardl74}, 
a sensitivity of $\sim4\,\mJy$\,beam$^{-1}$ is needed. Detecting 
such levels of polarized emission will be a challenging proposition 
with telescopes currently operating at these frequencies or with 
the SKA-LOW, and, if detected at all, they are expected to be highly 
sporadic in nature, and even more so in the presence of realistic 
noise. This will hamper the inference of any meaningful information 
on the turbulence properties of ICM magnetic fields through observations 
performed at frequencies below $1\ghz$. We find that detecting 
polarized emission from ICM below $1\ghz$ and inferring the 
magnetic field properties are already difficult in the absence of 
realistic observations noise. We will address additional complications 
introduced by noise in a follow-up study. We emphasize that the 
flux densities and the sensitivity presented in this work are normalized 
for a Coma-like nearby cluster. Therefore, our assumed total synchrotron 
flux density of 1~Jy at 1~GHz, and thereby the estimated sensitivity 
requirements above, are already significantly higher than that expected 
for distant galaxy clusters.

In contrast, the telescope beam does not significantly affect the 
polarized emission at frequencies above $5\ghz$; and, the median level 
of polarization increases to about 30 per cent (see Table~\ref{tab:stats}) 
and the diffuse polarized intensity is expected to be few tens of 
$\mJy$\,beam$^{-1}$. ICM observations with the Karl G. Jansky Very 
Large Array (VLA) in the $4 \textrm{--} 8\ghz$ frequency range can 
detect such levels of polarized intensity with rms noise of few 
$\mJy$\,beam$^{-1}$. However, a challenge is faced with the fact that 
the Stokes $Q$ and $U$ parameters maps could have diffuse large-scale 
structures that span up to about $200\kpc$. This means that, in our 
choice of angular scales (a pixel size of 1~arcsec corresponds to a 
total projected size of the simulated map of 8.5~arcmin), the diffuse 
structures in Stokes $Q$ and $U$ could have angular extent of 
$\sim 3 \textrm{--}5$~arcmin. Missing spacing for interferometric 
observations with the VLA would suffer from missing emission on large 
angular scales, especially for the total intensity synchrotron emission. 
To address this issue, long observations using the VLA are required to 
improve the $u-v$ coverage in the 4--8$\ghz$ band. This issue can be 
easily circumvented by the Band\,5 of the SKA which will have compact 
network of antennas in its core \citep{Dewdney16} and in the near future 
by the extended MeerKAT array in the 1.75--3.5\,GHz range \citep{kramer16}. 
The shortest projected baseline of $\sim20$\,m for the SKA and the 
MeerKAT will be sensitive to angular scales up to $\sim7.5$\,arcmin, 
which will be sufficient to capture diffuse synchrotron emission from 
nearby galaxy clusters up to $8\ghz$ in Band\,5. For clusters at higher 
redshifts, the problem of resolving out diffuse emissions will be lower. 
In such cases, however, to avoid beam depolarization when averaged 
over large spatial scales or distinguish correlated structures within the 
beam with that from intrinsic structures, sub-arcsec angular resolution of 
the SKA will be crucial.

We have assumed here a constant $n_{\rm CRE}$ throughout the simulated 
volume following a power-law energy spectrum with a constant energy index 
of $\gamma=-3$. CREs emitting at GHz frequencies undergo radiative cooling 
predominantly due to synchrotron losses and inverse-Compton (IC) scattering 
with cosmic microwave background (CMB) photons, which can cause a spectral 
steepening beyond a break frequency $\nu_{\rm br}$. For the $\umu$G strength 
fields and $\ghz$ frequencies considered here, this cooling time scale is typically 
less than $10^8$ yr \citep[e.g.][]{longa11}. In this time CREs diffuse away from 
their source regions only by a distance of the order of tens of parsec under 
Bohm diffusion \citep{Drury83, Bagchi+02}, where one assumes strong 
electromagnetic fluctuations on the Larmour radius of the CREs, and kpc scales 
even if the waves responsible for resonant scattering of the CREs at their 
Larmour radii make up only a small fraction say $\sim 10^{-4}$ of the total wave 
energy \citep{Bru_Jon14}. This length scale is much smaller than the Mpc scales 
associated with radio halos and thus CREs need to be reaccelerated away 
from their sources, perhaps by the same turbulence that also leads to dynamo
action. For cluster-wide turbulence, we expect variation of the energy going into 
CREs to vary on cluster scales of order Mpc, and thus the approximation of an 
almost constant $n_{\rm CRE}$ over turbulent eddy scales seems reasonable. 
Moreover, for clusters at redshift $z$ with magnetic fields smaller than 
$3.2\,(1+z)^2\,\umu$G, as in our simulations, IC cooling in presence of the 
nearly uniform CMB dominates synchrotron cooling, and $\nu_{\rm br}$ would 
be less sensitive to the local ICM magnetic field. Finally, the same ICM 
turbulence is also expected to efficiently mix CREs due to the larger turbulent 
diffusivity, again damping spatial variations, including those in $\nu_{\rm br}$, 
resulting from cooling. A more quantitative study of these effects requires one 
to solve the CRE transport equations as well, incorporating all the above effects. 
However, we do expect our assumption of constant spatial distribution of CRE 
to be reasonable and thus the results on shape of the power spectrum of 
synchrotron emission presented in Figs~\ref{fig:totIPS_sub} and 
\ref{fig:spec_pol_sub} are likely robust.

In this work, we have made use of turbulence in a box simulation to probe the
properties of polarized emission resulting from fluctuation-dynamo generated
magnetic fields in the context of galaxy clusters. On the other hand, over the
course of the last two decades there has developed a significant body of work 
on cosmological simulations of large-scale structure formation together with
the formation of massive galaxy clusters that also include magnetic fields
\citep{DBL99, Xu+09, Xu+11, Xu+12, Miniati14, Miniati15, Mari+18, Vazza+18, 
Domin+19}. While these global simulations have the advantage of accommodating 
a large range of scales from cluster radius and beyond, they are also limited in 
terms of the resolution required to resolve the much smaller turbulent eddy 
scales which are at the heart of the amplifying negligible seed magnetic fields 
through dynamo action. Moreover, these simulations are devoid of physical 
viscosity and resistivity and thus the diffusion of magnetic fields is completely 
governed by the numerical scheme. In view of these limitations our simulations 
although performed in an idealized setting, offer a complimentary route to address 
the important issues discussed in this work. 

Our work considers a small representative volume of the ICM of size $512\kpc$
where the initial density is assumed to uniform. However, it is well known 
that the ICM is stratified with radius. Moreover, continuous accretion of matter 
from filaments and major and minor mergers renders the density distribution 
more complex \citep[e.g.,][]{Shi+18,Roh+19}, decreasing away from the cluster 
core. This would be particularly important if one where to probe LOS through 
many such smaller representative volumes (as considered here) arranged over a 
large radial distance scale. It would be equally interesting to perform a detailed 
analysis on the effects of intermittency of the magnetic field on the observables 
discussed in this work. Apart from galaxy clusters, the tools and methodology 
used here can be applied to the study of magnetic fields in young galaxies in 
the high-redshift Universe \citep{berne08,farne14,MCS20} where fluctuation 
dynamos could be responsible for generating and maintaining fields of strengths 
comparable to those found in nearby spiral galaxies \citep{SBS18}. These topics 
will form the subject of our investigation in future work.

\section*{Acknowledgments}
We thank Rainer Beck for critical comments and suggestions that improved 
the presentation of the results. We also thank the anonymous referee for a 
timely and constructive report. SS acknowledges computing time awarded at 
the CDAC National Param supercomputing facility, India, under the grant 
`Hydromagnetic-Turbulence-PR' and the use of the HPC facilities of IIA. 
He also thanks the Science and Engineering Research Board (SERB) of the 
Department of Science \& Technology (DST), Government of India, for 
support through research grant ECR/2017/001535. AB acknowledges financial 
support by the German Federal Ministry of Education and Research (BMBF) 
under grant 05A17PB1 (Verbundprojekt D-MeerKAT). The software used in 
this work was in part developed by the DOE NNSA-ASC OASCR Flash 
Center at the University of Chicago. This research made use of Astropy,\footnote{http://www.astropy.org} a 
community-developed core Python package for Astronomy \citep{astropy:2013, 
astropy:2018}, NumPy \citep{numpy11} and Matplotlib \citep{matplotlib07}.

\section*{Data Availability}
The simulation data, synthetic observations, and the {\scriptsize COSMIC} 
package will be made publicly available, until which they will be shared 
on reasonable request to the authors.

\bibliographystyle{mnras}


\appendix

\section{Numerical calculations in {\scriptsize COSMIC}} \label{sec:Appcosmic}

\subsection{Polarization parameters} \label{sec:cosmic}

A detailed discussion on the numerical calculations performed by 
{\scriptsize COSMIC} can be found in \citet{basu19b}. Here, we 
summarize in brief the basic equations used for calculating various 
observables presented in this paper.

The total synchrotron emissivity, $\varepsilon_{\rm sync}$, at a 
frequency $\nu$ at a mesh-point ($i, j, k$) is computed from the 
magnetic field component in the plane of the sky $B_\perp$ in 
each cell as
\EQ
\varepsilon_{\mathrm{sync}, \nu}(i, j, k) = \widetilde{N}_0\,n_{\rm CRE} \, \left[B_\perp(i,j,k)\right]^{1-\alpha}
\, \nu^{\alpha}.
\label{eq:sync_em}
\EN
Here, $B_\perp = (B_x^2 + B_y^2)^{1/2}$, $n_{\rm CRE}$ is the 
number density of cosmic ray electrons (CRE), $\widetilde{N}_0$ 
is an arbitrary normalization factor, and $\alpha$ is the spectral 
index of the synchrotron emission. $B_x$ and $B_y$ are the 
magnetic field components in Cartesian coordinate system.
The total synchrotron intensity ($I_{\rm sync}$) map is obtained 
by integrating $\varepsilon_{\mathrm{sync}, \nu}(i, j, k)$ along the 
$z$-axis as
\EQ
\begin{split}
I_{\rm sync, \nu}(i, j) & = \sum_k\, \varepsilon_{\mathrm{sync}, \nu}(i, j, k)
\ l_{\rm cell} \\
      & = N_0\,n_{\rm CRE} \,\nu^{\alpha} \, \sum_k\, \left[B_\perp(i,j,k)\right]^{1-\alpha}
\end{split}
\label{eq:sync_intensity}
\EN
In this paper, we have assumed constant $n_{\rm CRE}$ and a constant 
value of $\alpha = -1$. Here, $N_0 \equiv \widetilde{N}_0 \ l_{\rm cell}$ is 
a normalization that allows for a choice of user-defined total synchrotron 
intensity from the simulation. In {\scriptsize COSMIC}, the normalization 
$N_0$ is chosen such that, at $\nu = 1\ghz$, $I_{\rm sync, 1\,GHz} = 1$~Jy 
over the entire map.

The Stokes $Q$ and $U$ emissivities ($\varepsilon_{Q, \nu}$ and
$\varepsilon_{U,\nu}$, respectively) at each cell at a frequency $\nu$ are
computed as
\EQ
\begin{split}
\varepsilon_{Q, \nu}(i, j, k) & = p_{\rm max} \, \varepsilon_{\mathrm{syn}c, \nu}(i, j, k) \,
\cos \left[2\,\theta_0(i,j,k)\right],\\
\varepsilon_{U, \nu}(i, j, k) & = p_{\rm max} \, \varepsilon_{\mathrm{sync}, \nu}(i, j, k) \,
\sin \left[2\,\theta_0(i, j, k) \right].
\end{split}
\label{eq:qu}
\EN
Here, $\theta_0$ is the intrinsic angle of the linearly polarized emission,
\EQ
\theta_0 = \dfrac{\upi}{2} + \arctan\left(\dfrac{B_y}{B_x}\right),
\label{eq:pol_angle}
\EN
and $p_{\rm max} = (1 - \alpha)/(5/3 - \alpha)$ is the maximum fractional
polarization of the synchrotron emission. For $\alpha = -1$, $p_{\rm max} = 0.75$, 
and for a spatially constant spectrum, we used $p_{\rm max}(i,j,k) = 0.75$ 
at all the mesh points of the simulation volume.

In the presence of Faraday rotation, the 2D-projected Stokes $Q$ and 
$U$ parameters at a frequency $\nu$ are
\EQ
\begin{split}
Q_\nu(i,j) & = \sum_k p_{\rm max} \, \varepsilon_{\mathrm{sync}, \nu}(i,j,k)\, l_{\rm cell} \, \cos \left[ 2\,\theta(i,j,k)\right],\\
U_\nu(i,j) & = \sum_k p_{\rm max} \, \varepsilon_{\mathrm{sync},  \nu}(i,j,k)\, l_{\rm cell} \, 
\sin \left[ 2\, \theta(i,j,k) \right]. \\
\end{split}
\label{eq:compQU}
\EN
Here, $\theta(i,j,k)  = \theta_0(i,j,k) + {\rm FD}^\prime (i,j,k)\,c^2/\nu^2$, ${\rm FD^\prime}(i,j,k)$ 
is the Faraday depth of the mesh point $(i, j, k)$ given by
\EQ
{\rm FD}^\prime(i, j, k) = \sum\limits_{k^\prime = 0}^k \, {\rm FD}_{\rm cell}(i, j, k^\prime) - \frac 12\, {\rm FD}_{\rm cell}(i, j, k),
\label{eq:fc}
\EN
and ${\rm FD_{cell}} = 0.812\, n_{\rm e}\, B_\|\, l_{\rm cell}$ is the 
Faraday depth produced in each cell. The parallel component of the 
magnetic field $B_\| \equiv B_z$ in the default coordinate system of 
{\scriptsize COSMIC} and $l_{\rm cell} = 1$~kpc is the separation 
between the mesh points.

The linearly polarized intensity (PI) map at a frequency $\nu$ is computed
from the Stokes $Q$ and $U$ parameters in equation (\ref{eq:compQU}) 
as
\EQ
PI_\nu(i,j) = \sqrt{Q_\nu^2(i,j) + U_\nu^2(i,j)}.
\label{eq:compPI}
\EN

\subsection{Gaussian smoothing}
\label{sec:kernel}

Images obtained using radio telescopes are restored using a point-spread
function that is well represented by a 2D Gaussian \citep[see
e.g.][]{thompsonchapter17}, called as the beam. Therefore, in {\scriptsize COSMIC},
we perform smoothing of an emission quantity $S$ by convolving the synthetic
images obtained at the native resolution of the simulated data with a 2D
Gaussian kernel of a user-defined size. Smoothing is performed in the 
Fourier space by applying Fast Fourier transform (FFT) as
\EQ
\begin{split}
S_{\rm smooth}(x,y) &= G(x,y) \otimes S(x,y)\\
    & = {\rm IFFT}\left[{\rm FFT}\{G(x,y)\}\, \times {\rm FFT}\{S(x,y)\right].
\end{split}
\EN
Here, IFFT represents inverse Fast Fourier transform, $S_{\rm smooth}$ is 
the smoothed quantity and $G(x,y)$ is the convolution kernel. Gaussian 
kernel of unit amplitude of the form
\EQ
G(x,y) = {\exp}\left[{-\frac 12\left(\dfrac{x_{\rm PA}^2}{\sigma_{x}^2} + \dfrac{y_{\rm PA}^2}{\sigma_{y}^2}\right)}\right],
\EN
is used for convolution. Here, $\sigma_x$ and $\sigma_y$ are the widths 
of the major and minor axes. $x_{\rm PA}$ and $y_{\rm PA}$ accounts for 
the rotational transformation of the Gaussian kernel by a user-given positional 
angle (PA). Since the full-width at half-maximum (FWHM) of the beam's 
major and minor axes is typically reported as the resolution of a radio-frequency 
image, a user provides the FWHMs of the desired kernel that are converted 
to corresponding $\sigma_{x}$ and $\sigma_{y}$ in {\scriptsize COSMIC}. 
Note that, since a input map $S(x,y)$ produced using {\scriptsize COSMIC} 
have units of Jy\,pixel$^{-1}$, an unit amplitude convolution kernel is sufficient 
to conserve the total flux density of a map.

\section{Power spectra of the Stokes $Q$ and $U$ parameters} \label{sec:spec_qu}

\begin{figure}
\centering
\begin{tabular}{c}
{\mbox{\includegraphics[width=\columnwidth]{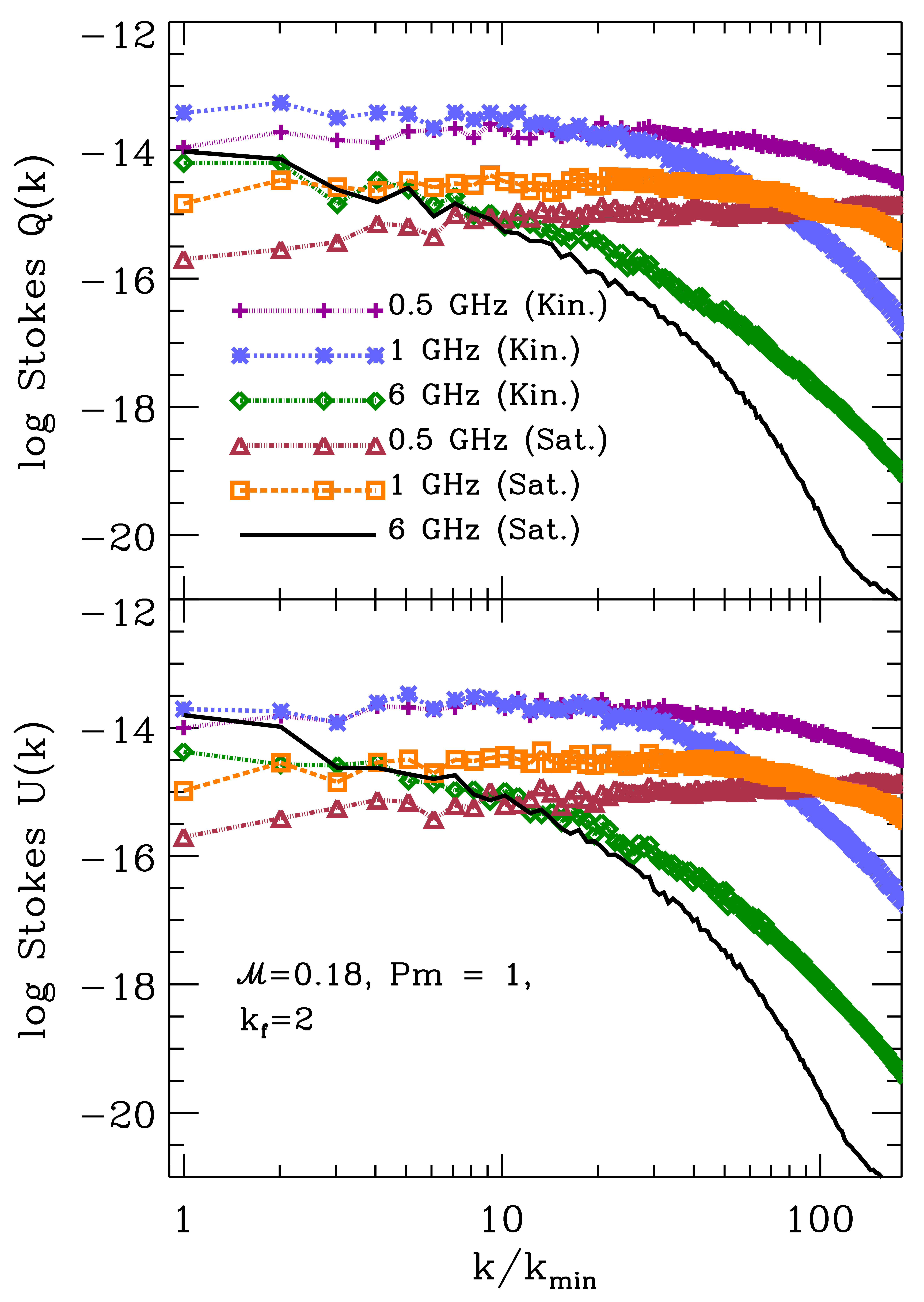}}}\\
\end{tabular}
\caption{Power spectra of the Stokes $Q$ (top) and $U$ (bottom) at 
0.5, 1 and 6~GHz in the kinematic ($t/t_{\rm ed} = 2$) and saturated 
phase ($t/t_{\rm ed} = 23$). The wavenumber is normalized in units of 
$\kmin = 2\upi/L$.}
\label{fig:spec_qu_sub}
\end{figure}

Here, we present the power spectra of the Stokes $Q$ and $U$ parameters 
in Fig.~\ref{fig:spec_qu_sub}, obtained at the native resolution of our 
simulations for snapshots during the kinematic and the saturated phases. 
The spectra for the saturated phase are obtained using maps of Stokes 
$Q$ and $U$ parameters at $0.5$, $1$ and $6\ghz$ shown in the top two 
rows of Fig.~\ref{fig:qu_subsonic}. Since Stokes $Q$ and $U$ parameters 
are sensitive to the orientation of magnetic fields, it is interesting to note 
that, although there is no mean field in our MHD simulations and the 
magnetic fields are turbulent, Stokes $Q$ and $U$ maps shows large-scale 
structures preserving their sign at higher frequencies ($\gtrsim 5\ghz$). 
This suggests that fields are locally ordered on somewhat large scales due 
to stretching and twisting by the fluctuation-dynamo action. Inferences 
about ordered magnetic fields from large-scale features in ${\rm PI}$ 
maps is not straightforward because it is positive definite. Signatures of 
large-scale features are readily seen as large power at $k \lesssim 10$ 
in the power spectra of Stokes $Q$ and $U$ parameters at $6\ghz$ for 
both the kinematic and the saturated stages.

At high frequencies, the power spectra of Stokes $Q$ and $U$ parameters
steepens significantly for wavenumbers above $\sim15$, for both kinematic
and saturated stages of field amplification. However, the slopes of the power 
spectra are different for the two stages. For the kinematic phase, the slope 
for $k > 15$ is $-3.7$, while, for the saturated phase, the slope is $-6.5$. 
Note that, in the presence of a telescope beam, structures are correlated 
within its scale, and the power spectrum will have a steeper slope by up to 
$\sim -4$ \citep{lee75} for $k > k_{\rm beam}$, where $k_{\rm beam}$ is 
the corresponding wavenumber of the Gaussian beam size. Comparatively 
steeper slope of power spectra of Stokes $Q$ and $U$ parameters obtained 
at high frequencies in the saturated phase mean that the evolutionary stage 
of field amplification in ICM can be investigated with sufficiently high angular 
resolution observations.

At frequencies below about $1.5\ghz$, Faraday depolarization increases in 
the presence of turbulent fields both in the plane of the sky and along the 
LOS giving rise to rapid spatial fluctuations in Stokes $Q$ and $U$ parameters 
(see Fig.~\ref{fig:qu_subsonic}). This flattens their power spectra in all spatial
scales for both the kinematic and saturated phases of field amplification.
Thus, extracting information on the dynamo action at low frequencies 
($\lesssim 1.5\ghz$) will be a challenging proposition.

\section{RM synthesis of synthetic spectro-polarimetric data} \label{sec:rmsynthApp}

Here we outline in detail the methodology adopted to compute FD from 
synthetic polarized synchrotron emission using RM synthesis. To this end, 
we used the {\scriptsize PYRMSYNTH} package\footnote{\url{ https://github.com/mrbell/pyrmsynth}.} 
to perform RM synthesis and {\scriptsize RM CLEAN} \citep{brent05, heald09} 
applied to fractional Stokes $Q$ and $U$ parameters, i.e. $Q(\nu)/I_{\rm sync}(\nu)$ 
and $U(\nu)/I_{\rm sync}(\nu)$ from the synthetic observations. Using these, 
we computed FD at a spatial pixel from RM synthesis ($\rm FD_{RM\,synth}$) 
following the standard method of fitting the highest peak of the Faraday depth 
spectrum by a parabola, wherein the location of the peak provides 
$\rm FD_{RM\,synth}$ and the amplitude provides the fractional 
polarization ($p_{\rm RM\,synth}$).

The frequency coverage and the channel resolution determines the sensitivity 
to structures observable in the Faraday depth spectrum. For the purpose of 
this work, we chose two frequency settings covering the frequencies (1) 
$1.5 \textrm{--} 7\ghz$ with 1024 channels of $\sim5.4$\,MHz width and 
(2) $4.5 \textrm{--} 7\ghz$ with 512 channels of $\sim4.88$\,MHz width. 
The polarization parameters from RM synthesis are obtained at a frequency 
that corresponds to the mean of the $\lambda^2$-coverage, i.e. at $3.25$ 
and $5.6\ghz$ for the frequency settings (1) and (2), respectively. 
Following the relations given in \citet{brent05}, both these frequency settings 
are sensitive to a similar maximum observable $\rm FD_{RM\,synth}$ of 
$\sim \pm 6\times10^5\radm$ and to similar structures extending up to 
$\sim1700\radm$ in the FD space. However, these two frequency settings 
have drastically different resolutions in FD space determined by the FWHM 
of the rotation measure spread function (RMSF). Frequency setting (1) has 
$\rm RMSF = 91\radm$, and setting (2) has $\rm RMSF = 1330\radm$.

\begin{figure}
\centering
\begin{tabular}{c}
{\mbox{\includegraphics[height=6cm, trim=1mm 2mm 2mm 0mm, clip]{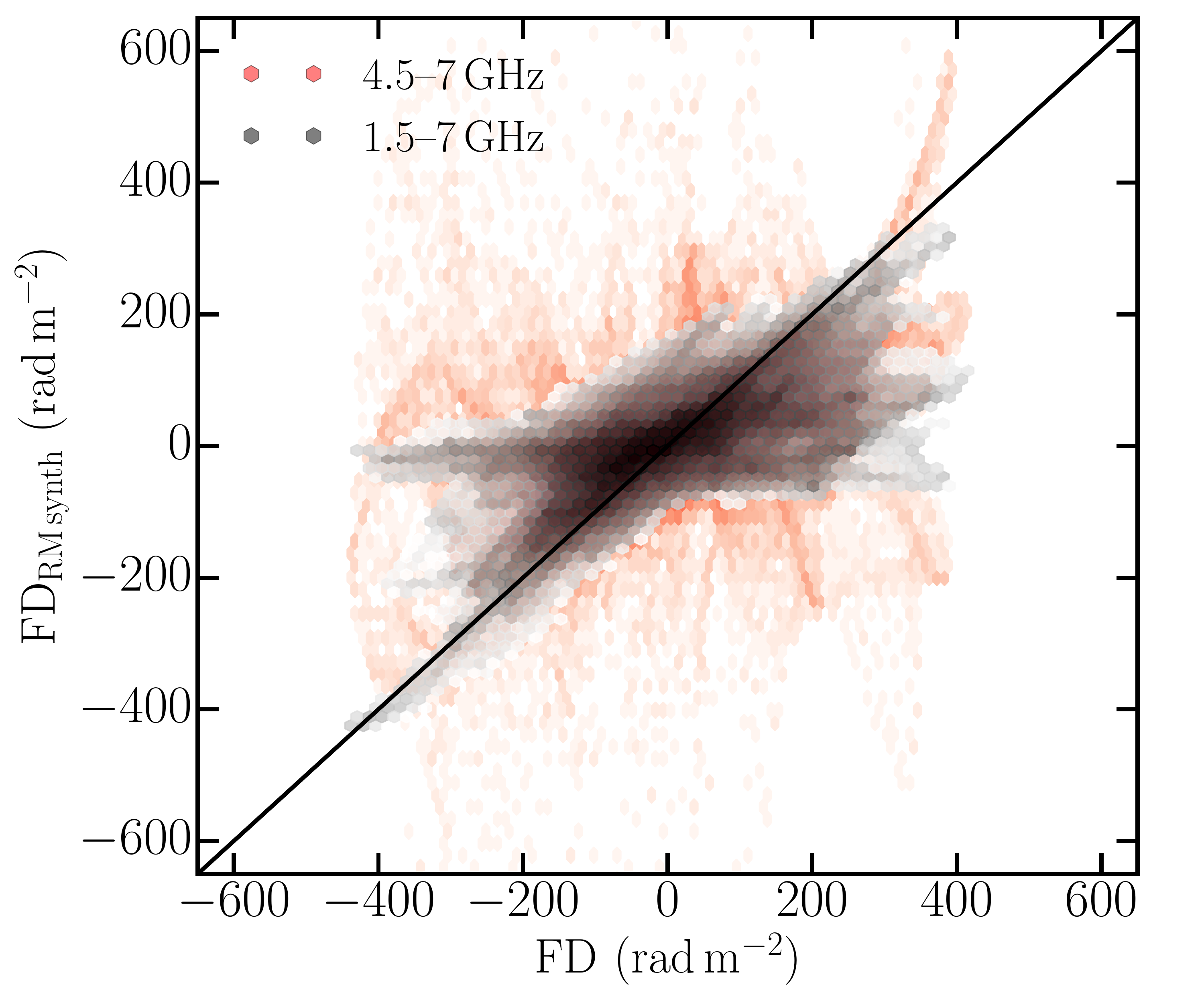}}}\\
{\mbox{\includegraphics[height=6cm, trim=1mm 2mm 2mm 0mm, clip]{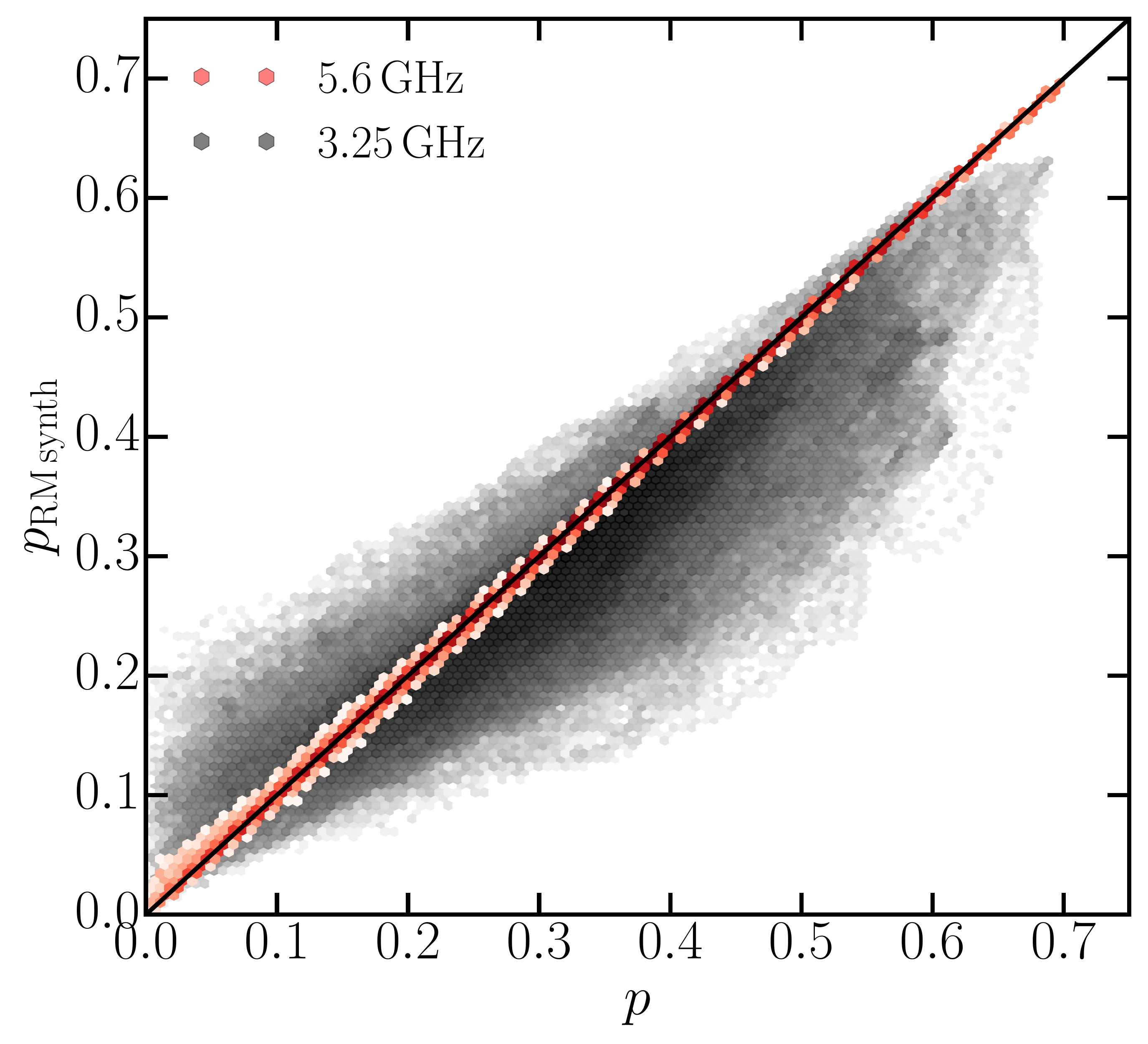}}}\\
\end{tabular}
\caption{Pixel-by-pixel comparison of maps obtained using RM synthesis with 
corresponding maps obtained directly from synthetic observations of the saturated 
phase ($t/t_{\rm ed}=23$). Top: Variation of $\rm FD_{RM\,synth}$ with 
$\rm FD_{MHD}$. Red and gray points are for RM synthesis applied to synthetic 
observations in the frequency ranges $4.5 \textrm{--} 7\ghz$, and $1.5  \textrm{--} 7\ghz$, 
respectively. Bottom: Variation of $p_{\rm RM\,synth}$ with $p_{\rm MHD}$ at 
$5.6\ghz$ (red points) and at $3.25\ghz$ (grey points). The solid line in both the 
panels is the 1:1 line.}
\label{fig:rmsynth}
\end{figure}

The top panel of Fig.~\ref{fig:rmsynth} shows the pixel-wise comparison between 
$\rm FD_{\rm RM\,synth}$ and FD obtained from our simulation in the saturated 
phase ($t/t_{\rm ed}=23$), i.e. the map shown in Fig.~\ref{fig:fd_subsonic}. It is 
clear that, due to the poor resolution in FD for the high frequency coverage, 
$4.5 \textrm{--} 7\ghz$, $\rm FD_{RM\,synth}$ is poorly determined (shown as 
red points). The estimated $\rm FD_{RM\,synth}$ lies in the range $-900$ to 
$+800\radm$, and have $\sigmafd\approx 77\radm$. For the $1.5 \textrm{--} 7\ghz$ 
coverage, although the FD resolution is higher than that of the higher-frequency 
coverage, a considerable difference between $\rm FD_{RM\,synth}$ and FD is 
seen (grey points). In this case, $\rm FD_{RM\,synth}$ lies in the range $-425$ 
to $+330\radm$ with $\sigmafd \approx 73\radm$. In contrast, 
$\sigmafd \approx 100\radm$ is obtained from FD computed directly by integrating 
along LOS. This difference is predominantly due to the fact that the Faraday 
depth spectra are partially resolved showing multiple peaks, and the location 
of the highest peak does not necessarily correspond to the FD along the entire 
LOS (see Section \ref{sec:rmsynth}). Thus the statistical properties of 
$\rm FD_{RM\,synth}$ for both the frequency coverage deviates significantly 
from the intrinsic values by up to $\pm300\radm$ and manifest as spurious 
structures in the maps of $\rm FD_{RM\,synth}$.  As a result, the power spectra 
computed from $\rm FD_{RM\,synth}$ maps shown in Fig.~\ref{fig:spec_rmsynth} 
deviates from the form determined by $M(k)/k$ in contrast to the power spectrum 
of FD map shown in Fig.~\ref{fig:fdPS_sub}.

The bottom panel of Fig.~\ref{fig:rmsynth} shows the pixel-by-pixel variation of 
fractional polarization $p_{\rm RM\,synth}$ computed from RM synthesis applied 
to the two frequency ranges with $p$ computed from synthetic observations at 
3.25 and $5.6\ghz$. At $5.6\ghz$, $p_{\rm RM\,synth}$ is in remarkable agreement 
with $p$ at better than the $\sim2$\,per\,cent level. This is a consequence of low 
Faraday rotation and depolarization at these frequencies; large RMSF means that 
complicated structures in the Faraday depth spectrum remains unresolved and all 
polarized structures in Faraday depth space are integrated within a single unresolved 
peak. As a result, polarization fractions are well recovered by a single peak. 
However, at $3.25\ghz$, $p_{\rm RM\,synth}$ versus $p$ has a much larger 
scatter and the agreement lies within a factor of about two. Notably, 
$p_{\rm RM\,synth}$ is in general underestimated as compared to $p$ at 
$3.25\ghz$. This is because, due to the relatively higher resolution in FD for the 
$1.5 \textrm{--} 7\ghz$ coverage, Faraday depth spectra are partially resolved. 
Therefore, the peak of the Faraday depth spectrum is somewhat lower than the 
expected $p$. Nonetheless, the offset of $p_{\rm RM\,synth}$ with respect to 
$p$ has negligible spatial dependence. Thus, the power spectra of $p_{\rm RM\,synth}$ 
for the two frequency ranges considered here matches the power spectra of 
actual $p$ very well, as seen in Fig.~\ref{fig:spec_rmsynth}.

\section{Smoothed Stokes $Q$ and $U$ parameters}
\label{sec:smooth_qu}

\begin{figure*}
\centering
\begin{tabular}{ccc}
\large{0.5~GHz} & \large{1~GHz} & \large{6~GHz} \\
 & & \\
 & \large{Stokes $Q$} &  \\
{\mbox{\includegraphics[height=4cm, trim=1mm 2mm 2mm 0mm, clip]{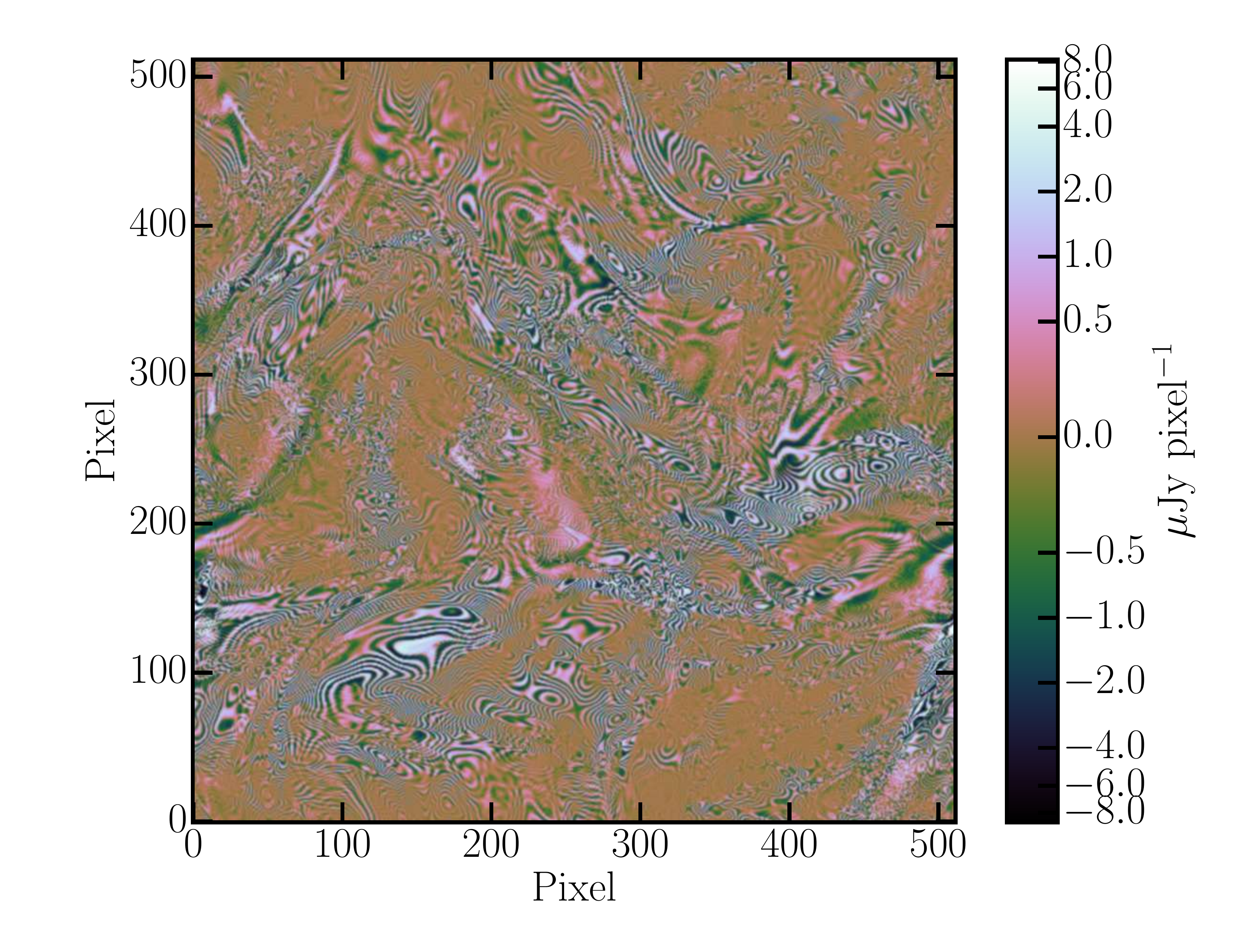}}}&
{\mbox{\includegraphics[height=4cm, trim=1mm 2mm 2mm 0mm, clip]{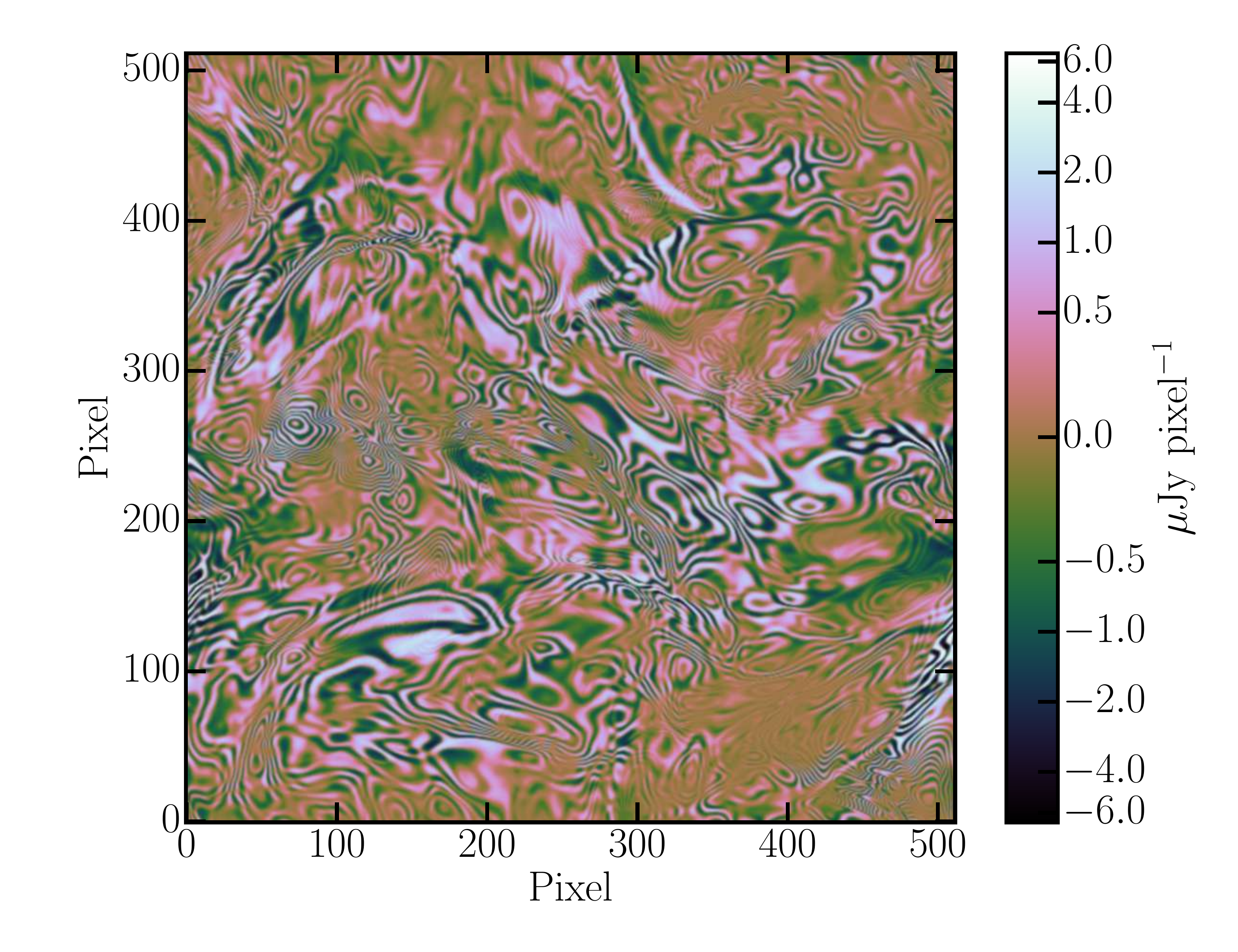}}}&
{\mbox{\includegraphics[height=4cm, trim=1mm 2mm 2mm 0mm, clip]{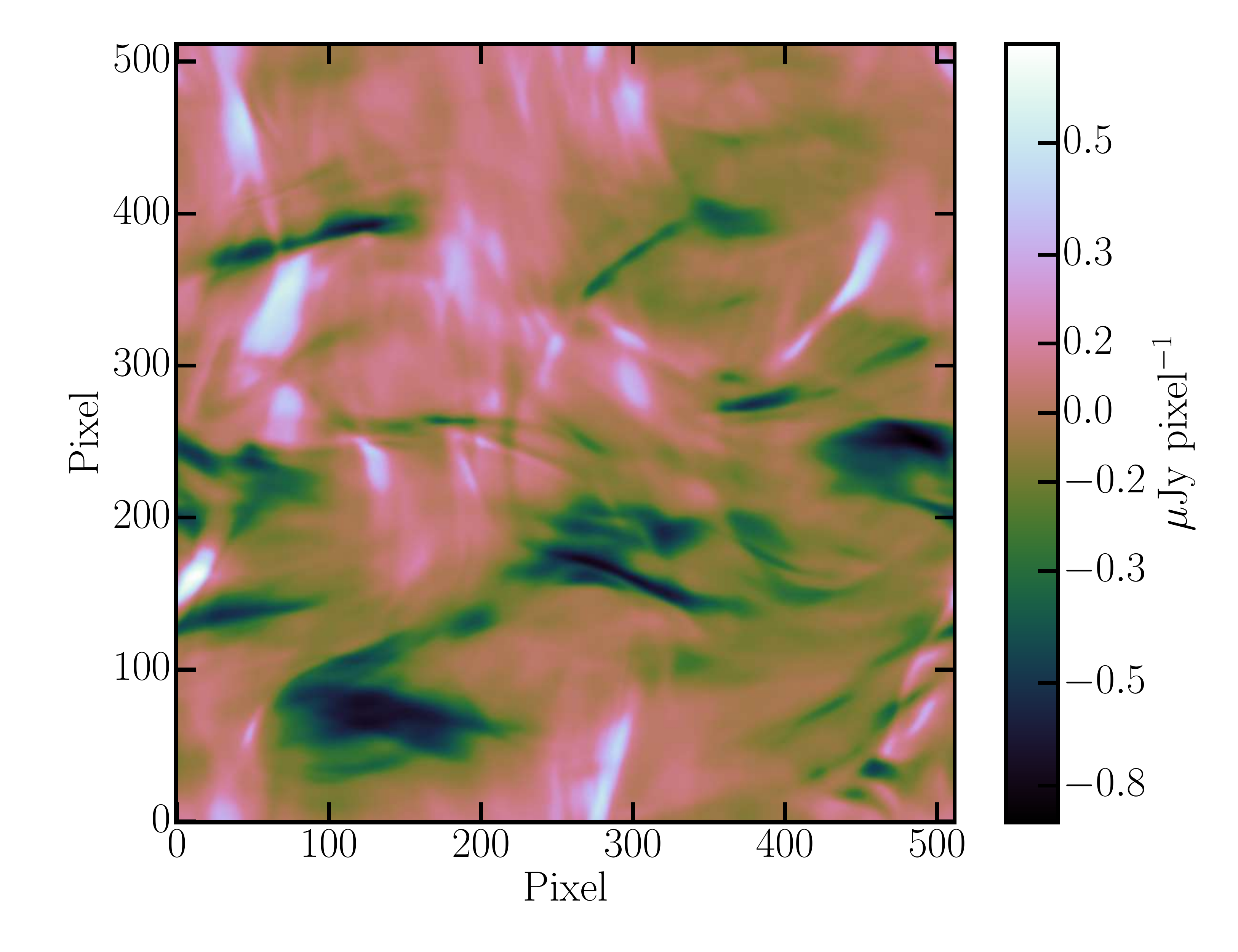}}}\\
 & & \\
 & \large{Stokes $U$} &  \\
{\mbox{\includegraphics[height=4cm, trim=1mm 2mm 2mm 0mm, clip]{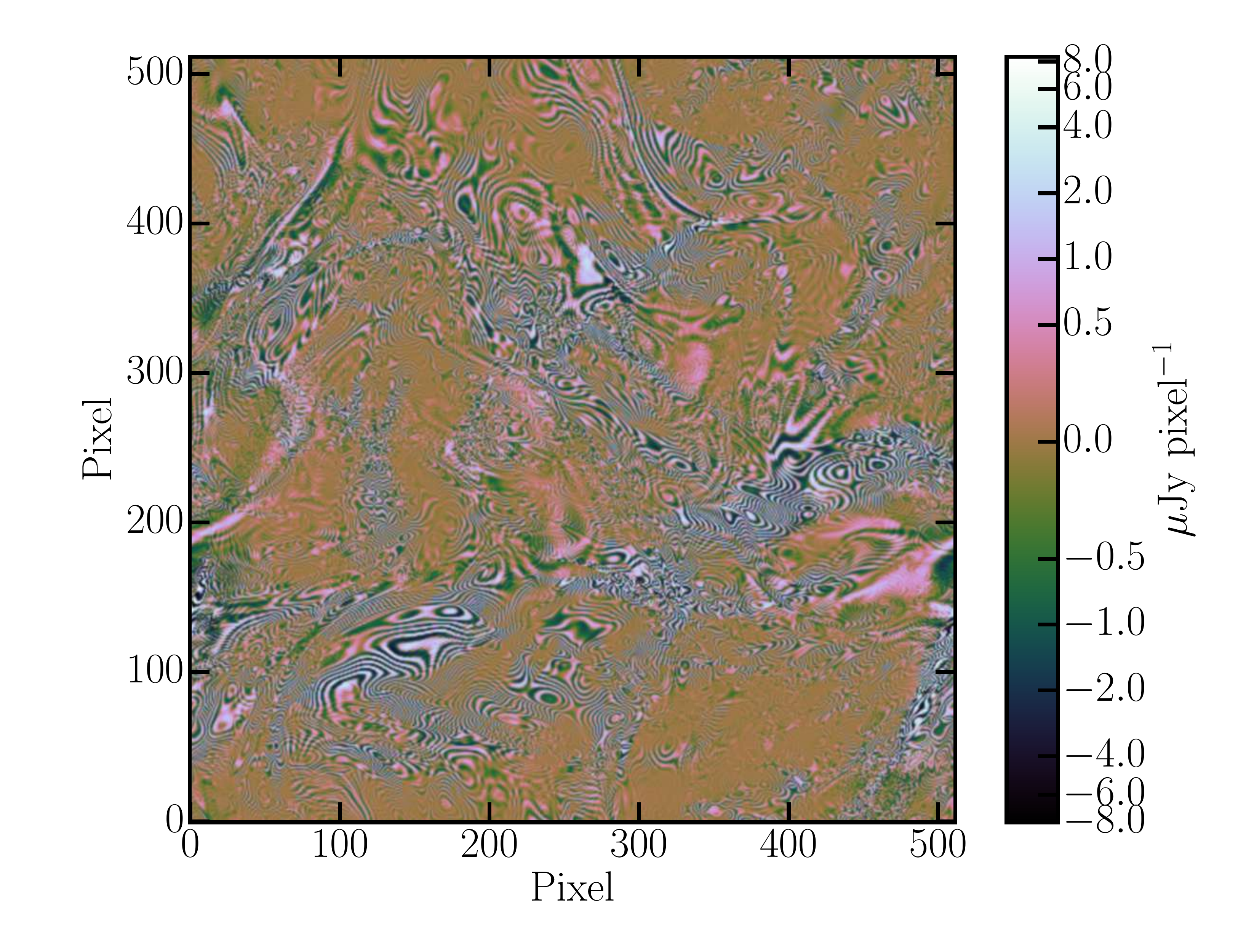}}}&
{\mbox{\includegraphics[height=4cm, trim=1mm 2mm 2mm 0mm, clip]{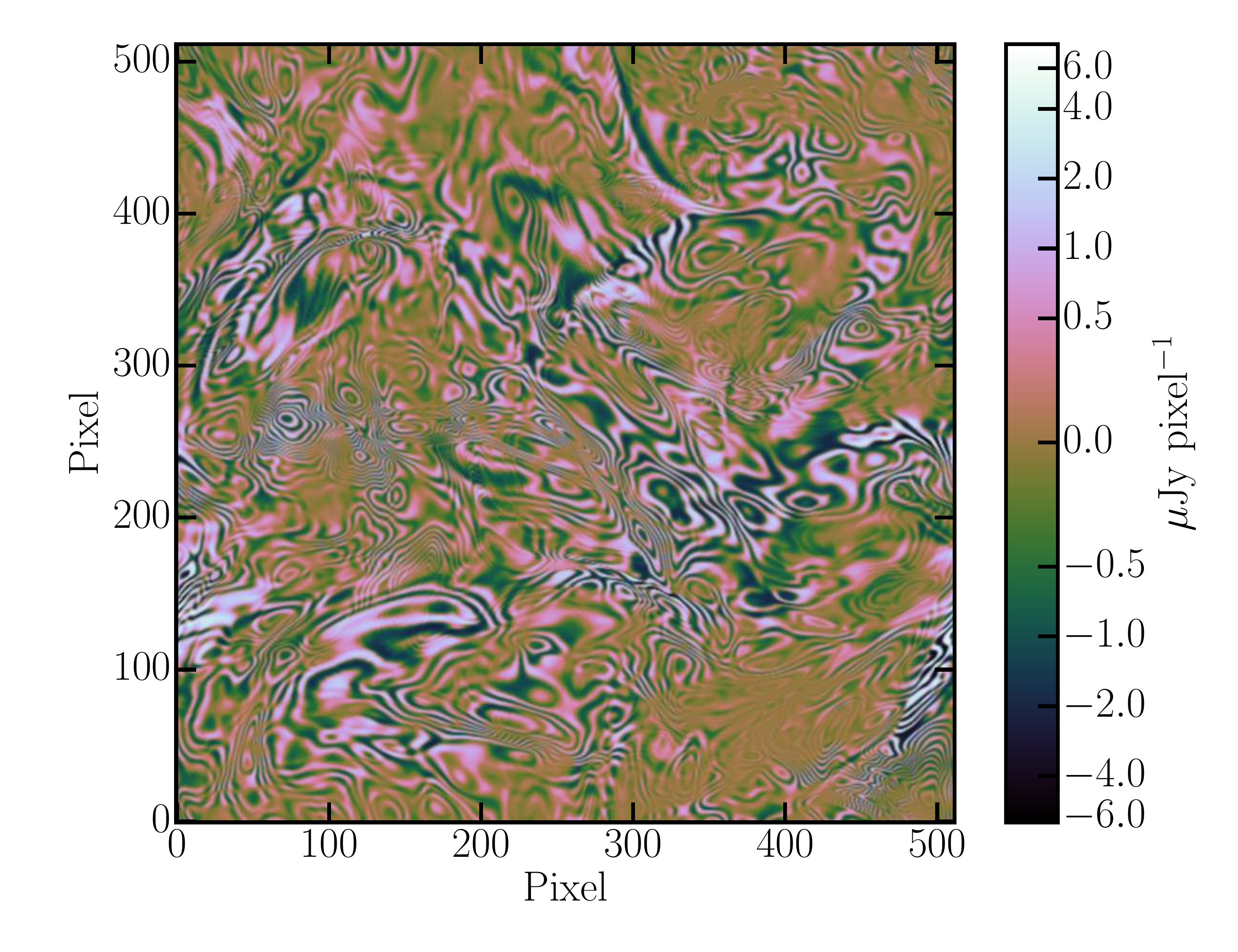}}}&
{\mbox{\includegraphics[height=4cm, trim=1mm 2mm 2mm 0mm, clip]{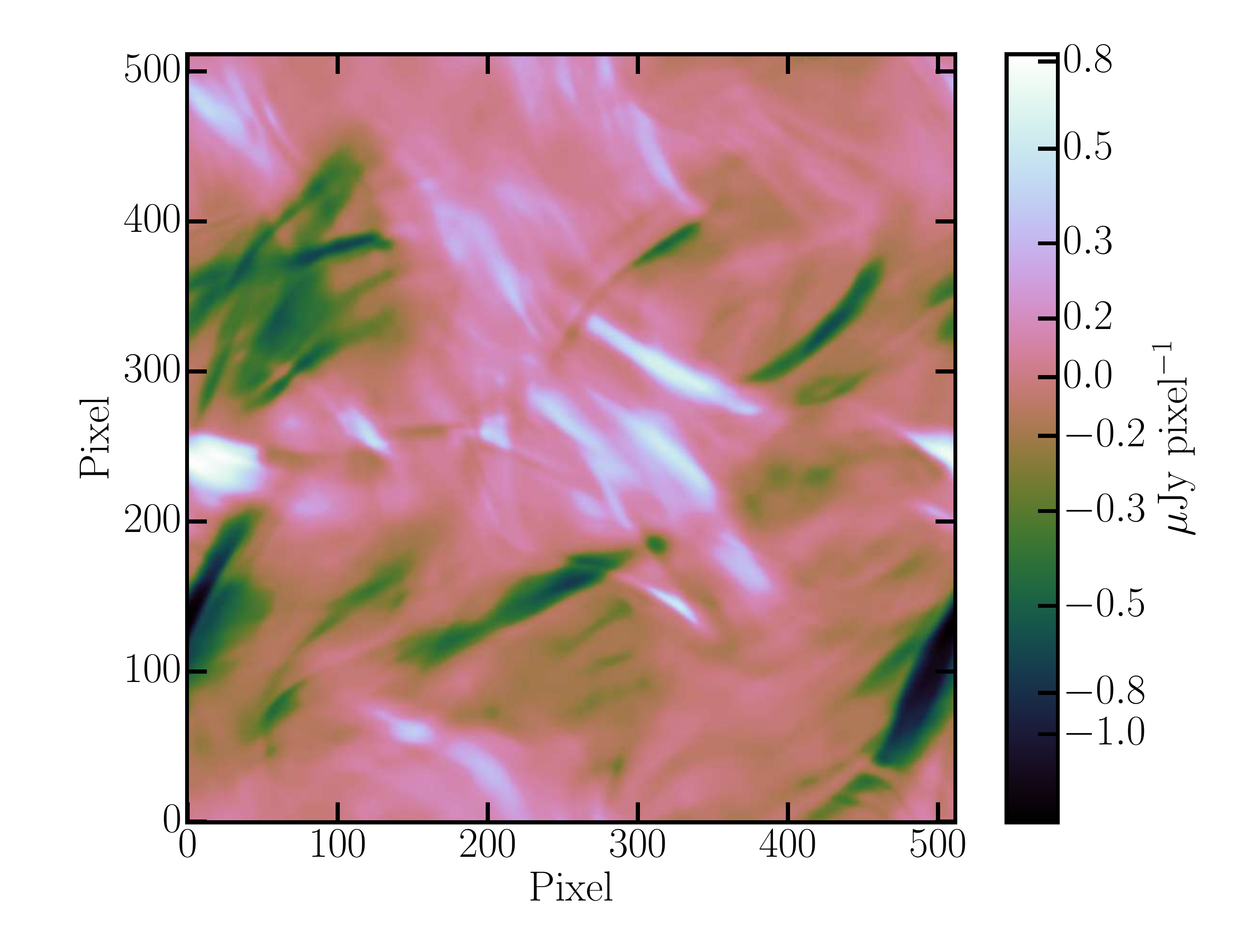}}}\\
 & \large{Stokes $Q$ (smoothed 10 pixels)} &  \\
{\mbox{\includegraphics[height=4cm, trim=1mm 2mm 2mm 0mm, clip]{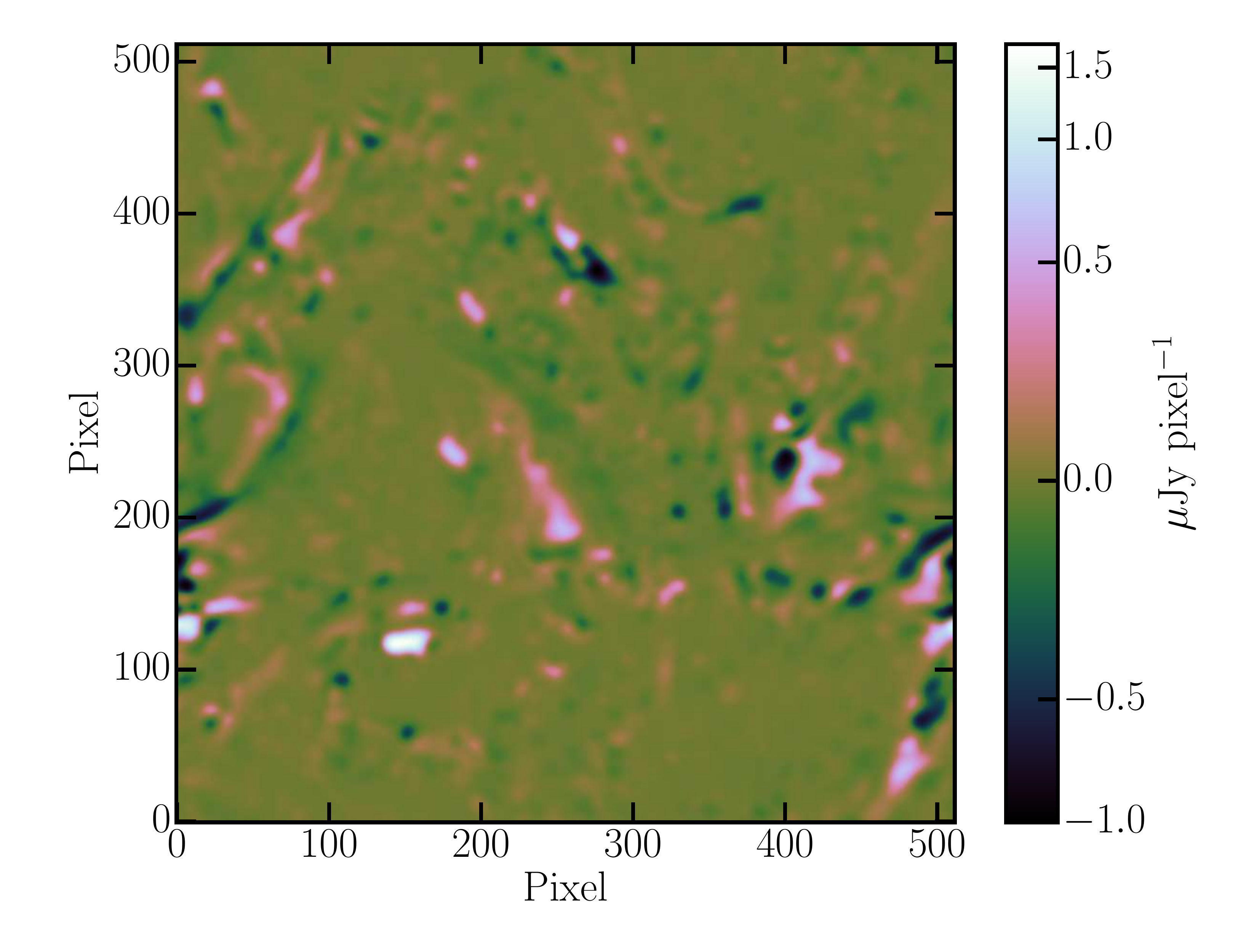}}}&
{\mbox{\includegraphics[height=4cm, trim=1mm 2mm 2mm 0mm, clip]{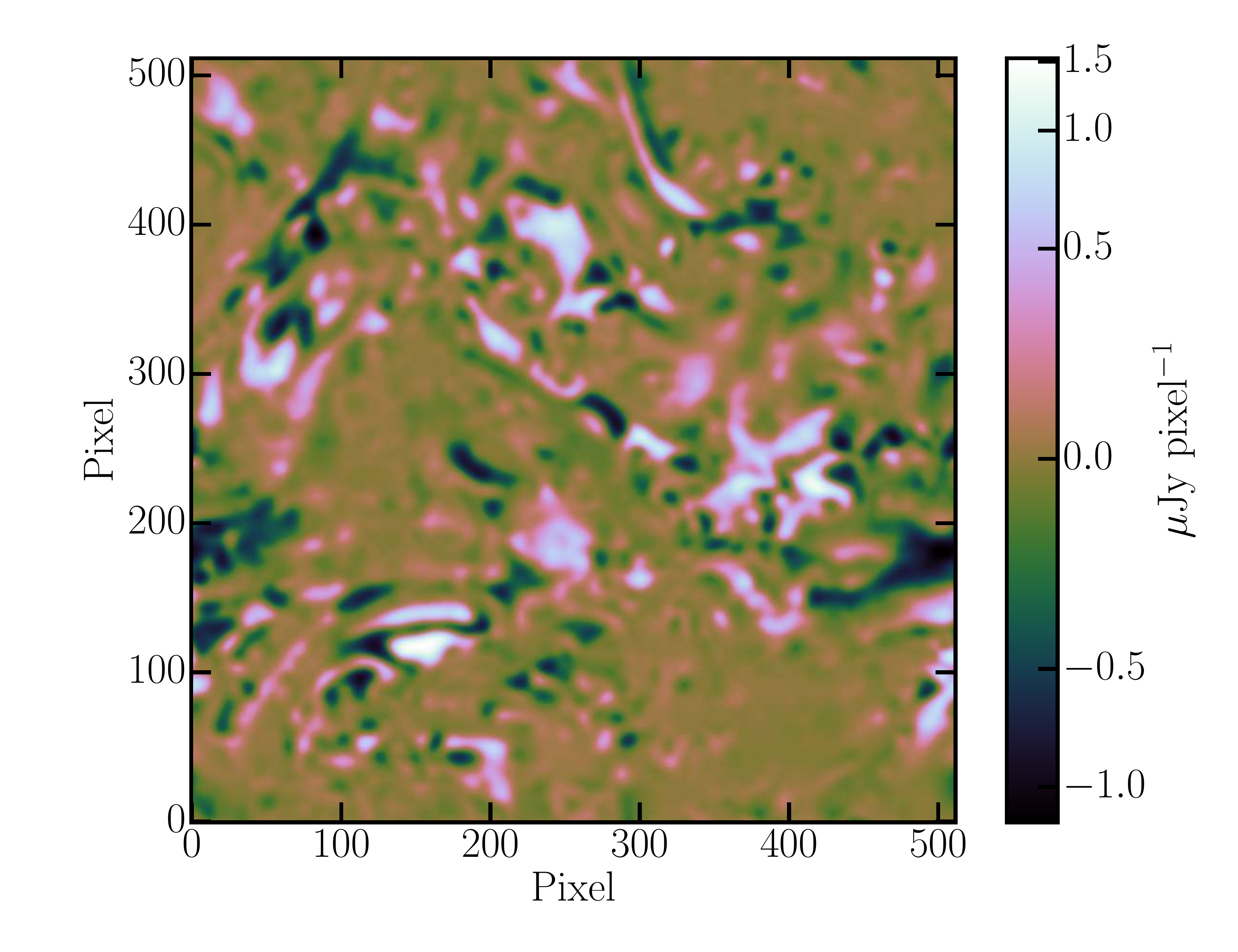}}}&
{\mbox{\includegraphics[height=4cm, trim=1mm 2mm 2mm 0mm, clip]{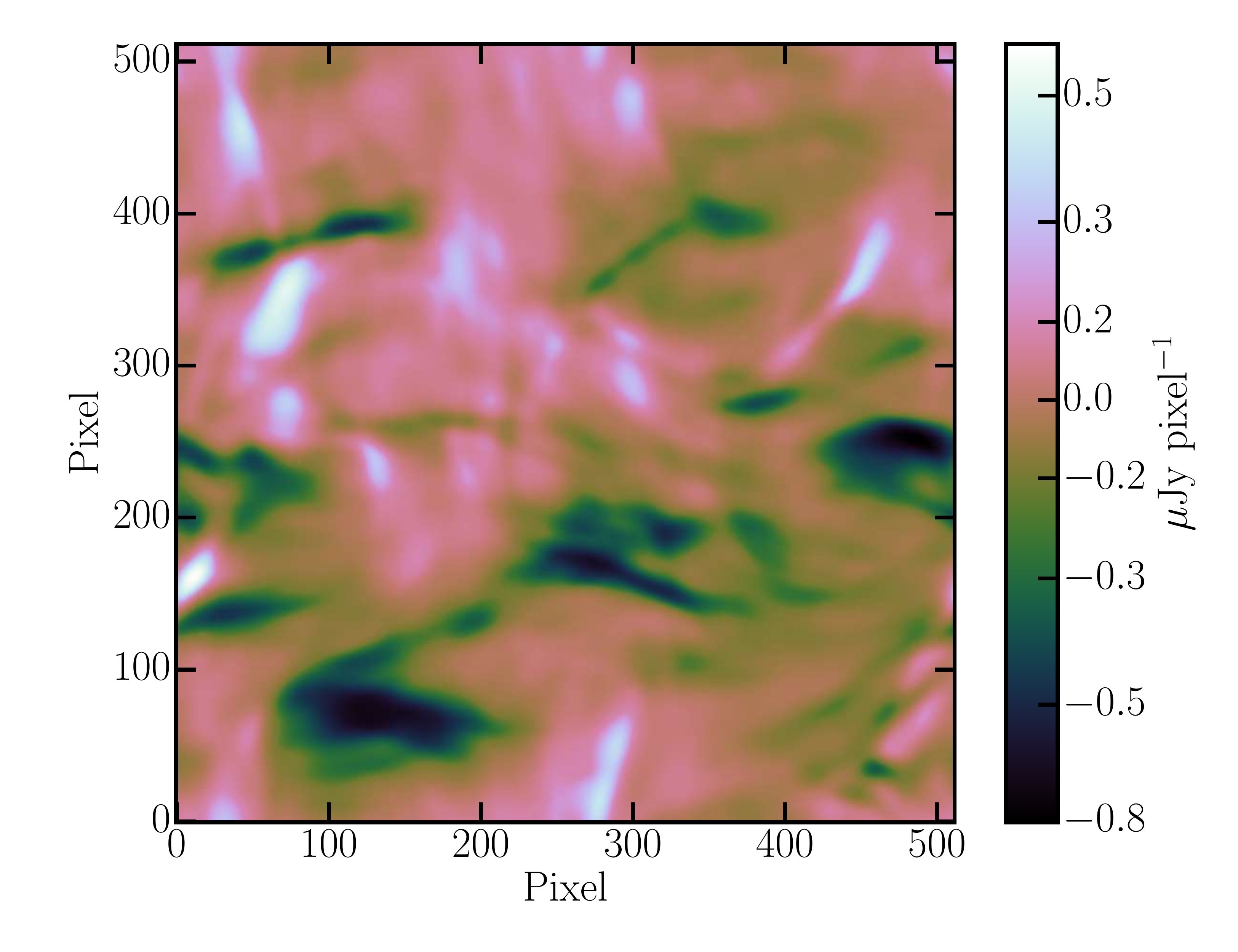}}}\\
 & & \\
 & \large{Stokes $U$ (smoothed 10 pixels)} &  \\
{\mbox{\includegraphics[height=4cm, trim=1mm 2mm 2mm 0mm, clip]{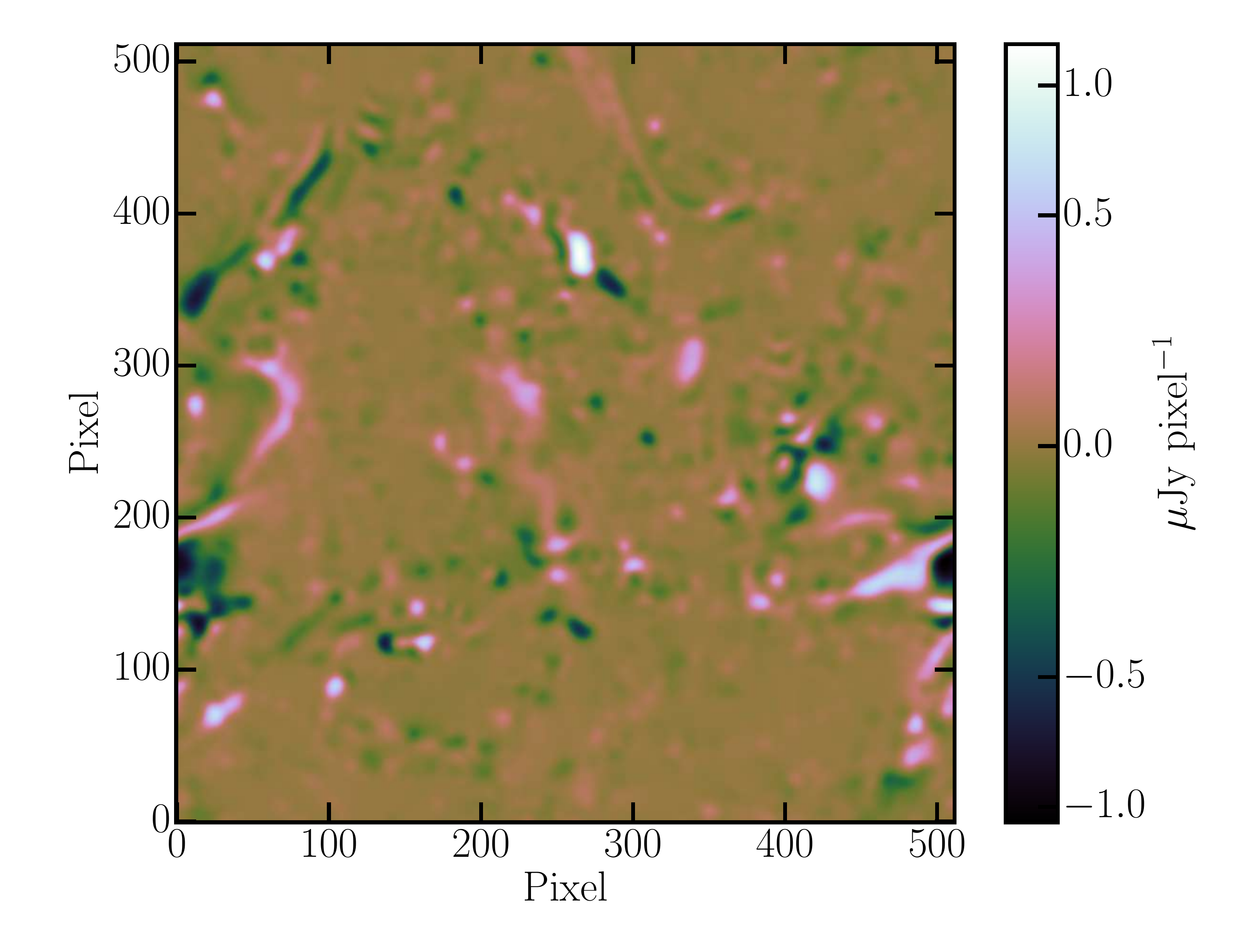}}}&
{\mbox{\includegraphics[height=4cm, trim=1mm 2mm 2mm 0mm, clip]{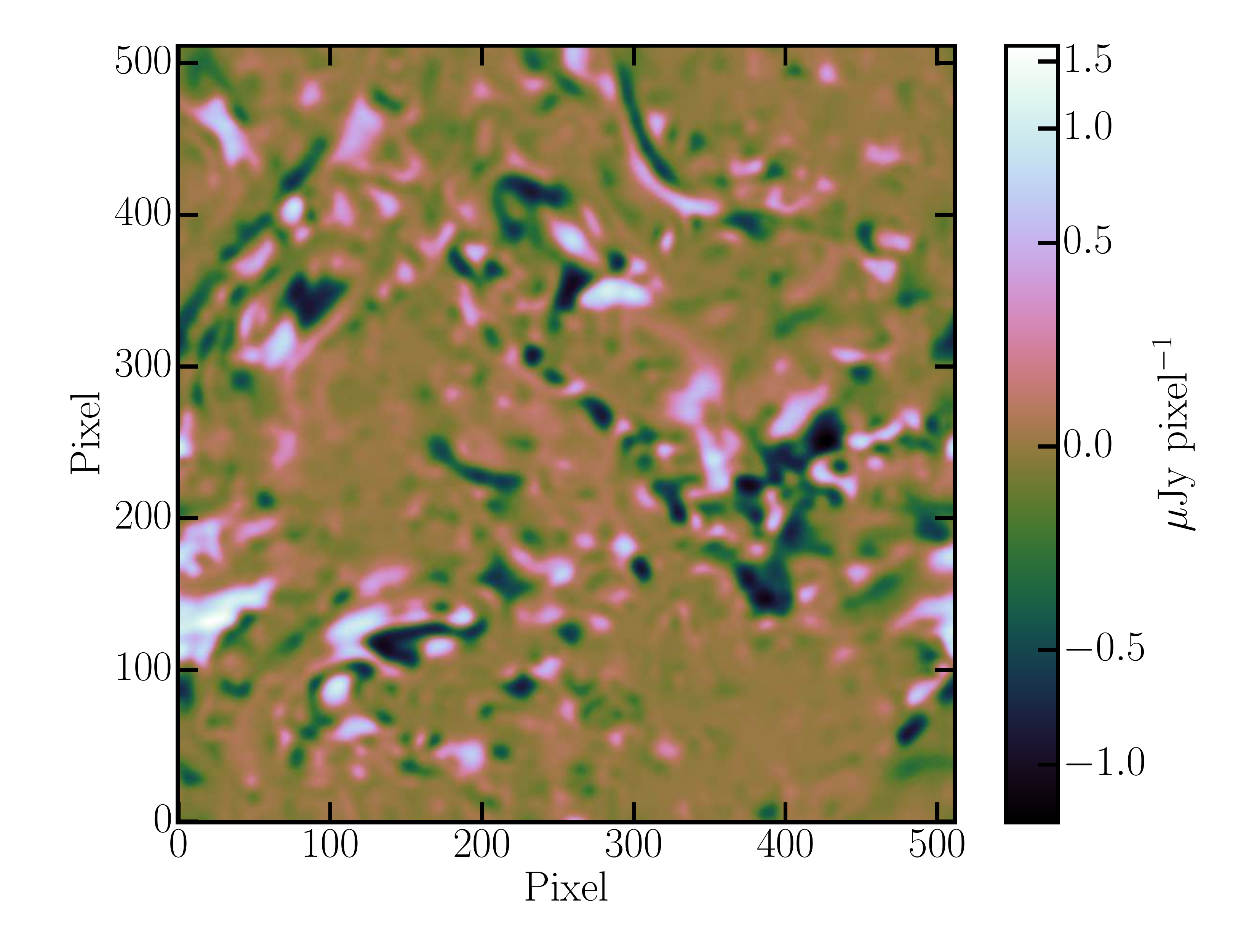}}}&
{\mbox{\includegraphics[height=4cm, trim=1mm 2mm 2mm 0mm, clip]{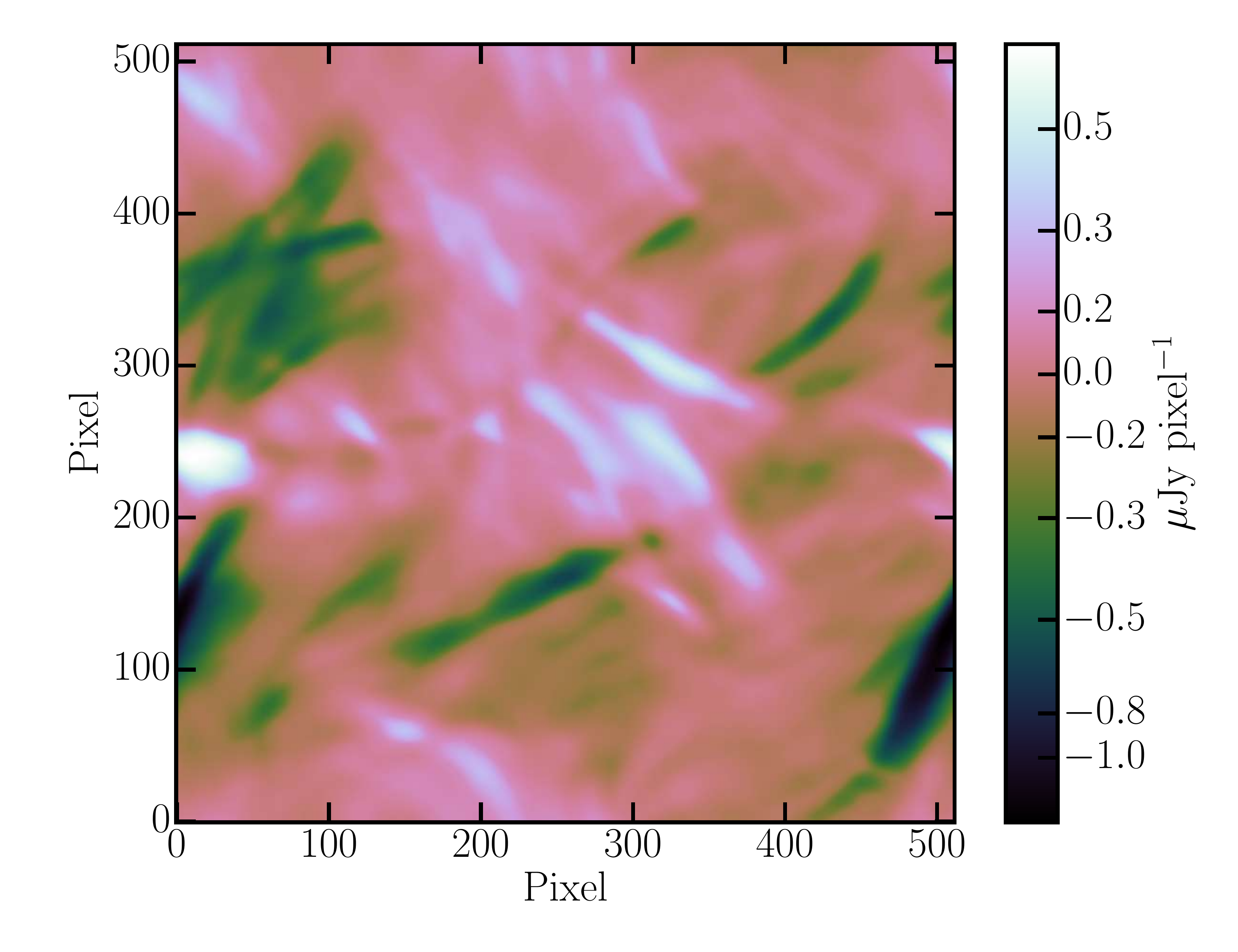}}}\\
\end{tabular}
\caption{First row: Stokes~$Q$ parameter maps in Jy~pixel$^{-2}$. Second row: 
Stokes~$U$ parameter maps in Jy~pixel$^{-2}$. Third and fourth rows show the 
quantities after smoothing by a Gaussian kernel with FWHM 10\,pixels. The left-hand, 
middle, and right-hand columns are at 0.5, 1, and 6~GHz, respectively, at $t/t_{\rm ed} = 23$.}
\label{fig:qu_subsonic}
\end{figure*}

In order to obtain the smoothed intensity map of linear polarization presented 
in Fig.~\ref{fig:pol_subsonic}, we first convolved the Stokes $Q$ and $U$
parameter maps with a symmetric Gaussian kernel having an FWHM of 
$10\times10$~pixel$^2$; the smoothed ${\rm PI}_\nu$ (${\rm PI}_{\rm \nu, smooth}$) 
is computed as
\EQ
PI_{\rm \nu,smooth} = \sqrt{Q_{\rm \nu, smooth}^2 + U_{\rm \nu, smooth}^2}.
\EN
The smoothed fractional polarization ($p_{\rm \nu, smooth}$) maps presented in
Fig.~\ref{fig:pol_subsonic} is obtained using $p_{\rm \nu, smooth} = {\rm PI}_{\rm
\nu, smooth}/I_{\rm \nu, smooth}$, where $I_{\rm \nu, smooth}$ is the smoothed
synchrotron intensity map. The chosen kernel size corresponds to Gaussian
smoothing on a 4.25~kpc spatial scale. Smoothed images of Stokes $Q$ and 
$U$ parameters, $Q_{\rm \nu, smooth}$ and $U_{\rm \nu, smooth}$, respectively, 
at 0.5, 1 and 6~GHz are presented in the bottom two rows of Fig.~\ref{fig:qu_subsonic}.

At $6\ghz$, smoothing by a telescope beam does not significantly affect the
structures observed in Stokes $Q$ and $U$ parameter images at native
resolution. This is due to the presence of large-scale features in the maps 
of Stokes $Q$ and $U$ parameters at frequencies above $\sim5\ghz$. 
However, at frequencies below $\sim1.5\ghz$, increased Faraday rotation 
due to turbulent fields leads to rapid spatial fluctuations in the sign of the 
Stokes parameters. This results in strong depolarization within the beam 
that increases with decreasing frequency (see left-hand and middle panels 
in the bottom two rows of Fig.~\ref{fig:qu_subsonic}). Hence, in the presence 
of noise, typical observations would be sensitive to the clumpy bright and 
dark patches only. These correspond to the clumpy structures observed in 
the ${\rm PI}_\nu$ and $p_\nu$ maps shown in the bottom two rows of
Fig.~\ref{fig:pol_subsonic}.

As seen in the left-hand panels of Fig.~\ref{fig:qu_subsonic}, large Faraday
rotation at frequencies near $0.5\ghz$ introduces fluctuations in Stokes $Q$
and $U$ parameters on scales of few pixels (also seen in
Fig.~\ref{fig:spec_qu_sub}). Therefore, the structures in Stokes $Q$ and 
$U$ maps that are recovered when observed with a finite telescope 
resolution at low frequencies are drastically different as compared to their 
intrinsic structures, e.g. those seen at native resolution or at frequencies 
above $5\ghz$.

From this study it is obvious that depolarization within the beam introduced by
Faraday rotation plays a very severe role in distorting the polarized intensity
structures in the ICM as compared to beam depolarization due to turbulent
magnetic fields. It should be noted that here we have investigated smoothing
on a spatial scale of only 4.25~kpc, which is much smaller than the driving
scale of turbulence of 256~kpc in these simulations. With currently available
radio interferometers, the spatial resolutions that are typically achieved in
radio continuum observations of the ICM are $\sim30\textrm{--}50$~kpc. 
This will further aggravate the difficulty in the detection of polarized emission 
from ICM at frequencies below $\sim1.5\ghz$. Hence, to maximize the chances 
of detecting polarized emission from ICM and robustly glean information on
the nature of turbulence in it, our study shows that it is imperative to reduce
the contamination from Faraday rotation by observing at frequencies higher 
than $5\ghz$.

\bsp

\label{lastpage}

\end{document}